\ifx\PhdMode\undefined  

\newcommand{\compileBody}{True}
\newcommand{\compileAppendix}{True}

\documentclass[10pt]{IEEEtran}  

\newcommand{\onlyphd}[1]{}
\newcommand{\onlypaper}[1]{#1}
\newcommand{\excluded}[1]{}
\newcommand{\onlyfull}[1]{#1}
\newcommand{\onlyconf}[1]{}

\usepackage{url}
\usepackage{amsmath,amsthm, amssymb, epsfig, verbatim, color, amsfonts} 
\usepackage{graphicx}
\usepackage{ifpdf} 
\usepackage{enumerate}

\usepackage{xr} 
\ifx\theorem\undefined
\theoremstyle{plain}
\newtheorem{theorem}{Theorem}
\newtheorem{lemma}{Lemma}
\newtheorem{corollary_in_theorem}{Corollary}[theorem]
\theoremstyle{definition}
\newtheorem{definition}{Definition}
\newtheorem{example}{Example}
\fi

\def\vr{\mathbf}

\def\Pr{\mathrm{Pr}}

\def\Ber{\mathrm{Ber}}

\def\const{\mathrm{const}}

\def\Remp{{R_{\mathrm{emp}}}}

\def\Ind{\mathrm{Ind}} 
\def\Ber{\mathrm{Ber}}

\def\half{\tfrac{1}{2}}

\def\msg{\mathrm{\mathbf{m}}} 
\def\msg{\mathbf{m}} 

\def\endofproof{\hspace{\stretch{1}}$\Box$}
\def\defeq{\triangleq} 

\def\unif{\mathbb{U}}

\newcommand{\tsubs}[1]{{\scriptscriptstyle \mathrm{#1}}}

\def\ntoinfty{\arrowexpl{n \to \infty}}

\newcommand{\arrowexpl}[1] {\underset{#1}{\textstyle \longrightarrow}}

\ifx\note\undefined
\newenvironment{note}{\comment}{\endcomment} 
\fi

\newcommand{\selector}[2]{\onlypaper{#1}\onlyphd{#2}} 

\newenvironment{inputpath}[1]
{ \let\origtexinput\input \renewcommand{\input}[1]{\origtexinput{#1/##1}}
\let\origtexincludegraphics\includegraphics \renewcommand{\includegraphics}[2][]{\origtexincludegraphics[##1]{#1/##2}} }
{ \let\input\origtexinput \let\includegraphics\origtexincludegraphics}

\newcommand{\flagTrue}{True}

\allowdisplaybreaks[1]

\def\Cifb{C_\tsubs{IFB}}
\title{Universal communication part I: modulo additive channels}

\author{Yuval Lomnitz and Meir Feder~\IEEEmembership{Fellow,~IEEE}}

\begin{document}
\maketitle

\begin{abstract}
Which communication rates can be attained over a channel whose output is an unknown (possibly stochastic) function of the input that may vary arbitrarily in time with no a-priori model? Following the spirit of the finite-state compressibility of a sequence, defined by Lempel and Ziv, a ``capacity'' is defined for such a channel as the highest rate achievable by a designer knowing the particular relation that indeed exists between the input and output for all times, yet is constrained to use a fixed finite-length block communication scheme without feedback, i.e. use the same encoder and decoder over each block. In the case of the modulo additive channel, where the output sequence is obtained by modulo addition of an unknown individual sequence to the input sequence, this capacity is upper bounded by a function of the finite state compressibility of the noise sequence. A universal communication scheme with feedback that attains this capacity universally, without prior knowledge of the noise sequence, is presented.
\end{abstract}


\fi 
\ifx\compileBody\flagTrue 

\onlypaper{
\section{Introduction}
Consider the problem of communicating over a channel, where the (possibly stochastic) relation between the input and output is unknown to the transmitter and the receiver and may be, in general, non stationary. In particular, no assumption is made that the channel behavior up to a certain point in time indicates anything about its expected behavior from this time on. The key characteristic of such a channel is that the channel law cannot be learned, i.e. it is impossible, using an asymptotically short measurement period, to obtain the channel probability law and use it during the rest of the transmission.

Clearly, communication over such an arbitrary channel is challenging. Furthermore, even the question what the limits of such communication are, is not well posed. To emphasize the fact that the relation between input and output is a function of the entire sequences, or vectors, this channel shall be termed a \emph{vector} channel.
A simple example of such a channel, which was discussed by Shayevitz and Feder \cite{Ofer_ModuloAdditive} is the modulo-additive channel with an individual noise sequence, defined by the relation $\vr y = \vr x + \vr z$ where $\vr x, \vr y, \vr z \in \mathcal{X}^n$ are $n$-length vectors, denoting the input, output and the noise sequence, $\mathcal{X}$ is a finite alphabet, the `$+$' denotes modulo addition over $\mathcal{X}$, and the sequence $\vr z$ is arbitrary and unknown. The main focus in the current paper is on this channel model. When the alphabet is $\mathcal{X}=\{0,1\}$ this channel is referred to as the binary additive channel.

In the general vector channel, when the conditional probability of the output vector given the input vector is \emph{known}, the classical Shannon capacity, i.e. the maximum communication rate achievable with an arbitrarily small error probability, is well defined. The Shannon capacity of the general causal vector channel was given by Han and Verd\'u \cite{HanVerdu}. When the channel is \emph{unknown}, the same communication rate is in many cases not attainable. In this case, the compound channel or arbitrarily varying channel (AVC) frameworks \cite{Lapidoth_AVC} may be used. In these frameworks, the capacity is defined as the maximum rate of transmission which guarantees robust communication over all possible channels. However these frameworks do not consider the ability to use feedback to adjust the communication parameters, and are therefore worst-case in nature. On the other hand, Shayevitz and Feder \cite{Ofer_ModuloAdditive} have shown that for the modulo-additive channel with an individual noise sequence, by using feedback to adapt the transmission rate to the actual channel occurrence, these worst case assumptions may be alleviated. These results were extended by the authors and others \cite{Eswaran, YL_individual_full}.

Since the channel is unknown, the target is to find a universal communication system that operates without knowing the channel. While there are known universal source encoders \cite{LZ78} and universal predictors \cite{FederMerhav98}, in the communication problem, the term ``universality'' had been used mainly with respect to decoders, competing against the maximum likelihood decoder in a compound channel \cite{Lapidoth_AVC,FederLapidoth_UnivDecod98}, and there is currently no notion of universality with respect to the complete communication system. This is since in the traditional AVC model, feedback is not considered and therefore the encoder is assumed to be fixed. On the other hand, in existing works that consider adaptation of the communication rate using feedback \cite{Ofer_ModuloAdditive, Eswaran, YL_individual_full}, the communication rates achieved do not have a strong justification. For example, these works define the rate using zero-order empirical distributions, and higher rates could be attained by considering empirical distributions with memory.

Let us denote by $P_\tsubs{Y|X}^{(\theta)}$ a conditional distribution of the channel output given the input defining a vector channel, where $\theta$ is an index belonging to a possibly infinite index set $\Theta$. Given a class of vector channels $\{P_\tsubs{Y|X}^{(\theta)}\}_{\theta \in \Theta}$, the objective is to assign a rate $C_{\theta}$ to each channel, such that on one hand $C_{\theta}$ has an operational meaning, for example the maximum rate achievable under certain constraints, and on the other hand, it would be possible to construct a universal system using feedback, that without knowledge of $\theta$, attains a rate of at least $C_{\theta}$ for all $\theta$. The difference from the AVC or compound channel models is that the communication rate depends on $\theta$. As shall be seen, the maximum rate achievable by block encoders and decoders that know $\theta$ is a reasonable target, that can be used as a definition for $C_{\theta}$. This target rate is universally achievable for the class of modulo-additive channels, as shown below. More generally, this target is universally achievable for arbitrary channels with fading memory, as shown in a follow up paper \cite{YL_UnivCommMemory}, using more elaborate tools. On the other hand, this rate is not universally achievable in general.

This main contributions of this paper are as follows. For the general problem of communication over any unknown vector channel, the first definition of competitive universality is given. In particular, the ``iterated finite block capacity'', $C_{\theta}=\Cifb(P_\tsubs{Y|X}^{(\theta)})$ is defined. A significant part of the paper is devoted to exploration of the problem boundaries: For which channel families can the IFB capacity be attained? Which other interesting definitions of the reference class can be given? The other contributions are specific to the modulo-additive channel with an individual noise sequence. A bound on the IFB capacity is given, a universal system that attains the target rate without knowing the channel is presented, and the redundancy in approaching this rate is analyzed. The converse part of the redundancy analysis holds also for larger families of channels that include the modulo-additive channel as a special case.

The paper is organized as follows: in Section~\ref{sec:motivation} the motivations for the definition of $\Cifb$ are explained. Section~\ref{sec:overview} is a high level overview of the results regarding the modulo-additive channel, and the main ideas behind the proofs. Section~\ref{sec:definitions} includes the detailed definitions with some discussion. Section~\ref{sec:univ_modulo_add_capacity} focuses on the modulo additive channel and includes the upper bound on $\Cifb$ and the universal system achieving it. The redundancy, i.e. the convergence rate, in achieving the IFB capacity is explored in Section~\ref{sec:ma_univ_redundancy}. Section~\ref{sec:ma_discussion} is devoted to discussion and comments and suggests some extensions and alternative definitions.
} 

\onlypaper{
\section{Motivation}\label{sec:motivation}

\onlypaper{Let us now discuss the motivations for the definitions of a target rate which is universally achievable. }
The inherent difficulty of defining the maximal communication rates over arbitrary vector channels can be appreciated by considering even the simple example of a binary additive channel $\vr y = \vr x \oplus \vr z$ with an individual noise sequence $\vr z$, where `$\oplus$' denotes modulo-2 addition and all vectors are of length $n$. For every specific individual noise sequence $\vr z$, the capacity of this channel is $1$ bit/use. On the other hand, if the noise sequence is arbitrary and unknown, the AVC capacity \cite{Lapidoth_AVC} is zero. It would initially seem that not much can be done, when the noise sequence is unknown; however it was shown \cite{Ofer_ModuloAdditive} \onlyphd{(see also \S\ref{sec:examples_modadditive}) }that, using feedback and common randomness, and by adapting the decoding rate, a communication rate of $R=1-\hat H(\vr z^n)$ can be achieved, where $\hat H(\cdot)$ denotes the empirical entropy of the noise sequence, i.e. the binary entropy of the empirical cross-over probability. The main idea is that if the empirical channel can be measured and the communication rate can be adapted, then rather than making a-priori pessimistic assumptions, one can opportunistically increase the rate when the noise sequence has a low empirical entropy.

A disturbing fact is that some arbitrariness exists in deciding on the rates to achieve per each channel: in the binary additive channel, given a sequence $\vr s$ of choice, one could also design a system that achieves the rate $1-\hat H(\vr z \oplus \vr s)$, by adding the sequence $\vr s$ to the channel output and then applying Shayevitz and Feder's scheme \cite{Ofer_ModuloAdditive}. Doing so, a rate of $1$ is obtained for the sequence $\vr z = \vr s$, where the original system's rate was $1-\hat H(\vr s)$, and a rate of $1-\hat H(\vr s)$ for the noiseless case $\vr z = 0$, so one may say that the noise sequence $\vr s$ is ``favored'' over $\vr 0$. This demonstrates the arbitrariness in determining which communication rates are possible. To remove this arbitrariness, a reasonable criterion is sought, to decide which channels (noise sequences, in the example) to favor over others.

This issue bears significant resemblance to issues tackled in universal source coding (compression) and in universal prediction. In universal compression, one would like to set a target for the compression rate of an individual sequence. As in the current problem, someone who knows the sequence can design an encoder which compresses it to 1 bit, whereas assuming the sequence is completely unknown and without favoring any sequence over another, no compression can be achieved. There are many possible fixed to variable encoders which are uniquely decodable, and the decision between them may seem arbitrary. One solution proposed by Lempel and Ziv \cite{LZ78} was to set as a target, the compression rates that are achievable by machines with limited capabilities, i.e. finite state machines (FSM). They defined the notion of \emph{finite state compressibility} for an infinite sequence, as the best compression rate that can be achieved by any information lossless FSM operating over the infinite sequence, and had shown that the LZ78 compression algorithm based on incremental parsing, defined there, achieves this compression rate universally for any sequence. This concept supplies a criterion to decide which sequences to favor over others, without assuming a probability law. A similar notion, i.e. that of comparing against the best machine out of a restricted class, is applied in universal prediction \cite{FederMerhav98,FederMerhav93}.

Following this lead, the comparison class is chosen to be the set of fixed finite-length block encoders and decoders, which repeatedly perform the same encoding and decoding operations over blocks of any fixed length (Figure~\ref{fig:iterative_mapping}). This class is a relatively simple one, while still yielding a reasonable criterion to set the communication rate. The \emph{iterated finite block capacity} of an infinite vector channel $\Cifb$ is defined as the supremum of all rates which are reliably achievable by encoders and decoders in the comparison class. This capacity value may be smaller, in general, than the Shannon capacity of the vector channel. This definition has operational significance, since many practical communication systems use block encoding, and therefore universally attaining the $\Cifb$ means that one can design a system which, without any prior knowledge of the channel, is essentially at least as good as any finite block code. The universal system itself does not belong to the comparison class -- it does not operate in fixed blocks, it modifies its behavior based on the past, and it uses feedback.
\onlypaper{Although achieving $\Cifb$ universally is possible for classes of vector channels wider than the modulo-additive channel \cite{YL_UnivCommMemory}, it is not possible to attain this rate for general unknown vector channels.
}

\onlyphd{
As will be seen below, achieving $\Cifb$ universally is not possible for general unknown vector channels (due to ``password'' channels \S\ref{sec:IFB_capacity_discussion}).
An alternative formulation, termed the ``arbitrary-finite-block'' (AFB) capacity is proposed as well. The AFB capacity can be attained for any channel. The rationale behind this definition is explained later on (\S\ref{sec:IFB_capacity_discussion}).
}

The \selector{IFB}{IFB/AFB} model presented here is simple and intuitive, however it has several drawbacks and alternative definitions and extensions can be proposed. These are discussed in \selector{Section~\ref{sec:alt_definitions}}{the next chapters (Sections~\ref{sec:alt_definitions},\ref{sec:um_alt_comparison_class}), after the implications are better understood}. Most notably, Misra and Weissman \cite{MisraPorosityISIT12} generalized the \selector{current results}{main results of Chapter~\ref{chap:univ_modadditive}} to finite-state communication systems with feedback. For the sake of simplicity, \selector{the paper}{this thesis} focuses on the basic model of reference systems using block coding.

Although the results are currently purely theoretical \selector{(see Section~\ref{sec:ma_univ_redundancy})}{(see \S\ref{sec:ma_univ_redundancy} and \S\ref{sec:um_discussion_asymptotics})}, they supply motivation for using competitive universality in communication.

}

\onlypaper{
\section{Overview of the main results}\label{sec:overview}

\onlyphd{
Section~\ref{sec:univ_modulo_add_capacity} shows that there exists an IFB-universal system for the modulo additive channel. In Section~\ref{sec:IFB_UB} it is shown that $\Cifb \leq (1 - \rho(\vr z)) \cdot  \log |\mathcal{X}|$, where $\rho(\vr z)$ is the finite state compressibility of $\vr z$ (as defined by Lempel and Ziv \cite{LZ78}). The main tool in the proof is the ``collapsed'' channel (Fig.\ref{fig:collapsed_channel_modadditive}), created by choosing the noise sequence over blocks uniformly. In Section~\ref{sec:ma_adaptive_rho} it is shown that, with common randomness and feedback, a universal system can asymptotically attain this rate without prior knowledge of the noise sequence. In order to show attainability, the result of Theorem~\ref{theorem:adaptive_L} from Section~\ref{sec:examples_compression_attanability} is used. It was shown there (when specializing the aforementioned result to the modulo-additive channel \S\ref{sec:examples_compression_modadditive}), that the rate function $\log | \mathcal{X} | - \frac{1}{n} L(\vr z)$ is asymptotically adaptively achievable, where $L(\vr z)$ is the compression length of the noise sequence. The result is obtained by plugging the encoding lengths $L(\vr z)$ determined by the LZ78 source encoder, whose the compression ratios asymptotically approach the finite state compressibility.\footnote{A more direct proof of this result, which does not go through the chain of theorems of Part-I, and a simplified intuitive explanation, are given in the paper \cite{YL_UnivModuloAdditive}.}
} 

\onlypaper{
This section provides an informal review of the main results and rough proof outlines. The purpose is to provide an understanding of the results without diving into mathematical detail. \onlypaper{The main results of this paper pertain to the modulo-additive channel with individual noise sequence.}
\selector{For this channel, it}{It} is shown in Section~\ref{sec:univ_modulo_add_capacity} that $\Cifb \leq (1 - \rho(\vr z)) \cdot  \log |\mathcal{X}|$, where $\rho(\vr z)$ is the finite state compressibility of the infinite sequence $\vr z$, as defined by Lempel and Ziv \cite{LZ78}. Assuming that common randomness exists and that there is a feedback link, a universal system employing feedback exists, which asymptotically attains this rate universally without prior knowledge of the noise sequence. In Section~\ref{sec:ma_univ_redundancy}, upper and lower bounds on the convergence rate are derived. Below, the main ideas in the proofs are described.

Let us begin with the upper bound on $\Cifb$. Suppose a given the reference system comprised of an encoder and a decoder, achieves the rate $R$ over $b$ blocks of size $k$ (Figure~\ref{fig:iterative_mapping}). During these $b$ blocks, the reference system ``sees'' $b$ different noise vectors of length $k$, namely $\vr z_{(i-1)k+1}^{ik}$, $i=1,\ldots,b$. Since the system is fixed during these $b$ blocks, this is equivalent to operating over a stochastic channel, where the noise vector $\tilde{\vr Z}$ is chosen uniformly from the set of these vectors, with probability $\frac{1}{b}$ for each. This random vector is termed the ``collapsed'' noise sequence, and the channel generated from it the ``collapsed'' channel (Figure~\ref{fig:collapsed_channel_modadditive}). The standard converse of the channel capacity theorem, without the assumption of a memoryless channel, can be applied to the collapsed channel, and yields an upper bound on $\Cifb$, which is roughly $\log |\mathcal{X}| - \frac{1}{k} H(\tilde{\vr Z})$. The entropy $H(\tilde{\vr Z})$ is lower bounded using the finite state compressibility of the sequence, since a finite state machine may achieve a compression rate close to the entropy by standard block-to-variable coding, where the code lengths are tuned to the statistics of the collapsed noise vector. Combining these bounds yields the result $\Cifb \leq (1 - \rho(\vr z)) \cdot  \log |\mathcal{X}|$ (Theorem \ref{theorem:Cifb_UB}).

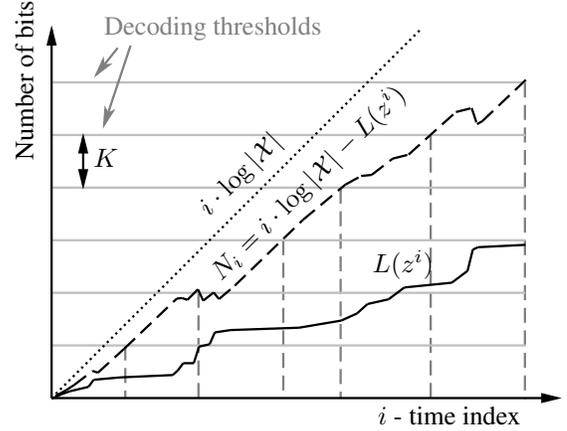
\begin{figure}
\centering
\ifpdf
  \setlength{\unitlength}{1bp}%
  \begin{picture}(216.44, 192.04)(0,0)
  \put(0,0){\includegraphics{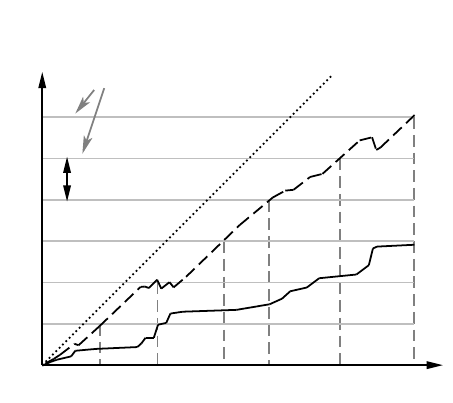}}
  \put(79.67,86.32){\rotatebox{45.00}{\fontsize{9.96}{11.95}\selectfont \smash{\makebox[0pt][l]{$i \cdot \log |\mathcal{X}|$}}}}
  \put(84.22,61.78){\rotatebox{45.00}{\fontsize{9.96}{11.95}\selectfont \smash{\makebox[0pt][l]{$N_i = i \cdot \log |\mathcal{X}| - L(z^i)$}}}}
  \put(141.32,65.36){\fontsize{9.96}{11.95}\selectfont $L(z^i)$}
  \put(143.30,7.81){\fontsize{9.96}{11.95}\selectfont  $i$ - time index}
  \put(13.45,107.59){\rotatebox{90.00}{\fontsize{9.96}{11.95}\selectfont \smash{\makebox[0pt][l]{Number of bits}}}}
  \put(40.12,154.65){\fontsize{9.96}{11.95}\selectfont \textcolor[rgb]{0.50196, 0.50196, 0.50196}{Decoding thresholds}}
  \put(36.15,105.04){\fontsize{9.96}{11.95}\selectfont $K$}
  \end{picture}%
\else
  \setlength{\unitlength}{1bp}%
  \begin{picture}(216.44, 192.04)(0,0)
  \put(0,0){\includegraphics{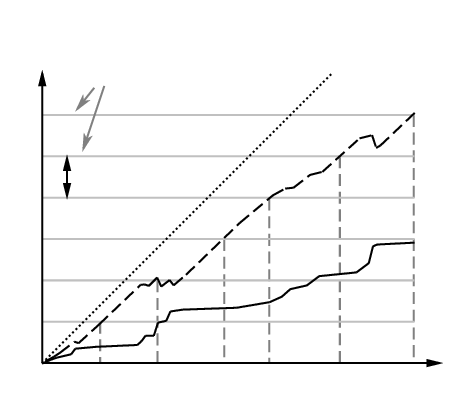}}
  \put(79.67,86.32){\rotatebox{45.00}{\fontsize{9.96}{11.95}\selectfont \smash{\makebox[0pt][l]{$i \cdot \log |\mathcal{X}|$}}}}
  \put(84.22,61.78){\rotatebox{45.00}{\fontsize{9.96}{11.95}\selectfont \smash{\makebox[0pt][l]{$N_i = i \cdot \log |\mathcal{X}| - L(z^i)$}}}}
  \put(141.32,65.36){\fontsize{9.96}{11.95}\selectfont $L(z^i)$}
  \put(143.30,7.81){\fontsize{9.96}{11.95}\selectfont  $i$ - time index}
  \put(13.45,107.59){\rotatebox{90.00}{\fontsize{9.96}{11.95}\selectfont \smash{\makebox[0pt][l]{Number of bits}}}}
  \put(40.12,154.65){\fontsize{9.96}{11.95}\selectfont \textcolor[rgb]{0.50196, 0.50196, 0.50196}{Decoding thresholds}}
  \put(36.15,105.04){\fontsize{9.96}{11.95}\selectfont $K$}
  \end{picture}%
\fi
\caption[The decoding rule of the compression based rate adaptive system]{An illustration of the decoding rule of the rate adaptive system. $L(z^i)$ is the compression length. Decoding thresholds with respect to $N_i=i \cdot \log |\mathcal{X}| - L(z^i)$ are depicted by horizontal lines.}\label{fig:rate_adaptive_nRemp_over_time}%
\end{figure}

Next, a communication scheme is demonstrated, that asymptotically attains the rate $\log | \mathcal{X} | - \frac{1}{n} L(\vr z^n)$, where $L(\vr z^n)$ is the compression length of the sequence $\vr z$ by a given sequential source encoder, and $n$ is the overall block length. The scheme is based on iterative application of rateless coding, sending $K$ bits in each block. Each codeword in the codebook of $\exp(K)$ words is chosen independently and distributed uniformly over $\mathcal{X}^n$. The transmitter sends symbols from the codeword matching the $K$ message bits, until a termination condition occurs at the receiver side. Then, the receiver indicates the end of the block through the feedback link and a new block begins. The termination condition is based on giving a rank to every possible noise sequence $\vr z_1^i$. This noise sequence has two parts: the part spanning previous blocks that had already been decoded is known with high probability, as both the channel input and the channel output are known. The part of the noise sequence from the beginning of the current block is unknown, but is related in a one-to-one relation to the unknown message, because the channel output is known. Therefore, there are $\exp(K)$ possible noise sequences, corresponding to the $\exp(K)$ possible codewords sent in the current block. The rank of each noise sequence is the code-length, or the number of bits in its representation by the given source encoder. The decoder terminates the block if for any codeword, this length is smaller than a threshold.

The proof of this scheme's performance is roughly as follows. Due to the random coding, most of the hypotheses, except the true one, yield random noise sequences. These sequences are incompressible, and therefore the number of bits representing the last block would be approximately $\log |\mathcal{X}|$ times the number of symbols in the block. It can be shown that setting the threshold approximately $K$ below this value, guarantees a small probability of exceeding the threshold for any of the $\exp(K)-1$ incorrect codewords, and therefore a small probability of error. It is convenient to define the ``porosity'' of the sequence up to time $i$ as $N_i = i \cdot \log|\mathcal{X}| - L(\hat{\vr z}_1^i)$, representing the gap between the compressibility of the hypothetical noise sequence, and the compressibility of a random sequence. The approximate termination condition may be interpreted as decoding when the value of $N_i$ increases by $K$ from the start of the current block. Since when this occurs, the system starts a new block, there is a correspondence between the increase in $N_i$ and the number of blocks and bits that are transmitted, i.e. the termination condition can be approximately interpreted as $N_i \geq K (b+1)$ where $b$ is the number of blocks so far. Therefore assuming by time $n$, $B$ blocks were transmitted, the number of transmitted bits is $K \cdot B \approx N_n = n \cdot \log|\mathcal{X}| - L(\hat{\vr z}_1^n)$. Assuming no errors occurred $\hat{\vr z}_1^n = \vr z_1^n$, and dividing by $n$ the desired result is obtained. This is depicted in Figure~\ref{fig:rate_adaptive_nRemp_over_time}, where the horizontal axis is the time $i$. The solid line presents $L(\vr z_1^i)$, and the dashed line $N_i$. The decoding thresholds $K b$ ($b=1,2,\ldots$) are depicted as horizontal lines, while the vertical lines depict the decoding times. Decoding occurs whenever $N_i$ crosses a threshold. A random hypothesized sequence in the current block implies that $N_i$ does not increase on average. It can be seen that the number of bits that will be sent is approximately $N_n$. In the full proof, various overheads that were neglected above are accounted for.

To obtain the universal system attaining $\Cifb$ (Theorem~\ref{theorem:adaptive_rho}), the scheme above is applied with the encoding lengths $L(\vr z^n)$ determined by the LZ78 source encoder, whose compression ratios asymptotically approach the finite state compressibility: asymptotically $\frac{1}{n} L(\vr z^n) \leq \rho(\vr z) \log |\mathcal{X}|$, therefore $\log | \mathcal{X} | - \frac{1}{n} L(\vr z^n) \geq (1-\rho(\vr z)) \log |\mathcal{X}|$, where all inequalities are up to asymptotically vanishing factors.
} 

Section~\ref{sec:ma_univ_redundancy} deals with the question of the redundancy, or how quickly the system converges to the rate attained by the best IFB system with a given block length $k$. Unfortunately, it is shown that $n$ must grow at least as fast as $|\mathcal{X}|^k$, approximately. The upper bound on redundancy is obtained by using a similar universal system employing a slightly more refined design: a universal probability assignment based on a mixture of Krichevsky-Trofimov distributions \cite{KrichevskyTrofimov81} is used instead of the LZ78 encoder. The lower bound is obtained by presenting a design of an IFB system together with a random channel, i.e. a distribution over noise sequences, such that the mutual information over the channel is smaller than the rate obtained by the IFB system. This is possible because the IFB system is designed together with the channel and can use the knowledge of the specific noise sequence. On the other hand, the rate obtained by any universal system with feedback is bounded by the mutual information, and this gap comprises the lower bound on redundancy. The upper bound and the lower bound on the redundancy agree in terms of the asymptotical growth rate of $n$ as function of $k$ (see Figure~\ref{fig:mod_additive_redundancy_fig2}).

}

\onlypaper{
\section{Channel model and definitions}\label{sec:definitions}
This section begins the formal presentation of the results, by presenting the channel model and the definitions of the capacity $\Cifb$, and discussing their implications.
} 

\onlypaper{
\subsection{Notation}
Vectors are denoted by boldface letters. Sub-vectors are defined by superscripts and subscripts: $\vr x_j^i \defeq [x_j, x_{j+1}, \ldots, x_i]$. $\vr x_j^i$ equals the empty string if $i<j$. The subscript is sometimes removed when it equals $1$, i.e. $\vr x^i \defeq \vr x_1^i$.

For a vector or random variable $\vr X$, $\vr X_i^{[k]} \defeq \vr X_{(i-1)k + 1}^{(i-1)k + k}$ denotes the $i$-th block of length $k$ in the vector. For brevity, vectors with similar ranges are sometimes joined together, for example, the notation $(\vr X \vr Y)_1^k$ is used instead of $\vr X_1^k \vr Y_1^k$.
Exponents and logs as well as information quantities are base 2. Random variables are distinguished from their sample values by capital letters. The indicator function $\Ind(E)$ where $E$ is a set or a probabilistic event is defined as $1$ over the set (or when the event occurs) and $0$ otherwise.

$h_b(p)$ denotes the binary entropy function, i.e. the entropy of a Bernully-$p$ random variable, and $\rho(\vr z^\infty)$ denotes the finite state compressibility \cite{LZ78} of $\vr z$, defined formally in Section~\ref{sec:IFB_UB}, \eqref{eq:ma444}-\eqref{eq:ma446}.
} 

\onlypaper{
\newcommand{\onlyumem}[1]{} 
\newcommand{\onlyumod}[1]{#1} 

\subsection{Channel model}\label{sec:def_ifb_afb_channel_model}
Let $\vr x$ and $\vr y$ be infinite sequences denoting the input and the output respectively, where each letter is chosen from the alphabets $\mathcal{X}, \mathcal{Y}$ respectively, $x_i \in \mathcal{X}, y_i \in \mathcal{Y}$. Throughout \selector{the current paper}{this part} the input and output alphabets are assumed to be finite. A channel $P_\tsubs{Y|X}$ is defined through the probabilistic relations
$P_\tsubs{Y|X}(\vr y^n | \vr x^\infty) = \Pr(\vr Y^n = \vr y^n | \vr X^\infty = \vr x^\infty)$ for $n=1,2,... \infty$. A finite length output sequence is considered in order to make the probability well defined.
\onlyfull{Sometimes, this probability will be informally referred to as $\Pr(Y_1^\infty | X_1^\infty)$, and should be understood as the sequence of these distributions for $n=1,2,\ldots$.}

\begin{definition}\label{def:causal_ch}
The channel defined by $\Pr(Y_1^n | X_1^\infty)$ is termed \emph{causal} if for all $n$:
\begin{equation}\label{eq:513}
\Pr(\vr Y_1^n | \vr X_1^\infty) = \Pr(\vr Y_1^n | \vr X_1^n)
.
\end{equation}
\end{definition}
All the definitions below (including IFB\onlyumem{/AFB} capacity) pertain to causal channels.
This characterization of a causal channel is similar to the definition used by Han and Verd\'u \cite{HanVerdu} (and references therein). \onlyfull{This definition is also limited in assuming the channel starts from a known state (at time 0). However this does not limit the current setting, because an arbitrary initial state can be modeled by considering the family of channels with all possible initial states. Note that non causality that consists of bounded negative delays can always be compensated by applying a delay to the output.}

\onlyumem{
\begin{definition}\label{def:fading_memory_ch}
The channel is termed a \emph{fading memory channel} if for any $h > 0$ there exists $L$ and a sequence of causal conditional vector distribution functions $\{P_n(\cdot | \cdot)\}$, such that for all $n$ and $m \geq n$:
\begin{equation}\label{eq:fading_memory_def}
\| \Pr(\vr Y_n^m | \vr X_1^\infty, \vr Y_1^{n-L-1}) - P_n(\vr Y_n^m | \vr X_{n-L}^\infty) \|_1 \leq h
,
\end{equation}
where the $L_1$ norm is calculated over $\vr Y_n^m$, and defined by $\| g(\vr Y | \cdot) \|_1 \defeq \sum_{\vr y} | g(\vr y | \cdot) |$
\end{definition}

\begin{figure}
\centering
\ifpdf
  \setlength{\unitlength}{1bp}%
  \begin{picture}(263.62, 124.72)(0,0)
  \put(0,0){\includegraphics{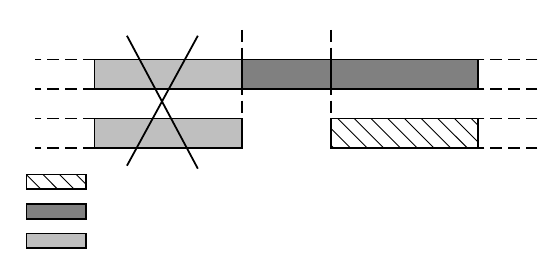}}
  \put(5.67,87.99){\fontsize{7.11}{8.54}\selectfont $\vr X$}
  \put(111.97,113.50){\fontsize{7.11}{8.54}\selectfont n-L}
  \put(157.32,113.50){\fontsize{7.11}{8.54}\selectfont n}
  \put(5.67,59.64){\fontsize{7.11}{8.54}\selectfont $\vr Y$}
  \put(42.52,21.37){\fontsize{7.11}{8.54}\selectfont - Condition is allowed to affect probability}
  \put(42.52,7.20){\fontsize{7.11}{8.54}\selectfont - Conditioning weakly affects probability}
  \put(42.52,35.55){\fontsize{7.11}{8.54}\selectfont - Part on which probability is evaluated in \eqref{eq:fading_memory_def}}
  \end{picture}%
\else
  \setlength{\unitlength}{1bp}%
  \begin{picture}(263.62, 124.72)(0,0)
  \put(0,0){\includegraphics{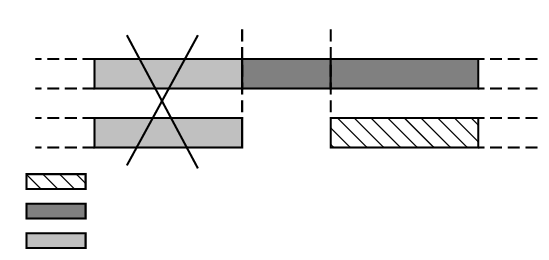}}
  \put(5.67,87.99){\fontsize{7.11}{8.54}\selectfont $\vr X$}
  \put(111.97,113.50){\fontsize{7.11}{8.54}\selectfont n-L}
  \put(157.32,113.50){\fontsize{7.11}{8.54}\selectfont n}
  \put(5.67,59.64){\fontsize{7.11}{8.54}\selectfont $\vr Y$}
  \put(42.52,21.37){\fontsize{7.11}{8.54}\selectfont - Condition is allowed to affect probability}
  \put(42.52,7.20){\fontsize{7.11}{8.54}\selectfont - Conditioning weakly affects probability}
  \put(42.52,35.55){\fontsize{7.11}{8.54}\selectfont - Part on which probability is evaluated in \eqref{eq:fading_memory_def}}
  \end{picture}%
\fi
\caption{\label{fig:illustration_fading_mem}%
 An illustration of the fading memory condition (Definition~\ref{def:fading_memory_ch}).}
\end{figure}

The difference between the terms on the LHS of \eqref{eq:fading_memory_def} is that $P_n$ does not include $(\vr X \vr Y)_1^{n-L-1}$ (see Fig.\ref{fig:illustration_fading_mem}), and thus the fading memory condition asserts that the dependence of the conditional distribution of future outputs, on the channel state at the far past, decays.
\onlyfull{Notice that the conditional distribution $\Pr(\vr Y_n^m | \vr X_1^\infty, \vr Y_1^{n-L-1})$ is completely defined by the channel, since it is conditioned on the entire input $\vr X_1^\infty$. On the other hand, the conditional distribution $\Pr(\vr Y_n^m | \vr X_{n-L}^\infty)$ may depend also on the input distribution (through the unspecified symbols $\vr X_1^{n-L-1}$). Therefore, the distribution $P_n$ in Definition~\ref{def:fading_memory_ch} is not identical to $\Pr(\vr Y_n^m | \vr X_{n-L}^\infty)$. On the other hand, \onlyfull{Proposition~\ref{prop:fading_memory_properties} shows}
\onlyconf{it can be shown}
that $\Pr(\vr Y_n^m | \vr X_{n-L}^\infty)$ obtained with any input distribution yields a legitimate $P_n$.
}

The fading memory condition does not imply stationarity or ergodicity. The memoryless arbitrary varying channel model considered in \selector{\cite{YL_PriorPrediction}}{Chapter~\ref{chap:prior_prediction}} is fading memory, and so are the FSC \cite[\S4.6]{Gallager_InfoTheoryBook} or compound-FSC models \cite{Lapidoth_Compound}, if the underlying FSC is indecomposable. An example of a non-homogeneous finite state channel with fading memory is presented in
\onlyfull{Section~\ref{sec:fading_mem_example}}\onlyconf{the full paper}.
} 

\subsection{IFB \onlyumem{and AFB}~capacity}\label{sec:def_ifb_afb_capacity}
\onlyfull{The following definitions lead to the definition\onlyumem{s} of \onlyumod{IFB capacity.}\onlyumem{IFB capacity and AFB capacity.}}

\begin{definition}[Reference encoder and decoder]\label{def:E_and_D}
A finite length encoder $E$ with block length $k$ and a rate $R$ is a mapping $E:\{1,\ldots,M\} \to \mathcal{X}^k$ from a set of $M \geq \exp(k R)$ messages to a set of input sequences $\mathcal{X}^k$. A respective finite length decoder $D$ is a mapping $D:\mathcal{Y}^k \to \{1,\ldots,M\}$ from the set of output sequences to the set of messages.
\end{definition}

\begin{definition}[IFB error probability]\label{def:mean_eps}
The \emph{average error probability in iterative mapping} of the $k$ length encoder $E$ and decoder $D$ to $b$ blocks over the channel $P_\tsubs{Y|X}$ is defined as follows: $b$ messages $\msg_1, \ldots, \msg_b$ are chosen as i.i.d. uniformly distributed random variables $\msg_i \sim U\{1,\ldots,M\}, i = 1,\ldots,b$. The channel input is set to $\vr X_i^{[k]} = E(\msg_i), i = 1,\ldots,b$, and the decoded message is $\hat \msg_i = D(\vr Y_i^{[k]})$ where $\vr Y$ is the channel output. The iterative mapping is illustrated in Fig.\ref{fig:iterative_mapping}. The average error probability is $P_e = \frac{1}{b} \sum_{i=1}^b \Pr(\hat \msg_i \neq \msg_i)$.
\end{definition}

\onlyphd{Recall that $\vr x_i^{[k]} \defeq \vr x_{(i-1)k + 1}^{(i-1)k + k}$ denotes the $i$-th block of length $k$ in the vector.}

\onlyumem{
\begin{definition}[AFB error probability]\label{def:mean_eps_afb}
The \emph{average error probability in arbitrary mapping} of the $k$ length encoder $E$ and decoder $D$ to $b$ blocks over the channel $P_\tsubs{Y|X}$ is defined as $P_e = \frac{1}{b} \sum_{i=1}^b P_e(i)$. $P_e(i)$ is the worst case per-block error probability, defined as:
 \begin{equation}\begin{split}\label{eq:181}
P_e(i)
&=
\max_{(\vr X \vr Y)_1^{(i-1)k}} \Big[
\onlypaper{\\&}
\Pr \left\{ D(\vr Y_i^{[k]}) \neq \msg \Big| \vr X_i^{[k]} = E(\msg), (\vr X \vr Y)_1^{(i-1)k} \right\} \Big]
,
\end{split}\end{equation}
where $\msg \sim U\{1,\ldots,M\}$.
\end{definition}
}

\begin{definition}[IFB\onlyumem{/AFB} achievability]\label{def:IFB_rate}
A rate $R$ is \emph{iterated-finite-block (IFB) \onlyumem{/ arbitrary-finite-block (AFB)} achievable (resp.)} over the channel $P_\tsubs{Y|X}$, if for any $\epsilon > 0$ there exist $k,b^* > 0$ such that for any $b > b^*$ there exist an encoder $E$ and a decoder $D$ with block length $k$ and rate $R$ for which the average error probability in
iterative\onlyumem{/arbitrary} mapping \onlyumem{(resp.)} of $E,D$ to $b$ blocks is at most $\epsilon$.
\end{definition}
This is equivalent to stating that the $\limsup$ of the average error probability with respect to $b$ is at most $\epsilon$.

\begin{definition}[IFB\onlyumem{/AFB} capacity]\label{def:IFB_capacity}
The \emph{IFB\onlyumem{/AFB} capacity} of the channel $P_\tsubs{Y|X}$ is the supremum of the set of IFB\onlyumem{/AFB} achievable rates, and is denoted $\Cifb$ \onlyumem{/$\Cafb$ (resp.)}.
\end{definition}

\onlyumem{
By definition, the AFB error probability is at least as large as the IFB error probability, and as a result, the AFB capacity is smaller than, or equal to the IFB capacity.
}

\onlyfull{
\begin{figure*}[t]
\setlength{\unitlength}{1mm}
\center
\begin{picture}(140, 60)
\multiput(10,55)(30,0){5}{\vector(0,-1){5}}
\put(12,53){$\msg_1$}
\put(42,53){$\msg_2$}
\put(72,53){$\msg_3$}
\put(102,53){$\msg_4$}
\put(132,53){$\msg_5$}
\multiput(0,50)(30,0){5}{\line(1,0){20}}\multiput(20,50)(30,0){5}{\line(0,-1){10}}
\multiput(20,40)(30,0){5}{\line(-1,0){20}}\multiput(0,40)(30,0){5}{\line(0,1){10}}
\multiput(3,44)(30,0){5}{Encoder}
\multiput(1,40)(2,0){10}{\vector(0,-1){5}}
\multiput(31,40)(2,0){10}{\vector(0,-1){5}}
\multiput(61,40)(2,0){10}{\vector(0,-1){5}}
\multiput(91,40)(2,0){10}{\vector(0,-1){5}}
\multiput(121,40)(2,0){10}{\vector(0,-1){5}}
\put(0,35){\line(1,0){140}}\put(140,35){\line(0,-1){10}}\put(140,25){\line(-1,0){140}}\put(0,25){\line(0,1){10}}
\put(60,29){Channel}
\put(1,30){$t=1$}
\put(125,30){$t=50$}
\multiput(1,25)(2,0){10}{\vector(0,-1){5}}
\multiput(31,25)(2,0){10}{\vector(0,-1){5}}
\multiput(61,25)(2,0){10}{\vector(0,-1){5}}
\multiput(91,25)(2,0){10}{\vector(0,-1){5}}
\multiput(121,25)(2,0){10}{\vector(0,-1){5}}
\multiput(0,20)(30,0){5}{\line(1,0){20}}\multiput(20,20)(30,0){5}{\line(0,-1){10}}
\multiput(20,10)(30,0){5}{\line(-1,0){20}}\multiput(0,10)(30,0){5}{\line(0,1){10}}
\multiput(3,14)(30,0){5}{Decoder}
\multiput(10,10)(30,0){5}{\vector(0,-1){5}}
\put(12,5){$\hat{\msg}_1$}
\put(42,5){$\hat{\msg}_2$}
\put(72,5){$\hat{\msg}_3$}
\put(102,5){$\hat{\msg}_4$}
\put(132,5){$\hat{\msg}_5$}
\end{picture}
\caption[An illustration of \textit{iterative mapping}]{An illustration of \textit{iterative mapping} used for the definition of average error probability (see Definition \ref{def:mean_eps}). The same encoder and decoder are used over each of the $b=5$ blocks of $k=10$ channel uses, and the average error probability is computed.}\label{fig:iterative_mapping}
\end{figure*}
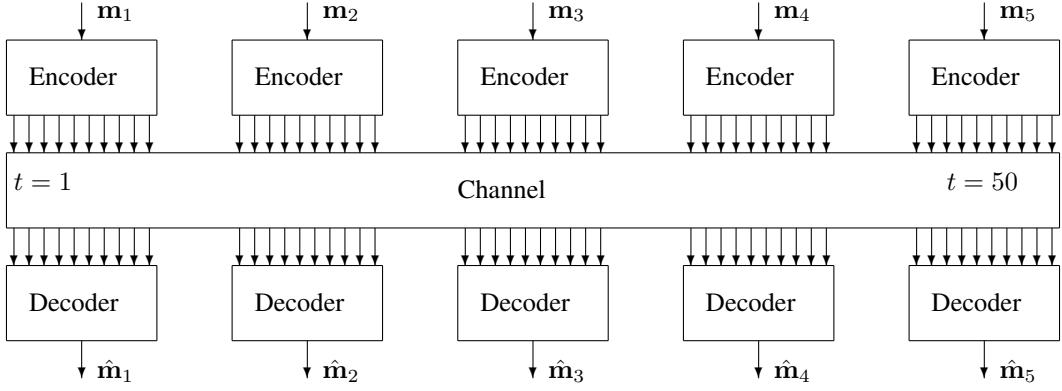
}
\onlyconf{
\begin{figure}[t]
\setlength{\unitlength}{0.5mm}
\scriptsize
\center
\begin{picture}(140, 60)
\multiput(10,55)(30,0){5}{\vector(0,-1){5}}
\put(12,53){$\msg_1$}
\put(42,53){$\msg_2$}
\put(72,53){$\msg_3$}
\put(102,53){$\msg_4$}
\put(132,53){$\msg_5$}
\multiput(0,50)(30,0){5}{\line(1,0){20}}\multiput(20,50)(30,0){5}{\line(0,-1){10}}
\multiput(20,40)(30,0){5}{\line(-1,0){20}}\multiput(0,40)(30,0){5}{\line(0,1){10}}
\multiput(3,44)(30,0){5}{Encoder}
\multiput(1,40)(2,0){10}{\vector(0,-1){5}}
\multiput(31,40)(2,0){10}{\vector(0,-1){5}}
\multiput(61,40)(2,0){10}{\vector(0,-1){5}}
\multiput(91,40)(2,0){10}{\vector(0,-1){5}}
\multiput(121,40)(2,0){10}{\vector(0,-1){5}}
\put(0,35){\line(1,0){140}}\put(140,35){\line(0,-1){10}}\put(140,25){\line(-1,0){140}}\put(0,25){\line(0,1){10}}
\put(60,29){Channel}
\put(1,30){$t=1$}
\put(125,30){$t=50$}
\multiput(1,25)(2,0){10}{\vector(0,-1){5}}
\multiput(31,25)(2,0){10}{\vector(0,-1){5}}
\multiput(61,25)(2,0){10}{\vector(0,-1){5}}
\multiput(91,25)(2,0){10}{\vector(0,-1){5}}
\multiput(121,25)(2,0){10}{\vector(0,-1){5}}
\multiput(0,20)(30,0){5}{\line(1,0){20}}\multiput(20,20)(30,0){5}{\line(0,-1){10}}
\multiput(20,10)(30,0){5}{\line(-1,0){20}}\multiput(0,10)(30,0){5}{\line(0,1){10}}
\multiput(3,14)(30,0){5}{Decoder}
\multiput(10,10)(30,0){5}{\vector(0,-1){5}}
\put(12,5){$\hat{\msg}_1$}
\put(42,5){$\hat{\msg}_2$}
\put(72,5){$\hat{\msg}_3$}
\put(102,5){$\hat{\msg}_4$}
\put(132,5){$\hat{\msg}_5$}
\end{picture}
\caption{An illustration of iterative mapping used for the comparison class, with $b=5,k=10$}\label{fig:iterative_mapping}
\end{figure}
}

\subsection{Competitive Universality}\label{sec:universal_mod_defs}

\onlypaper{\onlyfull{
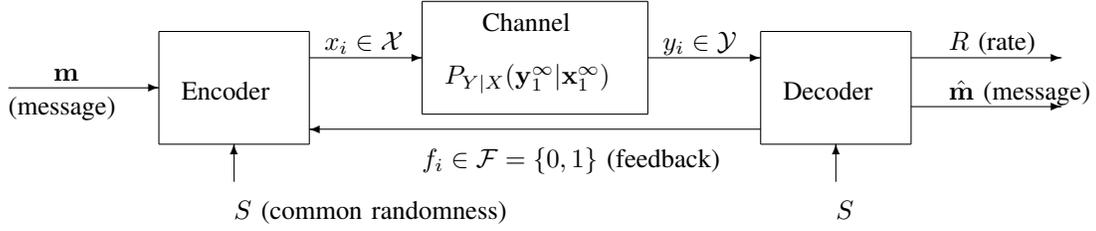
\begin{figure*}[t]
\setlength{\unitlength}{1mm}
\center
\begin{picture}(140, 30)
\put(23,16){Encoder}\put(20,10){\line(1,0){20}}\put(40,10){\line(0,1){15}}\put(40,25){\line(-1,0){20}}\put(20,25){\line(0,-1){15}}
\put(63,25){Channel} \put(58,18){$P_{Y|X}(\vr y_1^\infty | \vr x_1^\infty)$}
\put(55,14){\line(1,0){30}}\put(85,14){\line(0,1){15}}\put(85,29){\line(-1,0){30}}\put(55,29){\line(0,-1){15}}
\put(103,16){Decoder}\put(100,10){\line(1,0){20}}\put(120,10){\line(0,1){15}}\put(120,25){\line(-1,0){20}}\put(100,25){\line(0,-1){15}}
\put(0,17.5){\vector(1,0){20}}\put(6,18.5){$\msg$}\put(0,14){(message)}
\put(40,21.5){\vector(1,0){15}}\put(42,22.5){$x_i \in \mathcal{X}$}
\put(85,21.5){\vector(1,0){15}}\put(87,22.5){$y_i \in \mathcal{Y}$}
\put(100,12){\vector(-1,0){60}}\put(55,7){$f_i \in \mathcal{F} = \{0,1\}$ (feedback)}
\put(120,21.5){\vector(1,0){20}}\put(125,22.5){$R$ (rate)}
\put(120,15){\vector(1,0){20}}\put(125,16){$\hat{\msg}$ (message)}
\put(30,5){\vector(0,1){5}}\put(30,0){$S$ (common randomness)}
\put(110,5){\vector(0,1){5}}\put(110,0){$S$}
\end{picture}
\caption{Rate adaptive encoder-decoder pair with feedback, over an unknown channel}\label{fig:system_adaptive}
\end{figure*}
}} 

In the following, the properties of the adaptive system with feedback, and IFB\onlyumem{/AFB}-universality are defined.
A randomized rate-adaptive transmitter and receiver for block length $n$ with feedback are defined \onlyphd{formally as in Definition~\ref{def:sys_adaptive}. Let us shortly repeat the definition:}\onlypaper{as follows (see also formal definitions in \cite[\S5]{YL_PhdThesis}):} the transmitter is presented with a message expressed by an infinite bit sequence, and following the reception of $n$ symbols, the decoder announces the achieved rate $R$, and decodes the first $\lceil nR \rceil$ bits. An error means any of these bits differs from the bits of the original message sequence. Both encoder and decoder have access to a random variable $S$ (the common randomness) distributed over a chosen alphabet, and a causal feedback link allows the transmitted symbols to depend on previously sent feedback from the receiver. \onlyfull{The system is illustrated in Fig. \ref{fig:system_adaptive}. }

The following definition states formally the notion of IFB\onlyumem{/AFB}-universality for rate adaptive systems:
\begin{definition}[IFB\onlyumem{/AFB} universality]\label{def:IFBAFB_universality}
With respect to a set of channels $\{P_\tsubs{Y|X}^{(\theta)}\}, \theta \in \Theta$ (not necessarily finite or countable), a rate-adaptive communication system (possibly using feedback and common randomness) is called \emph{IFB\onlyumem{/AFB} universal} if for every channel in the family and any $\epsilon,\delta>0$ there is $n$ large enough such that when the system is operated over $n$ channel uses, then with probability $1-\epsilon$, the message is correctly decoded and the rate is at least
\onlyumod{$\Cifb(P_\tsubs{Y|X}) - \delta$.}\onlyumem{$\Cifb(P_\tsubs{Y|X}) - \delta$ or $\Cafb(P_\tsubs{Y|X}) - \delta$ (resp.).}
\end{definition}

Notice that the definitions above (and specifically Definitions~\ref{def:IFB_rate},\ref{def:IFBAFB_universality}) do not require uniform convergence with respect to the channel, i.e. the number of channels uses $n$ or blocks $b$ for which the requirements hold may be a function of the channel.

}

\onlypaper{
\subsection{A discussion on IFB capacity and universality}\label{sec:IFB_capacity_discussion}

Following are some comments regarding IFB capacity and IFB universality.
Note that the use of average error probability over time and messages (expressed in the assumed uniform distribution) rather than maximum error probability (over time or messages) reduces the requirements from $E,D$ and therefore increases $\Cifb$.

As noted, $\Cifb \leq C$, where $C$ is the Shannon capacity \cite{HanVerdu}. However for i.i.d. memoryless channels clearly $\Cifb = C$. The difference between $C$ and $\Cifb$ relates to the stability of the channel over time, and the ability to utilize channel structure which cannot be observed in finite time. Let us give two examples to sharpen this difference:

\begin{example}\label{example:blocking_channel}
Consider the binary product channel $y_i = x_i \cdot z_i$, and let the sequence $\vr z$ alternate between $0$ and $1$, in blocks of ever growing size, but such that the overall frequency of $0$ is $\half$, and the length of each block is negligible compared to the total length of previous blocks. For example, set $z_i$ to $0$ in $i \in \cup_{k=1}^{\infty} [2 k^2, (k+1)^2 + k^2]$.
For this channel $\Cifb=0$ while $C = \half$. The reason is that for every finite length encoder/decoder, ultimately as $m \to \infty$ half the blocks will fall on bursts of $z = 0$ and be in error. Note that if rate adaptation would have been allowed at the IFB decoder, this capacity would not be zero (see Section~\ref{sec:alt_definitions})
\end{example}

\begin{example}\label{example:growing_delay}
Consider a channel with ever growing delay: Suppose that $d_i$ is a sequence of slowly growing delays. For example, $d_i = \lfloor \log i \rfloor$, and the channel is $y_i = x_{i-d_i}$, where $x,y$ are binary. The capacity of this channel is $C=1$, whereas $\Cifb=0$. Here, the reason for the gap is the in-ability to utilize the channel structure with a finite block size.
\end{example}

Following these examples the choice of $\Cifb$ may be justified by two main reasons: one is its operational significance, i.e. that universally attaining $\Cifb$, means competing with every static block coding system, and the other is the rejection of ``pathological'' behaviors of the channel, as the ones mentioned in the examples above.

Note that although $\Cifb \leq C$, the universal \selector{system presented here}{systems presented in the following chapters} may opportunistically achieve rates above $C$. This means the communication rate may exceed $C$ in part of the time. Consider for example the binary non-ergodic channel that with probability $p$ has $\vr y = \vr x$, and with probability $1-p$ the output is independent of the input. While the capacity of this channel is $C = 0$ (and $\Cifb=0$), by adapting the rate, one could attain a rate of $1$ with probability $p$.

An interesting question is whether for a general vector channel, $\Cifb$ can be universally attained. Unfortunately, the answer is negative, and the reason is that, because the input sequences used by the reference encoder and by the universal system are different, infinite memory in the channel may cause the channel to get ``stuck'' in an unfortunate state. This phenomenon is nicknamed a ``password'' channel, since it is similar to a situation where a password is required at the beginning of transmission, otherwise the channel becomes useless. In this case, a reference system knowing the password may succeed and a universal system, having only one attempt to find the password, is bound to fail. More generally, given an encoder, a channel can be structured such that it will identify the specific encoder's codebook, and fail if any deviation from this codebook is observed. Here is a simple example:
\begin{example}[Password channel]\label{example:password_channel}
Consider a family of two binary channels. In the first channel, if $x_1 = 0$ then the channel will become clean, i.e. $\forall i \geq 2: y_i = x_i$, but if $x_1 = 1$, then it becomes blocked, i.e. $\forall i \geq 2: y_i = 0$. The second channel is the same, except the roles of $0,1$ are reversed. Clearly, for both channels $\Cifb=1$, since the only constraint required to avoid blocking is that the first symbol in each encoded block is constant 0 or 1, and therefore a rate of $\frac{k-1}{k}$ can be obtained with block size $k$. On the other hand, no universal system can guarantee any rate with a vanishing error probability, since any choice of the first symbol will lead to blocking in one of the two channels.
\end{example}

The conclusion from the above is that the concept of iterated finite block capacity is not as strong as the concept of finite state compressibility, which is truly universally attainable. This problem relates to a fundamental difficulty in universal communication compared to universal compression: in universal compression, the sequence is given and does not depend on the encoder's actions, while in communication, the encoder's actions (the input symbols) affect the channel behavior in an unexpected way. \onlyphd{This motivates the definition of fading memory channels above.}

One may be tempted to think that depriving the IFB class from its block-wise operation and limiting it to i.i.d. distributions would solve the ``password'' problem. However it is easy to devise a channel that would identify the input distribution of the reference encoder, while blocking the universal system. See Example~\ref{example:password_channel_iid_example} in Appendix~\ref{sec:password_channel_iid_example}. These difficulties exemplify the complexity of the universal communication problem.

\onlyphd{
The advantage of the IFB reference class which enables it to win over any universal system in this case, is its ability to determine such a codebook that will not only enable reliable transmission, but will also keep the channel in a favorable state, whereas the universal system does not know the long term effects of certain input symbols or distributions. In the alternative AFB class, the reference system is crippled, so that it cannot enjoy the ability to shape the past: the encoder and decoder operate over finite blocks, however the error probability is required to be small in the worst case channel state (history) prior to each block, and average over blocks. This models a situation where the reference encoder and decoder are ``thrown'' each time into a different location in time, where the past state might have been arbitrary. It is not required to have good performance in each of these events, but only on average. This reference class is less natural than the IFB, yet it enables releasing constraints on the channel. Refer to Section~\ref{sec:um_discussion_fading_wide_sense} for further discussion on the definition of AFB capacity and fading memory channels. }

}

\onlypaper{
\section{Universal communication over the modulo-additive channel}\label{sec:univ_modulo_add_capacity}
}\onlyphd{
\section{Introduction}
}
This \selector{section and the next, focus}{chapter focuses} on the modulo-additive channel with an individual noise sequence. It is shown that the IFB capacity of this channel is bounded by $(1-\rho(\vr z)) \log |\mathcal{X}|$ and that this rate is universally achievable. Upper and lower bounds on the convergence rates are given, which show that, unfortunately, the transmission length $n$ required to obtain universal communication grows exponentially with the block length $k$ of the competing system. \onlyphd{Although the results of Chapter~\ref{chap:univ_vectormem} are more general, the issue of channel memory adds a significant complexity. Therefore, it is interesting to explore first the example of the modulo-additive channel due to its simplicity and the ability to present the concepts and techniques at their pure form.}

\onlyphd{
The modulo additive channel is defined by the relation
\begin{equation}\label{eq:ma169}
\vr y = \vr x + \vr z
,
\end{equation}
where $\vr x, \vr y, \vr z \in \mathcal{X}^n$ are $n$-length vectors, denoting the input, output and the noise sequence, $\mathcal{X}$ is a finite alphabet, the '$+$' denotes modulo addition over $\mathcal{X}$, and the sequence $\vr z$ is arbitrary and unknown.
}

The modulo-additive channel is a relatively ``easy'' case because of two main reasons:
\begin{itemize}
\item It is memoryless in the input, and thus the ``password'' issue is avoided.
\item There is a single input prior, the uniform i.i.d. distribution, which attains capacity for any noise sequence, since it maximizes the output entropy. Therefore no adaptation of the prior is needed.
\end{itemize}

\onlypaper{
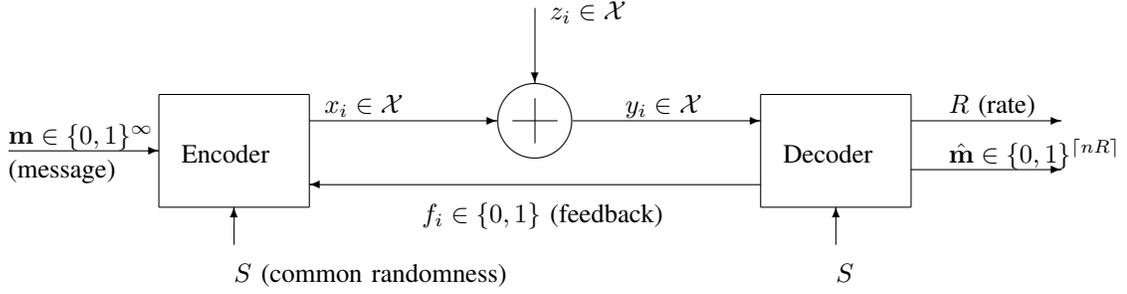
\begin{figure*}[t]
\center
\setlength{\unitlength}{1mm}
\begin{picture}(140, 30)
\put(23,16){Encoder}\put(20,10){\line(1,0){20}}\put(40,10){\line(0,1){15}}\put(40,25){\line(-1,0){20}}\put(20,25){\line(0,-1){15}}
\put(103,16){Decoder}\put(100,10){\line(1,0){20}}\put(120,10){\line(0,1){15}}\put(120,25){\line(-1,0){20}}\put(100,25){\line(0,-1){15}}
\put(0,17.5){\vector(1,0){20}}\put(0,18.5){$\msg \in \{0,1\}^{\infty}$}\put(0,14){(message)}
\put(40,21.5){\vector(1,0){25}}\put(42,22.5){$x_i \in \mathcal{X}$}
\put(75,21.5){\vector(1,0){25}}\put(82,22.5){$y_i \in \mathcal{X}$}
\put(70,21.5){\circle{10}}\put(67,21.5){\line(1,0){6}}\put(70,18.5){\line(0,1){6}}
\put(70,36.5){\vector(0,-1){10}}\put(72,35){$z_i \in \mathcal{X}$}
\put(100,13){\vector(-1,0){60}}\put(55,8){$f_i \in \{0,1\}$ (feedback)}
\put(120,21.5){\vector(1,0){20}}\put(125,22.5){$R$ (rate)}
\put(120,15){\vector(1,0){20}}\put(125,16){$\hat{\msg} \in \{0,1\}^{\lceil n R \rceil}$}
\put(30,5){\vector(0,1){5}}\put(30,0){$S$ (common randomness)}
\put(110,5){\vector(0,1){5}}\put(110,0){$S$}
\end{picture}
\caption{An adaptive system over the modulo-additive channel with feedback}\label{fig:modulo_additive_w_fb}
\end{figure*}
}

\onlyphd{
\section{Overview of this chapter}\label{sec:ma_overview}

\onlyphd{
Section~\ref{sec:univ_modulo_add_capacity} shows that there exists an IFB-universal system for the modulo additive channel. In Section~\ref{sec:IFB_UB} it is shown that $\Cifb \leq (1 - \rho(\vr z)) \cdot  \log |\mathcal{X}|$, where $\rho(\vr z)$ is the finite state compressibility of $\vr z$ (as defined by Lempel and Ziv \cite{LZ78}). The main tool in the proof is the ``collapsed'' channel (Fig.\ref{fig:collapsed_channel_modadditive}), created by choosing the noise sequence over blocks uniformly. In Section~\ref{sec:ma_adaptive_rho} it is shown that, with common randomness and feedback, a universal system can asymptotically attain this rate without prior knowledge of the noise sequence. In order to show attainability, the result of Theorem~\ref{theorem:adaptive_L} from Section~\ref{sec:examples_compression_attanability} is used. It was shown there (when specializing the aforementioned result to the modulo-additive channel \S\ref{sec:examples_compression_modadditive}), that the rate function $\log | \mathcal{X} | - \frac{1}{n} L(\vr z)$ is asymptotically adaptively achievable, where $L(\vr z)$ is the compression length of the noise sequence. The result is obtained by plugging the encoding lengths $L(\vr z)$ determined by the LZ78 source encoder, whose the compression ratios asymptotically approach the finite state compressibility.\footnote{A more direct proof of this result, which does not go through the chain of theorems of Part-I, and a simplified intuitive explanation, are given in the paper \cite{YL_UnivModuloAdditive}.}
} 

\onlypaper{
This section provides an informal review of the main results and rough proof outlines. The purpose is to provide an understanding of the results without diving into mathematical detail. \onlypaper{The main results of this paper pertain to the modulo-additive channel with individual noise sequence.}
\selector{For this channel, it}{It} is shown in Section~\ref{sec:univ_modulo_add_capacity} that $\Cifb \leq (1 - \rho(\vr z)) \cdot  \log |\mathcal{X}|$, where $\rho(\vr z)$ is the finite state compressibility of the infinite sequence $\vr z$, as defined by Lempel and Ziv \cite{LZ78}. Assuming that common randomness exists and that there is a feedback link, a universal system employing feedback exists, which asymptotically attains this rate universally without prior knowledge of the noise sequence. In Section~\ref{sec:ma_univ_redundancy}, upper and lower bounds on the convergence rate are derived. Below, the main ideas in the proofs are described.

Let us begin with the upper bound on $\Cifb$. Suppose a given the reference system comprised of an encoder and a decoder, achieves the rate $R$ over $b$ blocks of size $k$ (Figure~\ref{fig:iterative_mapping}). During these $b$ blocks, the reference system ``sees'' $b$ different noise vectors of length $k$, namely $\vr z_{(i-1)k+1}^{ik}$, $i=1,\ldots,b$. Since the system is fixed during these $b$ blocks, this is equivalent to operating over a stochastic channel, where the noise vector $\tilde{\vr Z}$ is chosen uniformly from the set of these vectors, with probability $\frac{1}{b}$ for each. This random vector is termed the ``collapsed'' noise sequence, and the channel generated from it the ``collapsed'' channel (Figure~\ref{fig:collapsed_channel_modadditive}). The standard converse of the channel capacity theorem, without the assumption of a memoryless channel, can be applied to the collapsed channel, and yields an upper bound on $\Cifb$, which is roughly $\log |\mathcal{X}| - \frac{1}{k} H(\tilde{\vr Z})$. The entropy $H(\tilde{\vr Z})$ is lower bounded using the finite state compressibility of the sequence, since a finite state machine may achieve a compression rate close to the entropy by standard block-to-variable coding, where the code lengths are tuned to the statistics of the collapsed noise vector. Combining these bounds yields the result $\Cifb \leq (1 - \rho(\vr z)) \cdot  \log |\mathcal{X}|$ (Theorem \ref{theorem:Cifb_UB}).

\begin{figure}
\centering
\ifpdf
  \setlength{\unitlength}{1bp}%
  \begin{picture}(216.44, 192.04)(0,0)
  \put(0,0){\includegraphics{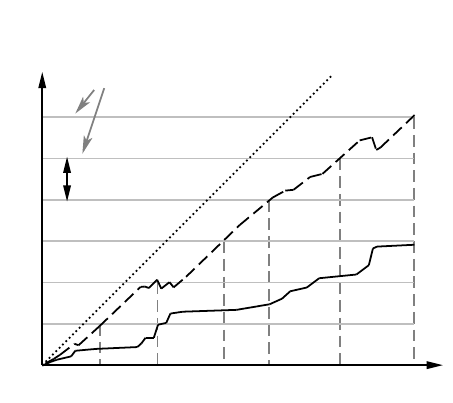}}
  \put(79.67,86.32){\rotatebox{45.00}{\fontsize{9.96}{11.95}\selectfont \smash{\makebox[0pt][l]{$i \cdot \log |\mathcal{X}|$}}}}
  \put(84.22,61.78){\rotatebox{45.00}{\fontsize{9.96}{11.95}\selectfont \smash{\makebox[0pt][l]{$N_i = i \cdot \log |\mathcal{X}| - L(z^i)$}}}}
  \put(141.32,65.36){\fontsize{9.96}{11.95}\selectfont $L(z^i)$}
  \put(143.30,7.81){\fontsize{9.96}{11.95}\selectfont  $i$ - time index}
  \put(13.45,107.59){\rotatebox{90.00}{\fontsize{9.96}{11.95}\selectfont \smash{\makebox[0pt][l]{Number of bits}}}}
  \put(40.12,154.65){\fontsize{9.96}{11.95}\selectfont \textcolor[rgb]{0.50196, 0.50196, 0.50196}{Decoding thresholds}}
  \put(36.15,105.04){\fontsize{9.96}{11.95}\selectfont $K$}
  \end{picture}%
\else
  \setlength{\unitlength}{1bp}%
  \begin{picture}(216.44, 192.04)(0,0)
  \put(0,0){\includegraphics{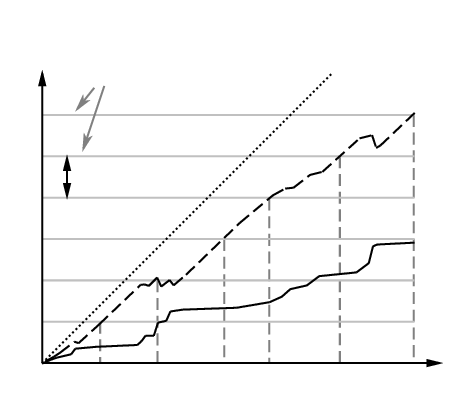}}
  \put(79.67,86.32){\rotatebox{45.00}{\fontsize{9.96}{11.95}\selectfont \smash{\makebox[0pt][l]{$i \cdot \log |\mathcal{X}|$}}}}
  \put(84.22,61.78){\rotatebox{45.00}{\fontsize{9.96}{11.95}\selectfont \smash{\makebox[0pt][l]{$N_i = i \cdot \log |\mathcal{X}| - L(z^i)$}}}}
  \put(141.32,65.36){\fontsize{9.96}{11.95}\selectfont $L(z^i)$}
  \put(143.30,7.81){\fontsize{9.96}{11.95}\selectfont  $i$ - time index}
  \put(13.45,107.59){\rotatebox{90.00}{\fontsize{9.96}{11.95}\selectfont \smash{\makebox[0pt][l]{Number of bits}}}}
  \put(40.12,154.65){\fontsize{9.96}{11.95}\selectfont \textcolor[rgb]{0.50196, 0.50196, 0.50196}{Decoding thresholds}}
  \put(36.15,105.04){\fontsize{9.96}{11.95}\selectfont $K$}
  \end{picture}%
\fi
\caption[The decoding rule of the compression based rate adaptive system]{An illustration of the decoding rule of the rate adaptive system. $L(z^i)$ is the compression length. Decoding thresholds with respect to $N_i=i \cdot \log |\mathcal{X}| - L(z^i)$ are depicted by horizontal lines.}\label{fig:rate_adaptive_nRemp_over_time}%
\end{figure}

Next, a communication scheme is demonstrated, that asymptotically attains the rate $\log | \mathcal{X} | - \frac{1}{n} L(\vr z^n)$, where $L(\vr z^n)$ is the compression length of the sequence $\vr z$ by a given sequential source encoder, and $n$ is the overall block length. The scheme is based on iterative application of rateless coding, sending $K$ bits in each block. Each codeword in the codebook of $\exp(K)$ words is chosen independently and distributed uniformly over $\mathcal{X}^n$. The transmitter sends symbols from the codeword matching the $K$ message bits, until a termination condition occurs at the receiver side. Then, the receiver indicates the end of the block through the feedback link and a new block begins. The termination condition is based on giving a rank to every possible noise sequence $\vr z_1^i$. This noise sequence has two parts: the part spanning previous blocks that had already been decoded is known with high probability, as both the channel input and the channel output are known. The part of the noise sequence from the beginning of the current block is unknown, but is related in a one-to-one relation to the unknown message, because the channel output is known. Therefore, there are $\exp(K)$ possible noise sequences, corresponding to the $\exp(K)$ possible codewords sent in the current block. The rank of each noise sequence is the code-length, or the number of bits in its representation by the given source encoder. The decoder terminates the block if for any codeword, this length is smaller than a threshold.

The proof of this scheme's performance is roughly as follows. Due to the random coding, most of the hypotheses, except the true one, yield random noise sequences. These sequences are incompressible, and therefore the number of bits representing the last block would be approximately $\log |\mathcal{X}|$ times the number of symbols in the block. It can be shown that setting the threshold approximately $K$ below this value, guarantees a small probability of exceeding the threshold for any of the $\exp(K)-1$ incorrect codewords, and therefore a small probability of error. It is convenient to define the ``porosity'' of the sequence up to time $i$ as $N_i = i \cdot \log|\mathcal{X}| - L(\hat{\vr z}_1^i)$, representing the gap between the compressibility of the hypothetical noise sequence, and the compressibility of a random sequence. The approximate termination condition may be interpreted as decoding when the value of $N_i$ increases by $K$ from the start of the current block. Since when this occurs, the system starts a new block, there is a correspondence between the increase in $N_i$ and the number of blocks and bits that are transmitted, i.e. the termination condition can be approximately interpreted as $N_i \geq K (b+1)$ where $b$ is the number of blocks so far. Therefore assuming by time $n$, $B$ blocks were transmitted, the number of transmitted bits is $K \cdot B \approx N_n = n \cdot \log|\mathcal{X}| - L(\hat{\vr z}_1^n)$. Assuming no errors occurred $\hat{\vr z}_1^n = \vr z_1^n$, and dividing by $n$ the desired result is obtained. This is depicted in Figure~\ref{fig:rate_adaptive_nRemp_over_time}, where the horizontal axis is the time $i$. The solid line presents $L(\vr z_1^i)$, and the dashed line $N_i$. The decoding thresholds $K b$ ($b=1,2,\ldots$) are depicted as horizontal lines, while the vertical lines depict the decoding times. Decoding occurs whenever $N_i$ crosses a threshold. A random hypothesized sequence in the current block implies that $N_i$ does not increase on average. It can be seen that the number of bits that will be sent is approximately $N_n$. In the full proof, various overheads that were neglected above are accounted for.

To obtain the universal system attaining $\Cifb$ (Theorem~\ref{theorem:adaptive_rho}), the scheme above is applied with the encoding lengths $L(\vr z^n)$ determined by the LZ78 source encoder, whose compression ratios asymptotically approach the finite state compressibility: asymptotically $\frac{1}{n} L(\vr z^n) \leq \rho(\vr z) \log |\mathcal{X}|$, therefore $\log | \mathcal{X} | - \frac{1}{n} L(\vr z^n) \geq (1-\rho(\vr z)) \log |\mathcal{X}|$, where all inequalities are up to asymptotically vanishing factors.
} 

Section~\ref{sec:ma_univ_redundancy} deals with the question of the redundancy, or how quickly the system converges to the rate attained by the best IFB system with a given block length $k$. Unfortunately, it is shown that $n$ must grow at least as fast as $|\mathcal{X}|^k$, approximately. The upper bound on redundancy is obtained by using a similar universal system employing a slightly more refined design: a universal probability assignment based on a mixture of Krichevsky-Trofimov distributions \cite{KrichevskyTrofimov81} is used instead of the LZ78 encoder. The lower bound is obtained by presenting a design of an IFB system together with a random channel, i.e. a distribution over noise sequences, such that the mutual information over the channel is smaller than the rate obtained by the IFB system. This is possible because the IFB system is designed together with the channel and can use the knowledge of the specific noise sequence. On the other hand, the rate obtained by any universal system with feedback is bounded by the mutual information, and this gap comprises the lower bound on redundancy. The upper bound and the lower bound on the redundancy agree in terms of the asymptotical growth rate of $n$ as function of $k$ (see Figure~\ref{fig:mod_additive_redundancy_fig2}).

}

\onlyphd{
\section{Universal communication over the modulo-additive channel}\label{sec:univ_modulo_add_capacity}
}

\subsection{A bound on the IFB capacity of the modulo-additive channel}\label{sec:IFB_UB}
\onlyfull{In this section, the following Theorem is proven:}
\onlyconf{The following theorem is proven in the full paper \cite{YL_UnivModuloAdditive}:}
\begin{theorem}\label{theorem:Cifb_UB}
The IFB-capacity of the modulo-additive channel $\vr y = \vr x + \vr z$ where $\vr x, \vr y, \vr z \in \mathcal{X}^{\infty}$ are infinite sequences denoting the channel input, output and noise sequence, satisfies
\begin{equation}
\Cifb \leq   ( 1 - \rho(\vr z) ) \cdot \log|\mathcal{X}|
,
\end{equation}
where $\rho(\vr z)$ is the finite state compressibility of $\vr z$.
\end{theorem}

\onlyfull{
For the sake of completeness let us shortly repeat the definition of finite state compressibility.
A finite state encoder $F$ with $s$ states is defined by a next state function $g:(\{1,\ldots,s\},\mathcal{X}) \to \{1,\ldots,s\}$, and an output function $f:(\{1,\ldots,s\},\mathcal{X}) \to \{ \{0,1\}^k \}_{k=0}^\infty$, where the output may be a bit sequence of any length, including the empty sequence. The encoder is said to be \emph{information lossless} if for any $\vr z_1^n$, the input $\vr z_1^n$ can be uniquely decoded from the output sequence $F(\vr z_1^n)$, given the initial and final states. Let $\mathcal{F}(s)$ denote the group of all finite state information lossless encoders with at most $s$ states. Let the length of the output sequence of encoder $F$ for an input sequence of length $n$ be denoted $|F(\vr z_1^n)|$, then the compression ratio of $\vr z_1^n$ by $F$ is defined as:
\begin{equation}\label{eq:ma444}
\rho_F(\vr z_1^n) \defeq \frac{1}{n \log |\mathcal{X}|} |F(\vr z_1^n)|
.
\end{equation}
The compression ratio of the best information lossless finite state encoder with at most $s$ states is denoted:
\begin{equation}\label{eq:ma445}
\rho_{\mathcal{F}(s)} (\vr z_1^n) \defeq \min_{F \in \mathcal{F}(s)} \rho_F(\vr z)
,
\end{equation}
and finally, the finite state compressibility of the infinite sequence $\vr z = \vr z_1^\infty$ is defined as:
\begin{equation}\label{eq:ma446}
\rho(\vr z) = \lim_{s \to \infty} \limsup_{n \to \infty} \rho_{\mathcal{F}(s)} (\vr z_1^n)
.
\end{equation}
Note that the order of limits is critical for this definition, since if the number of states is taken to infinity first, any sequence can be compressed to 1 bit by having the state machine ``remember'' and identify the particular sequence. The outer limit exists, since $\rho_{\mathcal{F}(s)}$ is non-increasing in $s$ and bounded from below.

\textit{Theorem~\ref{theorem:Cifb_UB} proof outline:}
Define $\tilde{\vr Z}_{b,k}$ as the random vector of length $k$ formed by selecting one vector from the set of $b$ vectors $(\vr z_i^{[k]})_{i=1}^b$, with uniform probability of $\frac{1}{b}$ for each. In other words, the probability distribution of $\tilde{\vr Z}_{b,k}$ equals the empirical distribution of the first $b$ blocks of length $k$ in $\vr z$. Similarly define the random variables $\tilde{\vr X}_{b,k}$ and $\tilde{\vr Y}_{b,k}$ derived from the sequences $\vr x, \vr y$.

Suppose a given $E,D$ achieve rate $R$ and average error probability $\epsilon$ over $b$ blocks of size $k$. This is equivalent to saying they achieve error probability $\epsilon$ when operating on the stochastic channel $\tilde{\vr Y}_{b,k} = \tilde{\vr X}_{b,k} + \tilde{\vr Z}_{b,k}$ (Figure~\ref{fig:collapsed_channel_modadditive}). Therefore the standard converse of the channel capacity theorem implies that the rate $R$ can be bounded by $R \stackrel{\approx}{\leq} \log |\mathcal{X}| - \frac{1}{k} H(\tilde{\vr Z}_{b,k})$. Then, the limit of $\frac{1}{k} H(\tilde{\vr Z}_{b,k})$ is related to the finite state compressibility $\rho(\vr z)$. The later relation is a variation of a result by Lempel and Ziv \cite[Theorem 3]{LZ78} on the convergence of the sliding-window empirical entropy measured over increasing block lengths to the finite state compressibility, whereas here the block-wise empirical entropy is used instead. The full proof is given in Appendix~\ref{sec:proof_theorem_Cifb_UB}.

\begin{figure*}
\centering
\ifpdf
  \setlength{\unitlength}{1bp}%
  \begin{picture}(392.70, 137.79)(0,0)
  \put(0,0){\includegraphics{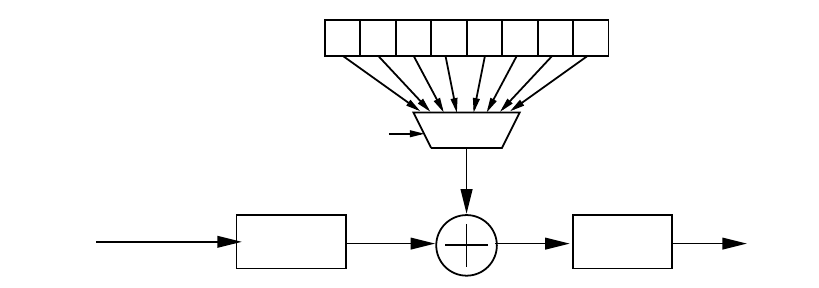}}
  \put(140.11,20.00){\fontsize{10.24}{12.29}\selectfont \makebox[0pt]{Encoder}}
  \put(296.58,20.00){\fontsize{10.24}{12.29}\selectfont \makebox[0pt]{Decoder}}
  \put(105.74,30.43){\fontsize{8.54}{10.24}\selectfont \makebox[0pt][r]{$\msg \sim U\{1,\ldots,M \}$}}
  \put(186.53,30.43){\fontsize{8.54}{10.24}\selectfont \makebox[0pt]{$\tilde{\vr X}_{b,k}$}}
  \put(249.46,30.43){\fontsize{8.54}{10.24}\selectfont \makebox[0pt]{$\tilde{\vr Y}_{b,k}$}}
  \put(232.45,54.24){\fontsize{8.54}{10.24}\selectfont $\tilde{\vr Z}_{b,k}$}
  \put(164.42,118.02){\fontsize{8.54}{10.24}\selectfont \makebox[0pt]{$\vr z_1^{[k]}$}}
  \put(329.49,27.87){\fontsize{8.54}{10.24}\selectfont $\hat{\vr \msg}$}
  \put(249.46,118.02){\fontsize{8.54}{10.24}\selectfont \makebox[0pt]{$\vr z_6^{[k]}$}}
  \put(266.47,118.02){\fontsize{8.54}{10.24}\selectfont \makebox[0pt]{$\vr z_7^{[k]}$}}
  \put(283.48,118.02){\fontsize{8.54}{10.24}\selectfont \makebox[0pt]{$\vr z_8^{[k]}$}}
  \put(181.43,118.02){\fontsize{8.54}{10.24}\selectfont \makebox[0pt]{$\vr z_2^{[k]}$}}
  \put(198.44,118.02){\fontsize{8.54}{10.24}\selectfont \makebox[0pt]{$\vr z_3^{[k]}$}}
  \put(215.44,118.02){\fontsize{8.54}{10.24}\selectfont \makebox[0pt]{$\vr z_4^{[k]}$}}
  \put(232.45,118.02){\fontsize{8.54}{10.24}\selectfont \makebox[0pt]{$\vr z_5^{[k]}$}}
  \put(300.95,119.13){\fontsize{16.63}{19.96}\selectfont ...}
  \put(193.43,77.20){\fontsize{8.54}{10.24}\selectfont \makebox[0pt][r]{$i \sim \unif\{1,\ldots,b\}$}}
  \end{picture}%
\else
  \setlength{\unitlength}{1bp}%
  \begin{picture}(392.70, 137.79)(0,0)
  \put(0,0){\includegraphics{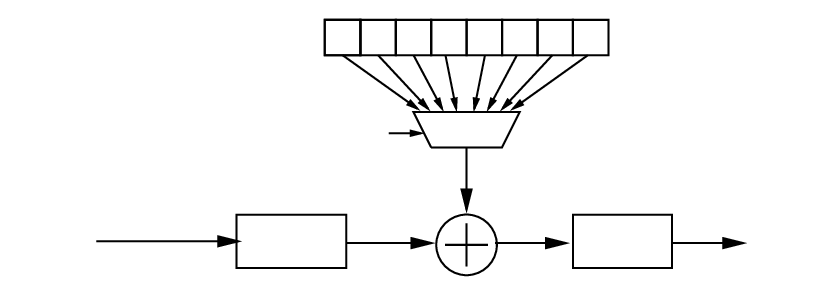}}
  \put(140.11,20.00){\fontsize{10.24}{12.29}\selectfont \makebox[0pt]{Encoder}}
  \put(296.58,20.00){\fontsize{10.24}{12.29}\selectfont \makebox[0pt]{Decoder}}
  \put(105.74,30.43){\fontsize{8.54}{10.24}\selectfont \makebox[0pt][r]{$\msg \sim U\{1,\ldots,M \}$}}
  \put(186.53,30.43){\fontsize{8.54}{10.24}\selectfont \makebox[0pt]{$\tilde{\vr X}_{b,k}$}}
  \put(249.46,30.43){\fontsize{8.54}{10.24}\selectfont \makebox[0pt]{$\tilde{\vr Y}_{b,k}$}}
  \put(232.45,54.24){\fontsize{8.54}{10.24}\selectfont $\tilde{\vr Z}_{b,k}$}
  \put(164.42,118.02){\fontsize{8.54}{10.24}\selectfont \makebox[0pt]{$\vr z_1^{[k]}$}}
  \put(329.49,27.87){\fontsize{8.54}{10.24}\selectfont $\hat{\vr \msg}$}
  \put(249.46,118.02){\fontsize{8.54}{10.24}\selectfont \makebox[0pt]{$\vr z_6^{[k]}$}}
  \put(266.47,118.02){\fontsize{8.54}{10.24}\selectfont \makebox[0pt]{$\vr z_7^{[k]}$}}
  \put(283.48,118.02){\fontsize{8.54}{10.24}\selectfont \makebox[0pt]{$\vr z_8^{[k]}$}}
  \put(181.43,118.02){\fontsize{8.54}{10.24}\selectfont \makebox[0pt]{$\vr z_2^{[k]}$}}
  \put(198.44,118.02){\fontsize{8.54}{10.24}\selectfont \makebox[0pt]{$\vr z_3^{[k]}$}}
  \put(215.44,118.02){\fontsize{8.54}{10.24}\selectfont \makebox[0pt]{$\vr z_4^{[k]}$}}
  \put(232.45,118.02){\fontsize{8.54}{10.24}\selectfont \makebox[0pt]{$\vr z_5^{[k]}$}}
  \put(300.95,119.13){\fontsize{16.63}{19.96}\selectfont ...}
  \put(193.43,77.20){\fontsize{8.54}{10.24}\selectfont \makebox[0pt][r]{$i \sim \unif\{1,\ldots,b\}$}}
  \end{picture}%
\fi
\caption{\label{fig:collapsed_channel_modadditive}%
 Collapsed channel: a probabilistic equivalence to iterative mapping}
\end{figure*}

Note that the upper bound of Theorem~\ref{theorem:Cifb_UB} can sometimes be strict, i.e. there are examples of sequences $\vr z$ for which $\Cifb <  ( 1 - \rho(\vr z) ) \log|\mathcal{X}|$, as shown in the following example. \selector{We}{I} do not have an expression for the IFB capacity.
\begin{example}\label{example:half_random_sequence}
Consider for the binary additive channel, the sequence $\vr z$ which consists of blocks with ever increasing size. The first half of each block is $0$, and the second half block is chosen randomly $Z_i \sim Ber(\half)$. With high probability, the finite state compressibility of the sequence is $\half$ (which can be attained, for example, by block-to-variable encoding, using one bit to denote the sequence of zeros). However, the IFB capacity of the channel is 0 with high probability, since for any encoder and decoder with large block size, approximately half of the blocks will be received in error. Therefore there exist sequences for which the inequality is strict.
\end{example}
} 

\subsection{Universally attaining the IFB capacity over the modulo-additive channel}\label{sec:ma_adaptive_rho}
\onlyphd{
In this section, a universal system for the modulo-additive channel with an unknown state sequence is presented. The basis is Theorem~\ref{theorem:adaptive_L} from Section~\ref{sec:examples_compression}, that shows that for a wide range of sequential source encoders, there is a communication scheme that asymptotically attains the rate $\log | \mathcal{X} | - \frac{1}{n} L(\vr z)$, where $L(\vr z)$ is the compression length of the sequence $\vr z$ by the source encoder, i.e. the number of bits used to encode the sequence. Substituting the compression length of the Lempel-Ziv (LZ78) algorithm, which satisfies the conditions of the theorem (see Section~\ref{sec:examples_compression_modadditive}) the finite state compressibility is obtained. This yields the following theorem:
} 
\onlypaper{
In this section, a universal system for the modulo-additive channel with an unknown state sequence is presented. It is first shown (see also \cite{YL_ModuloAdditiveEilat}\cite[\S 10.5]{YL_PhdThesis}), that for a wide range of sequential source encoders, there is a communication scheme that asymptotically attains the rate $\log | \mathcal{X} | - \frac{1}{n} L(\vr z^n)$, where $L(\vr z^n)$ is the compression length of the $n$-length sequence $\vr z$ by the source encoder, i.e. the number of bits used to encode the sequence.

Let us first define a class of sequential source encoders, for which Theorem~\ref{theorem:adaptive_L} below applies. Only source encoders that have the following structure are considered: The encoding algorithm is unaware and is not a function of the sequence length $n$. The encoder receives the sequence for compression letter by letter. From time to time, the encoder emits parts of the compressed sequence. After the last letter is entered into the source encoder, it receives an indication that the sequence has ended and may emit the final part of the compressed sequence.

For each sequence $\vr z$, of any given length, define $L_S(\vr z)$ as the unterminated coding length of the sequence, i.e. the length of the output of the encoder after the input $\vr z$ has been fed, but the sequence has not been terminated, i.e. the encoder did not receive an indication that the sequence ended and is expecting additional input. Define $L_T(\vr z) = L(\vr z)$ as the terminated coding length, i.e. the length of the output after the encoder received the termination indication. The sequence $\vr z$ is uniquely decodable from the $L_T(\vr z)$ bits of the terminated code, but not necessarily from the $L_S(\vr z)$ bits of the unterminated one. The difference $L_T(\vr z) - L_S(\vr z) \geq 0$ is the information stored in the encoder which has not been output yet. The class of source encoders is defined by the two assumptions below:
\begin{enumerate}[(A)]
\item The difference between the terminated and unterminated lengths is bounded by an asymptotically negligible value: $\frac{1}{n} (L_T(\vr z) - L_S(\vr z)) \leq \ \frac{1}{n} \Delta_L(n) \arrowexpl{n \to \infty} 0$ \\
This can be considered an embodiment of the limitation to ``sequential'' encoders and precludes encoders that process the entire sequence before producing outputs.
\item The encoding length does not decrease when the sequence is extended: $L_T(z_1^i) \geq L_T(z_1^{i-1})$. This is a technical requirement intended to simplify the analysis.
\end{enumerate}

\begin{theorem}\label{theorem:adaptive_L}
Given a sequential source coding scheme with input symbols from alphabet $\mathcal{X}$ that satisfies assumptions (A),(B), and assigns a codeword length of $L(\vr z)$ to the sequence $\vr z \in \mathcal{X}^n$, then for any $\epsilon>0$ there exists a sequence of adaptive-rate encoders and decoders using common randomness and feedback, for increasing block lengths $n$ over the channel $\vr y = \vr x + \vr z$ ($\vr x, \vr y, \vr z \in \mathcal{X}^n$), in which for any individual noise sequence $\vr z$ with probability at least $1-\epsilon$, the message is correctly decoded with rate of at least
\begin{equation}
R \geq \Remp(\vr z) - \delta_n
,
\end{equation}
where
\begin{equation}\label{eq:ma6281}
\Remp(\vr z) = \log | \mathcal{X} | - \frac{1}{n} L(\vr z)
,
\end{equation}
\begin{equation}
\delta_n = 3 \sqrt { \frac{\log |\mathcal{X}|}{n} \cdot \left[ \log \left(\frac{n \cdot |\mathcal{X}|}{\epsilon}\right) +  \Delta_L^{\max}(n) \right]} \arrowexpl{n \to \infty} 0
,
\end{equation}
and $\Delta_L^{\max}(n) = \max \{\Delta_L(i) \}_{i=1}^n$.
\end{theorem}

The communication scheme and the proof of Theorem~\ref{theorem:adaptive_L} appear in Appendix~\ref{sec:modadditive_scheme_and_theorem} (see also the proof outline in Section~\ref{sec:overview}). As shown in Appendix~\ref{sec:proof_rho}, both assumptions are satisfied by Lempel-Ziv algorithms LZ77 \cite{LZ77} and LZ78 \cite{LZ78}. Note the similarity between the rate expression \eqref{eq:ma6281} and the capacity of an ergodic stochastic modulo-additive channel, which is also attained with a uniform prior, $C = \overline I(X^{\infty};Y^{\infty})=\overline H(Y^{\infty})- \overline H(Y^{\infty}|X^{\infty}) =  \log | \mathcal{X} |-\overline H(Z^{\infty})$.  $\frac{1}{n} L(\vr z)$ can be considered a generalized empirical measure of the noise entropy rate. In this sense, Theorem~\ref{theorem:adaptive_L} is a generalization of Shayevitz and Feder's result \cite{Ofer_ModuloAdditive}.

Substituting the compression length of Lempel and Ziv's LZ78 algorithm, the finite state compressibility is obtained. This yields the following theorem:
} 
\begin{theorem}\label{theorem:adaptive_rho}
When the system of Theorem~\ref{theorem:adaptive_L} is used in conjunction with LZ78 source encoder, over the modulo additive channel, then the following holds:
For every infinite noise sequence $\vr z^\infty$ and every $\epsilon,\delta>0$ there is $n$ large enough so that when the system is operated over $n$ channel uses, then with probability $1-\epsilon$, the message is correctly decoded and the rate is at least $( 1 - \rho(\vr z) )\log|\mathcal{X}|  - \delta$.
\end{theorem}
\begin{corollary_in_theorem}\label{corollary:univ}
The system defined above is IFB-universal.
\end{corollary_in_theorem}
\begin{corollary_in_theorem}\label{corollary:ergodic}
The system attains the Shannon capacity of every modulo-additive channel with a stationary ergodic noise sequence.\excluded{\footnote{Note there is an error in our paper \cite{YL_ModuloAdditiveEilat} where it was claimed the system only attains the mutual information.}}
\end{corollary_in_theorem}

The proof of the theorem and its corollaries is given in Appendix~\ref{sec:proof_rho}, and its main point is to show that LZ78 satisfies the assumptions of Theorem~\ref{theorem:adaptive_L}.

Theorems~\ref{theorem:adaptive_L},\ref{theorem:adaptive_rho} are finite horizon, i.e. the system is designed for a given transmission length $n$, and because $n$ needs to grow for the overhead $\delta$ vanish, the asymptotic universality is obtained by a series of systems rather than a single one, as is standard in information theory. However, it is possible to design horizon-free systems in which the transmission length is not limited and redundancy vanishes with time \onlypaper{\cite[\S 8.6]{YL_PhdThesis}}. \onlyphd{In Section~\ref{sec:rate_adaptive_inf_horizon} it is shown how to do this for a rather general case, and this holds for the particular cases of Theorems~\ref{theorem:adaptive_L},\ref{theorem:adaptive_rho}, which are all derivatives of Theorem~\ref{theorem:framework}.}

The results of this section rely on LZ compression algorithm and stress the relations between channel coding rates and compression ratios, and between IFB capacity and finite state compressibility. This relation is intuitively appealing and the resulting system is relatively simple. On the other hand, the modified universal system presented in the next section yields better bounds on the convergence of the overhead terms, which also hold \emph{uniformly} in $\vr z$.\footnote{Notice that because Theorem~\ref{theorem:adaptive_rho} essentially indicates convergence to the IFB \emph{capacity}, the convergence cannot hold uniformly in $\vr z$, as the IFB capacity may be obtained by competing systems of ever growing complexity, depending on the noise sequence. In the next section, the IFB system and the universal system are compared directly for finite $n,k$ without referring to the asymptotic value of the IFB capacity, thereby making uniform convergence possible.}

\section{The redundancy of the universal system}\label{sec:ma_univ_redundancy}
Let us now consider the redundancy of the universal system and how fast it converges to zero as the block length increases, under the context of the modulo-additive channel. The interesting question is how large the transmission length size $n$ needs to be, in order to successfully compete with an IFB system of a given block size $k$. Unfortunately, $n$ must grow at least as fast as $|\mathcal{X}|^k$, approximately. Thus, even considering reference systems of relatively small block sizes compared to standard block codes, for instance $k=100$, the competition becomes infeasible.

\subsection{A definition of redundancy}
Before giving a definition of the redundancy, some considerations for the definition are provided. The finite state compressibility $\rho(\vr z)$ of the infinite sequence $\vr z$, used in Theorems~\ref{theorem:Cifb_UB},\ref{theorem:adaptive_rho} is irrelevant for the analysis of convergence. This is because $\rho(\vr z)$ is an asymptotical value, and the performance of the best block encoder or finite state machine encoder on any finite block of $n$ symbols, does not indicate anything about the final finite state compressibility. In other words, there is no guarantee on the rate of convergence of the $\limsup$ in \eqref{eq:ma446}. Consider as example a sequence $\vr z$ which is incompressible up to time $n_1$ and then all zero to infinity, or vice versa. Note that incompressible sequences must exist, by Kraft's inequality. Therefore, instead of considering the convergence of the rates obtained by the best IFB system and the universal system to $(1-\rho) \log |\mathcal{X}|$, the comparison is between the rate obtained by the best IFB system of block size $k$, with a universal system, at time $n$.

While the asymptotic results of Theorems~\ref{theorem:Cifb_UB},\ref{theorem:adaptive_rho} require the error probability of both systems to tend to zero with $n$, at a finite block length, a certain non-zero error probability would exist. In the two systems, error probabilities have different meanings: the IFB system's error probability is block-wise and the universal system's error probability is measured on the entire transmission. Therefore, for a fair comparison, and in order to remove the dependence on the error probability from the results, let us consider the following definition of an effective rate, for a system operating over block of size $k$ with rate $R$ and error probability $\epsilon$:
\begin{equation}\label{eq:ma468}
R^* = (1-\epsilon) R - \frac{1}{k} h_b(\epsilon)
.
\end{equation}
This definition is motivated by Fano's inequality; see for example \eqref{eq:maR_UB}. While the first factor is usually termed the good-put, i.e. the number of error free bits\onlyphd{~(see Section~\ref{sec:good_put_bound})}, the second factor compensates for the uncertainty in knowing whether there is an error or not. For example, a system delivering $R=1$ bit per channel use with error probability $\epsilon=\half$ per block of size $k=1$, i.e. transmits no information, would have $(1-\epsilon) R = \half$ but $R^* = 0$. Equivalently, $R^*$ may be interpreted as a bound on the normalized mutual information between the input message and the decoded message, given the parameters $R,\epsilon$ and $k$. Notice that $R^* \arrowexpl{\epsilon \to 0} R$. Regardless of the interpretation of $R^*$, the results below yield meaningful bounds on the actual rates $R$ by referring to $R^*$.

Another issue is how to compare a universal system with transmission length $n$ and an IFB system whose block length $k$ does not divide $n$. For a worst-case comparison, let us give the IFB system the luxury of using the last block that possibly extends beyond the $n$-th symbol, i.e. $l = \lceil \frac{n}{k} \rceil$ blocks overall, while letting the noise sequence on these symbols $\vr z_{n+1}^{kl}$ take the values which are best for the IFB system.

A definition of the minimax redundancy is given below. Let $E,D$ define an IFB system with block length $k$ and rate $R_\tsubs{IFB}$ (Definition~\ref{def:E_and_D}), which is iteratively mapped to the channel $P_\tsubs{Y|X}^{(\theta)}, \theta \in \Theta$, over $kl$ symbols, where $l = \lceil \frac{n}{k} \rceil$, and yield average error probability $\epsilon_\tsubs{IFB}$ (Definition~\ref{def:mean_eps}). Similarly, on the same channel over $n$ symbols, an adaptive system $U$ with feedback and common randomness (Section~\ref{sec:universal_mod_defs}), whose design must not depend on $\theta$, guarantees a rate of at least $R_\tsubs{U} = R_\tsubs{U}(\theta)$ with an error probability of at most $\epsilon_\tsubs{U}$. As in Definition~\ref{def:IFBAFB_universality}, $\epsilon_\tsubs{U}$ includes both the probability of error and the probability that the system's rate falls below $R_\tsubs{U}$. While $R_\tsubs{U}$ is allowed to depend on the channel index $\theta$, $\epsilon_\tsubs{U}$ is required to be fixed. Let $R_\tsubs{IFB}^* = R_\tsubs{IFB} \cdot (1 - \epsilon_\tsubs{IFB}) - \frac{1}{k} h_b(\epsilon_\tsubs{IFB})$ and $R_\tsubs{U}^* = R_\tsubs{U} \cdot (1 - \epsilon_\tsubs{U}) - \frac{1}{n} h_b(\epsilon_\tsubs{U})$. The rate and error probability for each system, are defined given the channel and the system. The values related to the IFB system, $R_\tsubs{IFB}$, $\epsilon_\tsubs{IFB}$ and $R_\tsubs{IFB}^*$ depend implicitly on $(n,k,E,D,\theta)$, while the values related to the rate adaptive system, $R_\tsubs{U}$ and $R_\tsubs{U}^*$ depend implicitly on $(n,U,\theta)$.

The minimax redundancy for finite $n,k$ is defined as follows:
\begin{equation}\label{eq:madef_IFB_redundancy_finite_nk}
\Delta^*(n,k) = \min_{U} \max_{\theta \in \Theta} \left[ \max_{E,D} \left( R_\tsubs{IFB}^* \right) - R_\tsubs{U}^* \right]
.
\end{equation}
In other words, it is the minimal gap $R_\tsubs{IFB}^* - R_\tsubs{U}^*$ that can be universally guaranteed by a single system $U$ over all channels. Note that the definition allows the universal system to depend on $k$ but this relaxation is not used by the universal system achieving the bounds below. For the special case of the modulo additive channel, the channel index $\theta$ is replaced by the noise sequence $\vr z_1^{kl}$.

\subsection{The minimax redundancy for the modulo-additive channel class}
The minimax redundancy of a universal system compared to the IFB system over the modulo-additive channel is bounded below. Let us begin with the main asymptotical result which formalizes the notion that, the minimum transmission length behaves asymptotically like $|\mathcal{X}|^k$:

\begin{theorem}\label{theorem:mod_additive_redundancy_asymp}
For a given $k$ and $\delta > 0$, let $n^*=n^*(k,\delta)$ be the minimum $n$ such that for the modulo additive channel, $\Delta^*(n,k) \leq \delta \log |\mathcal{X}|$, then:
\begin{equation}\label{eq:ma553k}
\lim_{\delta \to 0} \lim_{k \to \infty} \frac{\log n^*(k,\delta)}{k \log |\mathcal{X}|} = 1
.
\end{equation}
\end{theorem}

Theorem \ref{theorem:mod_additive_redundancy_asymp} is an immediate consequence of the explicit bounds given in the remainder of this section. Theorem~\ref{theorem:mod_additive_redundancy} below specifies bounds on $\Delta^*(n,k)$, and its Corollary~\ref{corollary:mod_additive_redundancy} specifies bounds on the minimum transmission length $n^*$ defined above.

\begin{theorem}\label{theorem:mod_additive_redundancy}
The minimax redundancy \eqref{eq:madef_IFB_redundancy_finite_nk} for the channel $\vr y = \vr x + \vr z$ ($\vr x, \vr y, \vr z \in \mathcal{X}^n$) satisfies:
\begin{equation}\label{eq:ma502}
\Delta_{-} \leq \Delta^*(n,k) \leq \Delta_{+}
,
\end{equation}
where
\begin{equation}\begin{split}\label{eq:ma1063b}
\Delta_{-}
&=
\begin{cases}
\left\lfloor  \log \left( k \tau \right)\frac{1}{ \log |\mathcal{X}|} \right\rfloor \frac{\log |\mathcal{X}|}{2 k} & \tau > \frac{ |\mathcal{X}| }{k} \\
\frac{\log |\mathcal{X}|}{2 |\mathcal{X}|} \cdot \tau  & \tau \leq \frac{ |\mathcal{X}| }{k}
\end{cases}
,
\end{split}\end{equation}
and for $\tau \leq 1$:
\begin{equation}\label{eq:ma506}
\Delta_{+} = \frac{\tau}{2} \log \left( \frac{1}{\tau} \right) + \left( \frac{k}{4} \tau^2 + \tau \right) \log e + \delta_n^* + \frac{k}{n} \log (e |\mathcal{X}|)
.
\end{equation}
The parameters are defined as follows:
\begin{equation}\begin{split}\label{eq:ma1133j}
\tau &= \frac{|\mathcal{X}|^k}{n}
\\
\delta_n^* &= 4 \sqrt{\frac{\log |\mathcal{X}| \cdot \log \left( n^2 |\mathcal{X}| \right)}{n}}
.
\end{split}\end{equation}
Furthermore, the universal system attaining the upper bound $\Delta_{+}$ does not depend on $k$.
\end{theorem}
The theorem is proven in the next section. Note that both bounds require $\tau$ to be small, and thus $n$ to be large, in order to achieve a small redundancy. While the lower bound is linear for $\tau \leq \frac{ |\mathcal{X}| }{k}$, for large values, it increases significantly more slowly, like $\log \tau$. This is because of the in-efficiency of the IFB system used in the lower bound, at high rates. The value of $\Delta_{-}$ in the range $\tau \leq \frac{|\mathcal{X}|}{k}$, is limited to $\frac{\log |\mathcal{X}|}{2 k}$, i.e. a rate offset of half a symbol per block. The bound for the range $\tau > \frac{ |\mathcal{X}| }{k} $ is useful, in showing that even if one is satisfied with a redundancy of more than $\frac{\log |\mathcal{X}|}{2 k}$, $\tau$ must be kept small. Fig.~\ref{fig:mod_additive_redundancy_fig1} illustrates the bounds of Theorem~\ref{theorem:mod_additive_redundancy} as function of the transmission length $n$, for a constant value of $k$. The logarithmic and quantized behavior of the lower bound for small values of $n$ can be observed. Fig.~\ref{fig:mod_additive_redundancy_fig2} presents $n^*(k,\delta)$, i.e. the minimum $n$ required to obtain $\Delta^*(n,k) \leq \delta \cdot \log |\mathcal{X}|$, according to the bounds of Theorem~\ref{theorem:mod_additive_redundancy}, as a function of $k$. The gap between the upper and lower bounds is significant: a little more than an order of magnitude. However, their trend is similar. This observation is formalized by Corollary~\ref{corollary:mod_additive_redundancy} below, concerning the asymptotical behavior of $n$:

\begin{figure}
  \includegraphics[width=8cm]{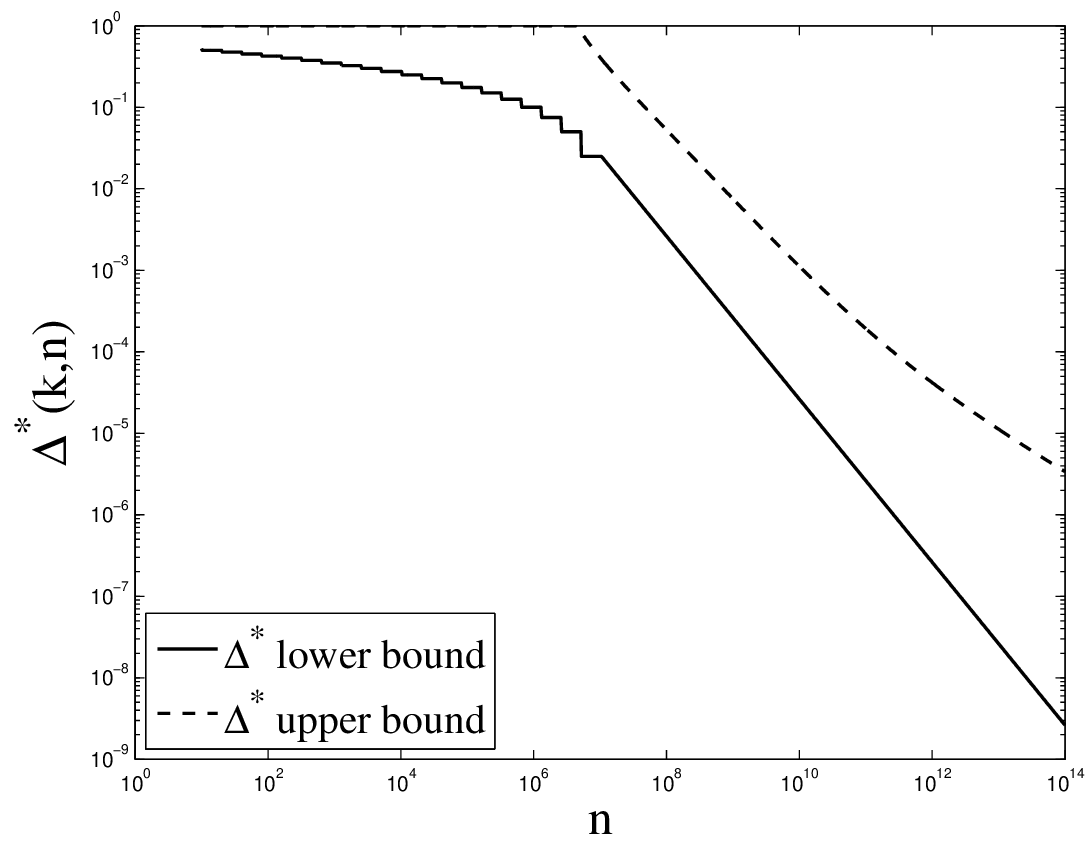}\\
  \caption[Theorem~\ref{theorem:mod_additive_redundancy} bounds on redundancy]{The upper and lower bound on the redundancy $\Delta^*(n,k)$ of universal systems given by Theorem~\ref{theorem:mod_additive_redundancy} for $k=20, |\mathcal{X}|=2$.}\label{fig:mod_additive_redundancy_fig1}
\end{figure}
\begin{figure}
  \includegraphics[width=8cm]{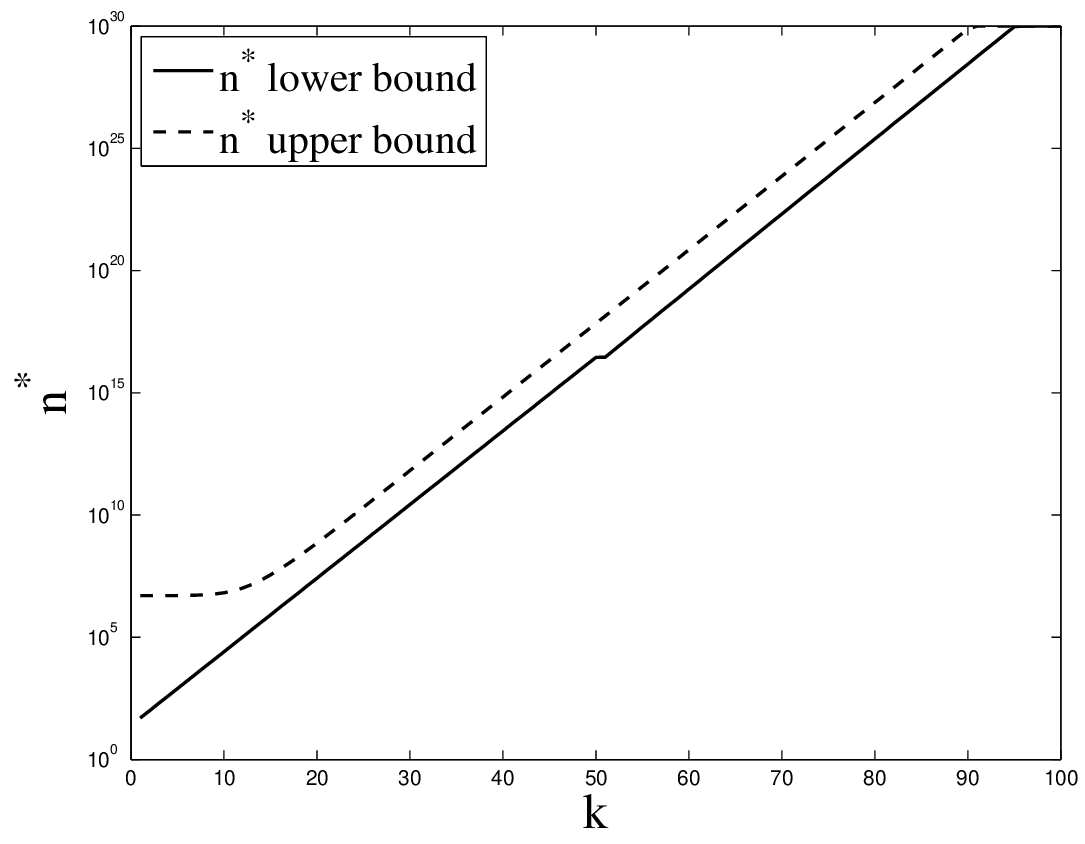}\\
  \caption[Theorem~\ref{theorem:mod_additive_redundancy} bounds on minimum transmission length]{The minimum transmission length $n$ required to obtain a minimax redundancy $\Delta^*(n,k) \leq \delta \cdot \log |\mathcal{X}|$, according to the bounds of Theorem~\ref{theorem:mod_additive_redundancy} as function of the IFB block size $k$, for $|\mathcal{X}|=2, \delta = 0.01$.}\label{fig:mod_additive_redundancy_fig2}
\end{figure}

\begin{corollary_in_theorem}\label{corollary:mod_additive_redundancy}
\begin{equation}\label{eq:ma546}
\frac{k}{|\mathcal{X}|} \cdot |\mathcal{X}|^{(1-2\delta) k} \leq n^* \leq \frac{k}{\min \left[ T(k,\delta,|\mathcal{X}|), 1 \right]} \cdot |\mathcal{X}|^k
,
\end{equation}
where $g(\tau) = \tau \log \left( \frac{1}{\tau} \right)$ and
\begin{equation}\label{eq:ma565}
T(k,\delta,|\mathcal{X}|) = k \cdot g^{-1} \left( \tfrac{1}{3} (\delta - 12 \cdot |\mathcal{X}|^{-k/2}) \cdot \log |\mathcal{X}| \right)
.
\end{equation}
\end{corollary_in_theorem}

For large $k$ and fixed $\delta$, $k^{-1} \cdot T(k,\delta,|\mathcal{X}|) \arrowexpl{k \to \infty} g^{-1} \left( \tfrac{1}{3} \delta \cdot \log |\mathcal{X}| \right) = \const$, and thus for large enough $k$, $T > 1$ and does not dominate the upper bound \eqref{eq:ma546}. For a small value of $\delta$ both bounds of Corollary~\ref{corollary:mod_additive_redundancy} behave approximately like $|\mathcal{X}|^{k}$. Corollary~\ref{corollary:mod_additive_redundancy} results from a technical simplification of the bounds of Theorem~\ref{theorem:mod_additive_redundancy} and is proven in Appendix~\ref{sec:proof_of_corollary_mod_additive_redundancy}. Most important is the lower bound on $n^*$ which indicates the minimum rate at which $n^*$ must grow. Finally, Theorem~\ref{theorem:mod_additive_redundancy_asymp} is an immediate consequence of Corollary~\ref{corollary:mod_additive_redundancy}.

Note that the system attaining the upper bound of Theorem~\ref{theorem:mod_additive_redundancy} yields a stronger type of universality than claimed in Theorem~\ref{theorem:adaptive_rho}, because for each value of $n$, the overheads are uniformly bounded for any noise sequence $\vr z$, whereas previously, while the overheads are guaranteed to tend to zero asymptotically with $n$, this convergence is not necessarily uniform with respect to $\vr z$.

Unlike other results in this \selector{paper}{part} where the IFB system is used merely as a converse, in the proof for the lower bound $\Delta_{-}$, it is required to devise a specific IFB system. Here, the simplicity of the IFB system, which makes the other results intuitive and simple to derive, complicates the proof. The collapsed channel capacity, which upper bounds the IFB system rate, is usually not achievable by a finite block encoder, and a specific channel has to be devised in order for the IFB system to operate provably better than any universal system. It seems that richer classes of reference systems, e.g. systems using feedback as considered in \cite{MisraPorosityISIT12}, may result in simpler and tighter lower bounds.

\subsection{Proof of Theorem~\ref{theorem:mod_additive_redundancy}}\label{sec:proof_theorem_mod_additive_redundancy}

\subsubsection{Lower bound (reverse part)}\label{sec:proof_theorem_mod_additive_redundancy_reverse}
In order to show that the redundancy must be at least $O \left( \frac{|\mathcal{X}|^k}{n} \right)$ an example random channel is constructed, in the following way. First, the encoder $E$ is defined. Then, a way to generate noise sequences $\vr z$ is defined, such that the noise sequences belong to a sub-set of all possible sequences $\vr z \in \mathbb{Z}_d$, and it is possible to decode the given code with zero error probability for any noise sequence in the set. The IFB decoder $D$ is specified only after the noise sequence has been chosen. The sequence $\vr z$ is drawn in a randomized way, thus creating a stochastic ``test'' channel. It is shown that there exists a noise sequence for which the rate of the universal system is bounded by the normalized mutual information over the test channel. Asymptotically, as there are certain constraints on the choice of the noise sequence, this normalized mutual information tends to the rate of the IFB encoder. However, at the beginning of the sequence, the entropy of the sequence is a little higher than its the long-term average, and thus the mutual information is a little lower than its asymptotic value, which equals the rate of the IFB encoder. Thus, the rate of the universal system is bounded by a value lower than the rate of the IFB system.

Let us first describe the IFB encoder. The encoder sends $d$ symbols from the alphabet $\mathcal{X}$ over $k$ channel uses, and therefore has a rate
\begin{equation}\label{eq:ma648}
R_\tsubs{IFB} = \frac{d}{k} \log |\mathcal{X}|
.
\end{equation}
The encoding is simple: the first $k-d$ symbols (prefix) are constant and the rest $d$ symbols (suffix) contain the message. The decoder would be able to know the value of the noise sequence over the prefix symbols, and knows a list of all possible noise sequences. Assuming that there is no more than one noise sequence with any given prefix, then zero error probability is possible: the decoder finds the noise sequence from the prefix symbols, and cancels it on the suffix to find the message.
\begin{figure}[h]
\centering
\ifpdf
  \setlength{\unitlength}{1bp}%
  \begin{picture}(184.17, 75.32)(0,0)
  \put(0,0){\includegraphics{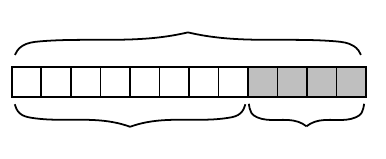}}
  \put(12.76,35.25){\fontsize{7.11}{8.54}\selectfont \makebox[0pt]{$0$}}
  \put(26.93,35.25){\fontsize{7.11}{8.54}\selectfont \makebox[0pt]{$0$}}
  \put(41.10,35.25){\fontsize{7.11}{8.54}\selectfont \makebox[0pt]{$0$}}
  \put(55.28,35.25){\fontsize{7.11}{8.54}\selectfont \makebox[0pt]{$0$}}
  \put(69.45,35.25){\fontsize{7.11}{8.54}\selectfont \makebox[0pt]{$0$}}
  \put(83.62,35.25){\fontsize{7.11}{8.54}\selectfont \makebox[0pt]{$0$}}
  \put(97.80,35.25){\fontsize{7.11}{8.54}\selectfont \makebox[0pt]{$0$}}
  \put(111.97,35.25){\fontsize{7.11}{8.54}\selectfont \makebox[0pt]{$0$}}
  \put(126.14,35.25){\fontsize{7.11}{8.54}\selectfont \makebox[0pt]{$m_1$}}
  \put(140.31,35.25){\fontsize{7.11}{8.54}\selectfont \makebox[0pt]{$m_2$}}
  \put(154.49,35.25){\fontsize{7.11}{8.54}\selectfont \makebox[0pt]{$m_3$}}
  \put(168.66,35.25){\fontsize{7.11}{8.54}\selectfont \makebox[0pt]{$m_4$}}
  \put(73.19,64.10){\fontsize{7.11}{8.54}\selectfont $k$ symbols}
  \put(147.56,7.20){\fontsize{7.11}{8.54}\selectfont \makebox[0pt]{suffix: $d$ symbols}}
  \put(61.54,7.20){\fontsize{7.11}{8.54}\selectfont \makebox[0pt]{prefix: $k-d$ symbols}}
  \end{picture}%
\else
  \setlength{\unitlength}{1bp}%
  \begin{picture}(184.17, 75.32)(0,0)
  \put(0,0){\includegraphics{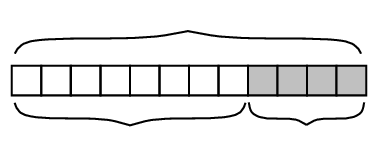}}
  \put(12.76,35.25){\fontsize{7.11}{8.54}\selectfont \makebox[0pt]{$0$}}
  \put(26.93,35.25){\fontsize{7.11}{8.54}\selectfont \makebox[0pt]{$0$}}
  \put(41.10,35.25){\fontsize{7.11}{8.54}\selectfont \makebox[0pt]{$0$}}
  \put(55.28,35.25){\fontsize{7.11}{8.54}\selectfont \makebox[0pt]{$0$}}
  \put(69.45,35.25){\fontsize{7.11}{8.54}\selectfont \makebox[0pt]{$0$}}
  \put(83.62,35.25){\fontsize{7.11}{8.54}\selectfont \makebox[0pt]{$0$}}
  \put(97.80,35.25){\fontsize{7.11}{8.54}\selectfont \makebox[0pt]{$0$}}
  \put(111.97,35.25){\fontsize{7.11}{8.54}\selectfont \makebox[0pt]{$0$}}
  \put(126.14,35.25){\fontsize{7.11}{8.54}\selectfont \makebox[0pt]{$m_1$}}
  \put(140.31,35.25){\fontsize{7.11}{8.54}\selectfont \makebox[0pt]{$m_2$}}
  \put(154.49,35.25){\fontsize{7.11}{8.54}\selectfont \makebox[0pt]{$m_3$}}
  \put(168.66,35.25){\fontsize{7.11}{8.54}\selectfont \makebox[0pt]{$m_4$}}
  \put(73.19,64.10){\fontsize{7.11}{8.54}\selectfont $k$ symbols}
  \put(147.56,7.20){\fontsize{7.11}{8.54}\selectfont \makebox[0pt]{suffix: $d$ symbols}}
  \put(61.54,7.20){\fontsize{7.11}{8.54}\selectfont \makebox[0pt]{prefix: $k-d$ symbols}}
  \end{picture}%
\fi
\caption{\label{fig:redundancy_LB_ref_encoder}%
 The reference encoder for the converse of Theorem~\ref{theorem:mod_additive_redundancy}}
\end{figure}

\label{page:definition_of_test_channel}
Next the test channel is defined. The set $\mathbb{Z}_d$ of allowed noise sequences are simply those sequences for which each prefix $\vr z_{k \cdot (i-1) + 1}^{k \cdot (i-1) + k - d}$ ($i=1,2,\ldots$) uniquely determines the respective suffix $\vr z_{k \cdot (i-1) + k-d+1}^{k \cdot (i-1) + k}$. The random noise sequence is generated as follows: at each block of $k$ symbols, the prefix of $k-d$ symbols is chosen randomly, uniformly over all possible $|\mathcal{X}|^{k-d}$ prefixes, and independently of the past noise sequence. Then, if the prefix had appeared before, the suffix equals the suffix of the noise sequence that already appeared. Otherwise, the suffix is chosen randomly, uniformly over all possible $|\mathcal{X}|^{d}$ suffixes.\footnote{An alternative way of generating the noise sequence, which yields the maximum entropy, is by uniform drawing over the set of all possible $k$-length sequences that satisfy the unique prefix condition. However this complicates the bound.}

The choice of the first sequence $\vr z_1^{[k]}$ is uniform over all possible sequences, and therefore the entropy of the noise sequence in the first block is maximal, $\log (|\mathcal{X}|^k)$. The choice of the noise sequences narrows with time, and after a long while, all possible prefixes would have been chosen, with one noise sequence per prefix. In this case, the choice of the suffix is determined by the prefix, and the entropy per $k$-block is $\log (|\mathcal{X}|^{k-d})$. This is the minimum entropy per block attained. The behavior of the entropy $H(\vr Z^n)$ in this channel is shown in Fig~\ref{fig:test_channel_noise_entropy}.

\begin{figure}
\centering
\ifpdf
  \setlength{\unitlength}{1bp}%
  \begin{picture}(184.02, 197.56)(0,0)
  \put(0,0){\includegraphics{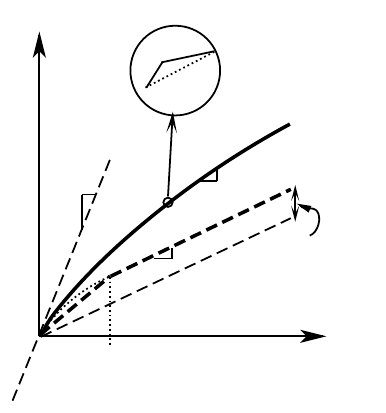}}
  \put(12.34,141.26){\rotatebox{90.00}{\fontsize{8.54}{10.24}\selectfont \smash{\makebox[0pt][l]{$H(\vr Z^n)$}}}}
  \put(144.19,25.85){\fontsize{8.54}{10.24}\selectfont $n$}
  \put(27.25,86.04){\rotatebox{66.25}{\fontsize{6.83}{8.19}\selectfont \smash{\makebox[0pt][l]{$\log|\mathcal{X}|$}}}}
  \put(65.40,76.34){\rotatebox{23.87}{\fontsize{8.54}{10.24}\selectfont \smash{\makebox[0pt][l]{Lower bound of \eqref{eq:ma1045}}}}}
  \put(89.65,98.33){\rotatebox{25.97}{\fontsize{6.83}{8.19}\selectfont \smash{\makebox[0pt][l]{$\to \overline H_1$}}}}
  \put(85.81,72.32){\rotatebox{25.22}{\fontsize{6.83}{8.19}\selectfont \smash{\makebox[0pt][l]{$\overline H_1$}}}}
  \put(30.58,36.20){\rotatebox{37.41}{\fontsize{6.83}{8.19}\selectfont \smash{\makebox[0pt][l]{$\overline H_0$}}}}
  \put(33.18,21.84){\fontsize{6.83}{8.19}\selectfont $n=k |\mathcal{X}|^{k-d}$}
  \put(92.04,116.70){\rotatebox{32.14}{\fontsize{8.54}{10.24}\selectfont \smash{\makebox[0pt][l]{$H(\vr Z^n)$}}}}
  \put(104.59,185.23){\fontsize{8.54}{10.24}\selectfont Entropy during a}
  \put(105.22,168.14){\fontsize{8.54}{10.24}\selectfont  noise sequence}
  \put(105.49,176.29){\fontsize{8.54}{10.24}\selectfont $k$-length}
  \put(147.44,84.12){\rotatebox{25.24}{\fontsize{6.83}{8.19}\selectfont \smash{\makebox[0pt][r]{Bound on $\Delta$ \eqref{eq:ma1063k}}}}}
  \end{picture}%
\else
  \setlength{\unitlength}{1bp}%
  \begin{picture}(184.02, 197.56)(0,0)
  \put(0,0){\includegraphics{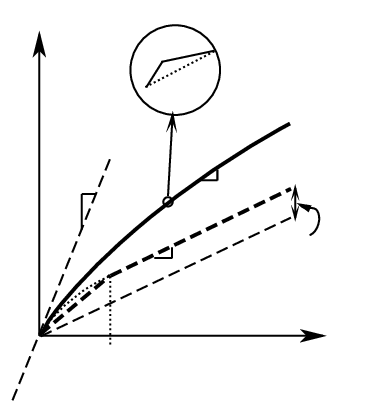}}
  \put(12.34,141.26){\rotatebox{90.00}{\fontsize{8.54}{10.24}\selectfont \smash{\makebox[0pt][l]{$H(\vr Z^n)$}}}}
  \put(144.19,25.85){\fontsize{8.54}{10.24}\selectfont $n$}
  \put(27.25,86.04){\rotatebox{66.25}{\fontsize{6.83}{8.19}\selectfont \smash{\makebox[0pt][l]{$\log|\mathcal{X}|$}}}}
  \put(65.40,76.34){\rotatebox{23.87}{\fontsize{8.54}{10.24}\selectfont \smash{\makebox[0pt][l]{Lower bound of \eqref{eq:ma1045}}}}}
  \put(89.65,98.33){\rotatebox{25.97}{\fontsize{6.83}{8.19}\selectfont \smash{\makebox[0pt][l]{$\to \overline H_1$}}}}
  \put(85.81,72.32){\rotatebox{25.22}{\fontsize{6.83}{8.19}\selectfont \smash{\makebox[0pt][l]{$\overline H_1$}}}}
  \put(30.58,36.20){\rotatebox{37.41}{\fontsize{6.83}{8.19}\selectfont \smash{\makebox[0pt][l]{$\overline H_0$}}}}
  \put(33.18,21.84){\fontsize{6.83}{8.19}\selectfont $n=k |\mathcal{X}|^{k-d}$}
  \put(92.04,116.70){\rotatebox{32.14}{\fontsize{8.54}{10.24}\selectfont \smash{\makebox[0pt][l]{$H(\vr Z^n)$}}}}
  \put(104.59,185.23){\fontsize{8.54}{10.24}\selectfont Entropy during a}
  \put(105.22,168.14){\fontsize{8.54}{10.24}\selectfont  noise sequence}
  \put(105.49,176.29){\fontsize{8.54}{10.24}\selectfont $k$-length}
  \put(147.44,84.12){\rotatebox{25.24}{\fontsize{6.83}{8.19}\selectfont \smash{\makebox[0pt][r]{Bound on $\Delta$ \eqref{eq:ma1063k}}}}}
  \end{picture}%
\fi
\caption[Noise entropy in the test channel of Section~\ref{sec:proof_theorem_mod_additive_redundancy_reverse}]{The entropy of the noise in the test channel $H(\vr Z^n)$ over time, and the lower bounds of \eqref{eq:ma1045}, \eqref{eq:ma971}}\label{fig:test_channel_noise_entropy}%
\end{figure}

Now, because $\epsilon_\tsubs{IFB} = 0$,
\begin{equation}\label{eq:ma1056}
R^*_\tsubs{IFB} = (1-\epsilon_\tsubs{IFB}) R_\tsubs{IFB} - \frac{1}{k} h_b(\epsilon_\tsubs{IFB}) =  R_\tsubs{IFB} = \frac{d}{k} \log |\mathcal{X}|
.
\end{equation}
Therefore:
\begin{equation}\begin{split}\label{eq:ma6710}
\Delta^*(n,k)
& \stackrel{\eqref{eq:madef_IFB_redundancy_finite_nk}}{\geq}
\min_{U} \max_{\vr z_1^n \in \mathbb{Z}_d} \left[ \max_{E,D} \left( R_\tsubs{IFB}^*(E,D) \right) - R_\tsubs{U}^* \right]
\\& \geq
R_\tsubs{IFB}^* - \min_{U} \min_{\vr z_1^n \in \mathbb{Z}_d} \left[  R_\tsubs{U}^* \right]
,
\end{split}\end{equation}
where $R_\tsubs{IFB}^*$ denotes the value defined in \eqref{eq:ma1056} for the specific reference system described.

The universal system guarantees error probability $\epsilon_\tsubs{U}$ for any $\vr z$. By definition, for any $\vr z_1^n \in \mathbb{Z}_d$, $R_\tsubs{U} \geq R_0 = \min_{\vr z_1^n \in \mathbb{Z}_d} \left[  R_\tsubs{U} \right]$. Therefore if $\vr z_1^n$ is drawn randomly in $\mathbb{Z}_d$, then the universal system yields a rate of at least $R_0$, with error probability at most $\epsilon_\tsubs{U}$ over the test channel, and can be converted to a fixed-rate system with feedback with rate $R_0$ over the same channel. Using Fano's inequality, which holds also in the case of feedback (see \eqref{eq:maR_UB} in the proof of Theorem~\ref{theorem:Cifb_UB}, and \eqref{eq:ma8812}),
\begin{equation}\label{eq:ma8812k}
R_0 (1 - \epsilon_\tsubs{U}) - \frac{1}{n} h_b(\epsilon_\tsubs{U}) \leq \frac{1}{n} I(\vr X^n; \vr Y^n) \leq \log |\mathcal{X}| - \frac{1}{n} H(\vr Z^n)
,
\end{equation}
and therefore for any universal system $U$:
\begin{equation}\begin{split}\label{eq:ma689}
\min_{\vr z_1^n \in \mathbb{Z}_d} \left[  R_\tsubs{U}^* \right]
&=
(1 - \epsilon_\tsubs{U}) \underbrace{\min_{\vr z_1^n \in \mathbb{Z}_d} \left[  R_\tsubs{U} \right]}_{R_0} - \frac{1}{n} h_b(\epsilon_\tsubs{U})
\\& \leq
\log |\mathcal{X}| - \frac{1}{n} H(\vr Z^n)
,
\end{split}\end{equation}
which yields the bound:
\begin{equation}\begin{split}\label{eq:ma1063k}
\Delta^*
& \stackrel{\eqref{eq:ma6710}, \eqref{eq:ma689}}{\geq}
R^*_\tsubs{IFB} - \log |\mathcal{X}| + \frac{1}{n} H(\vr Z^n)
\\& \stackrel{\eqref{eq:ma1056}}{=}
\frac{1}{n} H(\vr Z^n) -  \frac{k-d}{k}  \log |\mathcal{X}|
.
\end{split}\end{equation}
Asymptotically, $\frac{1}{n} H(\vr Z^n) \ntoinfty \frac{k-d}{k}  \log |\mathcal{X}|$, and thus the bound above tends to zero. The main point of the proof is to bound the convergence rate of $\frac{1}{n} H(\vr Z^n)$.

It may appear surprising, that while it will be shown that the mutual information over the channel is slightly lower than $R^*_\tsubs{IFB}$, the IFB system transmits rate $R^*_\tsubs{IFB}$ with zero error over this channel. This is explained by the fact that the decoder is designed knowing the specific noise sequence, and therefore its effective rate is not limited by the mutual information.

The next step is to bound $H(\vr Z^n)$. This rather technical derivation is deferred to Appendix~\ref{sec:bound_on_HZ_test_channel}, where the following lemma is proven:
\begin{lemma}\label{lemma:bound_on_HZ_test_channel}
For the distribution of $\vr Z^n$ of the test channel defined above, the entropy satisfies:
\begin{equation}\begin{split}\label{eq:ma1045}
H(\vr Z^{n}) \geq n \cdot \overline H_1 + \min(n, k |\mathcal{X}|^{k-d}) (\overline H_0 - \overline H_1)
.
\end{split}\end{equation}
where
\begin{equation}\begin{split}\label{eq:664}
\overline H_0 & \defeq \frac{k-d/2}{k} \log |\mathcal{X}| \\
\overline H_1 & \defeq  \frac{k-d}{k} \log |\mathcal{X}|
\end{split}\end{equation}
are the initial slope of the bound for small $n$, and the asymptotical entropy rate per symbol, respectively.
\end{lemma}

Substituting the above in \eqref{eq:ma1063k} yields:
\begin{equation}\begin{split}\label{eq:ma1063}
\Delta^*
&\geq
\frac{1}{n} H(\vr Z^n) -  \frac{k-d}{k}  \log |\mathcal{X}|
\\& \stackrel{\eqref{eq:ma1045}}{\geq}
\overline H_1 + \min \left(1, \frac{k |\mathcal{X}|^{k-d}}{n} \right) (\overline H_0 - \overline H_1) - \overline H_1
\\& \stackrel{\eqref{eq:ma840}}{=}
\min \left(1, \frac{k |\mathcal{X}|^{k-d}}{n} \right) \cdot \frac{d}{2 k} \cdot \log |\mathcal{X}|
.
\end{split}\end{equation}

The bound is true for every $d \in \{1,\ldots, k\}$. Let us find a value of $d$ that approximately maximizes the bound for given $n,k$. Starting from $d=k$ and decreasing $d$, each decrease of $1$ doubles the first term in the RHS of \eqref{eq:ma1063}, as long as $\frac{k |\mathcal{X}|^{k-d}}{n} \leq 1$, and only linearly decreases the second term. Therefore it is beneficial to decrease $d$ as long as $\frac{k |\mathcal{X}|^{k-d}}{n} \leq 1$, and no more than one additional step. For simplicity let us always take the additional step and determine $d$ as the maximum $d \in \{1,\ldots, k\}$ so that $\frac{k |\mathcal{X}|^{k-d}}{n} \geq 1$, or $d=1$ if no such $d$ exists, i.e.
\begin{equation}
d
=
\max \left( \left\lfloor  \log \left( \frac{k |\mathcal{X}|^k}{n} \right)\frac{1}{ \log |\mathcal{X}|} \right\rfloor, 1 \right)
.
\end{equation}
If $n \geq k |\mathcal{X}|^{k-1}$, then $d=1$, and $\min \left( 1 , \frac{k |\mathcal{X}|^{k-d}}{n} \right) = \frac{k |\mathcal{X}|^{k-1}}{n}$. In this case \eqref{eq:ma1063} yields:
\begin{equation}\label{eq:ma1105}
\Delta^* \geq \half \frac{|\mathcal{X}|^{k-1}}{n} \cdot \log |\mathcal{X}|
.
\end{equation}
Otherwise, $\min \left( 1 , \frac{k |\mathcal{X}|^{k-d}}{n} \right) = 1$, and  \eqref{eq:ma1063} yields:
\begin{equation}\begin{split}\label{eq:ma1105b}
\Delta^*
\geq
\half \left\lfloor  \log \left( \frac{k |\mathcal{X}|^k}{n} \right)\frac{1}{ \log |\mathcal{X}|} \right\rfloor \frac{\log |\mathcal{X}|}{k}
.
\end{split}\end{equation}
Equations \eqref{eq:ma1105}, \eqref{eq:ma1105b} are represented in a compact form in \eqref{eq:ma1063b} above. This proves the lower bound of Theorem~\ref{theorem:mod_additive_redundancy}.
\endofproof

\subsubsection{Upper bound (direct part)}\label{sec:proof_theorem_mod_additive_redundancy_direct}
The purpose is to show the existence of a universal system that attains a small redundancy with respect to the reference system, i.e. referring to \eqref{eq:madef_IFB_redundancy_finite_nk}, it is desired to show that there exists a universal system $U$ such that:
\begin{equation}\label{eq:madef_IFB_redundancy_finite_nkx}
\max_{\vr z_1^n} \left[ \max_{E,D} \left( R_\tsubs{IFB}^* \right) - R_\tsubs{U}^* \right] \leq \Delta_{+}
.
\end{equation}
The desired result is similar to the one of Theorem~\ref{theorem:adaptive_rho}, however to reach the desired overheads, a slightly different design of the universal system, and a more careful analysis of the overheads is required.

Following the same logic as the proof of Theorems~\ref{theorem:Cifb_UB},\ref{theorem:adaptive_rho}, the difference between the good-put of the two systems is bounded by the following relations:
\begin{enumerate}[(a)]
\item The relation between $R_\tsubs{U}^*$ and the ideal $\Remp$ target of the rate adaptive system (i.e. the overhead term of Theorem~\ref{theorem:adaptive_L}).
\item The relation between $\Remp$ and the collapsed channel capacity, or equivalently the collapsed noise entropy $H(\tilde{\vr Z}_{l,k})$.
\item The relation between $R_\tsubs{IFB}^*$ and $H(\tilde{\vr Z}_{l,k})$ obtained using Fano's inequality (as in the proof of Theorem~\ref{theorem:Cifb_UB}).
\end{enumerate}
Considering the scheme that was described for the achievability result of Theorems~\ref{theorem:adaptive_L},\ref{theorem:adaptive_rho}, the largest overhead is due to step (b). This large overhead is in some sense unavoidable, as the converse shows, however it is especially large due to the use of LZ78 algorithm which has a slow $O(1/\log n)$ convergence rate. Specifically, using \cite[Thm 1,2]{LZ78}, this term, i.e. the bound on $\frac{1}{n} L_{78}(\vr z^n) - \frac{1}{k} H(\tilde{\vr Z}_{b,k})$ behaves like $O \left( \frac{\log \left( |\mathcal{X}|^{2k} \right)}{\log n} \right)$, i.e. in order for this term to be small, it is required that $n \gg |\mathcal{X}|^{2k}$, and any small improvement in the overhead requires an ever growing increase in $n$: improving the overhead by a factor of $2$ requires squaring $n$.

To obtain a tighter bound, a more general result from \selector{\cite{YL_PhdThesis}}{Part-I} can be applied. Theorem~\selector{8.2 there}{\ref{theorem:adaptive_causal_distribution}} shows that for every causal probability distribution $P(\vr x | \vr y)$, i.e. satisfying for all $i \leq n$: $P(\vr x^i | \vr y^n) = P(\vr x^i | \vr y^i)$, the rate function $\Remp = \frac{1}{n} \log \frac{P(\vr x^n | \vr y^n)}{Q(\vr x^{n})}$ is adaptively achievable with overhead of $\delta_n = 3 \sqrt{\frac{\log q_{\min}^{-1} \cdot (\log \frac{n}{\epsilon_U} + \log q_{\min}^{-1})}{n}}$, where $q_{\min}$ is the minimum non-zero value of $Q(x_i | \vr x_{i-1})$.\footnote{Substituting $d_\tsubs{FB}=1, D=0$ in the parameters of the theorem.}

Substitute as $Q$ the uniform distribution $Q(\vr x^i) = |\mathcal{X}|^{-i}$ having $q_{\min}^{-1} = |\mathcal{X}|$. Take $P(\vr x | \vr y) = P_Z(\vr x - \vr y)$, for some probability distribution $P_Z(\vr z)$. This choice satisfies the causality condition and yields
\begin{equation}\label{eq:ma512}
\Remp = \log |\mathcal{X}| + \frac{1}{n} \log P_Z(\vr x^n - \vr y^n)
,
\end{equation}
with $\delta_n = 3 \sqrt{\frac{\log |\mathcal{X}| \cdot \log \left( \frac{n |\mathcal{X}|}{\epsilon_U} \right)}{n}}$.

Let us begin by analyzing the relation between $\Remp$ and $R_\tsubs{U}^*$ in step (a) above. While the convergence of $\delta_n \ntoinfty 0$ requires $\epsilon_U$ to decay subexponentially with $n$, the choice of $\epsilon_U$ will lead to a reduction of $\epsilon_U \Remp \leq \epsilon_U \log |\mathcal{X}|$ in rate. For simplicity let us choose $\epsilon_U = \frac{1}{n}$ as this factor is insignificant. In other words, the exists a system with $\epsilon_U = \frac{1}{n}$ which with probability $1-\epsilon_U$ transmits a rate $\Remp - \delta_n$ without error over the channel. Therefore
\begin{equation}\begin{split}\label{eq:ma520k}
R_\tsubs{U}^*
& =
R_\tsubs{U} (1 - \epsilon_\tsubs{U}) - \frac{1}{n} h_b(\epsilon_\tsubs{U})
\\ & \geq
(\Remp - \delta_n) (1 - \epsilon_\tsubs{U}) - \frac{1}{n} h_b(\epsilon_\tsubs{U})
\\ & \stackrel{\Remp \leq \log |\mathcal{X}|, \eqref{eq:ma512}}{\geq}
\Remp - \delta_n - \epsilon_\tsubs{U} \log |\mathcal{X}| - \frac{1}{n} h_b(\epsilon_\tsubs{U})
\\& \geq
\Remp - 3 \sqrt{\frac{\log |\mathcal{X}| \cdot \log \left( n^2 |\mathcal{X}| \right)}{n}} - \frac{1}{n} \log |\mathcal{X}| - \frac{1}{n}
\\& \geq
\Remp - \underbrace{4 \sqrt{\frac{\log |\mathcal{X}| \cdot \log \left( n^2 |\mathcal{X}| \right)}{n}}}_{\defeq \delta_n^*}
,
\end{split}\end{equation}
where in the last step, for simplification of the bound, it was assumed that $n \geq \log |\mathcal{X}|$ -- otherwise, $\delta_n$ is large. Equation \eqref{eq:ma520k} yields the desired relation for step (a) above. Next, the relation between $\Remp$ and the collapsed noise sequence entropy in step (b) is considered.

If one is interested in competing with an IFB system with block length $k$, it would make sense to treat each $k$ symbols of the noise sequence as a single super-symbol, and take as $P_Z$ the universal distribution defined by Krichevsky and Trofimov \cite{KrichevskyTrofimov81} over these super-symbols. This distribution is universal in the sense that up to a small overhead, $-\frac{1}{n} \log P_Z(\vr z) \approx \hat H(\vr z)$, i.e. the probability matches the empirical entropy of the sequence, which in the current case is $H(\tilde{\vr Z}_{b,k})$. Furthermore, this holds with a redundancy close to the minimum possible. It is possible to construct a universal distribution $P_Z$ that compares well with all distributions over the $n$ symbols which are i.i.d. over $k$-length blocks, by a weighted average of Krichevsky-Trofimov distributions.

Let $\pi_k(\vr z^k)$ denote a distribution over the $k$-letter $\vr z^k$, where $k$ is not assumed to divide $n$. This defines also a distribution on the partial sequence of length $i<k$ by taking the marginal $\pi_k(\vr z^i) = \sum_{\vr z_{i+1}^k} \pi_k(\vr z^k)$. The distribution over $n$ length vectors, associated with $\pi_k$ is defined as the i.i.d. extension of $\pi_k$, where the marginal distribution is used for the remainder that does not divide by $k$. This $n$-length distribution will be denoted by the same symbol:
\begin{equation}\label{eq:ma520}
\pi_k(\vr z) \defeq \prod_{i=1}^{\lfloor n/k \rfloor} \pi_k(\vr z_{(i-1)k+1}^{(i-1)k+k}) \cdot \pi_k(\vr z_{\lfloor n/k \rfloor k + 1}^n)
.
\end{equation}

Then, by weighting Krichevsky-Trofimov distributions it is possible to obtain the following result:
\begin{lemma}\label{lemma:univ_distribution_multiple_k}
There exists a distribution $P_Z(\vr z^n), \vr z^n \in \mathcal{X}^n$, such that for all $k$ for which $\tau \defeq \frac{|\mathcal{X}|^k}{n} \leq 1$:
\begin{equation}\label{eq:ma527l}
\forall \pi_k: \frac{1}{n} \log \pi_k(\vr z^{n}) \leq \frac{1}{n} \log P_Z(\vr z^n) + \Delta_\pi(k,n)
,
\end{equation}
where
\begin{equation}\label{eq:ma532l}
\Delta_\pi = \frac{\tau}{2} \log \left( \frac{1}{\tau} \right) + \left( \frac{k}{4} \tau^2 + \tau + \frac{k}{n} \right) \log e
.
\end{equation}\end{lemma}
The detailed derivation and proof appears in Appendix~\ref{sec:proof_lemma_univ_distribution_multiple_k}. The next stage is to relate $\pi_k(\vr z^{n})$ to $H(\tilde{\vr Z}_{l,k})$. Let $\vr z_{i}^{[k]} \defeq \vr z_{(k-1)i+1}^{(k-1)i+k}$ be the $i$-th $k$-block of $\vr z^n$. Recall that $l = \lceil \tfrac{n}{k} \rceil$ is the number of $k$-blocks that cover the $n$ symbols, and $\tilde{\vr Z}_{l,k}$ is a random variable generated by uniform selection out of $\vr z_{1}^{[k]},\ldots, \vr z_{l}^{[k]}$. Let $P_{\tilde{\vr Z}_{l,k}}$ be the distribution of $\tilde{\vr Z}_{l,k}$ which is the empirical distribution of $\vr z_{1}^{[k]},\ldots, \vr z_{l}^{[k]}$.
\begin{equation}\begin{split}\label{eq:ma603}
H(\tilde{\vr Z}_{l,k})
& \defeq
-\sum_{\vr a \in \mathcal{X}^k} P_{\tilde{\vr Z}_{l,k}}(\vr a) \log P_{\tilde{\vr Z}_{l,k}}(\vr a)
\\& =
- \frac{1}{l} \sum_{i=1}^{l} \log P_{\tilde{\vr Z}_{l,k}}(\vr z_{i}^{[k]})
\\& \stackrel{(a)}{=}
- \frac{1}{l} \max_{\pi} \log \pi_k(\vr z_{1}^{k \cdot l})
\\& \stackrel{(b)}{\geq}
- \frac{1}{l} \max_{\pi} \log \pi_k(\vr z_{1}^n)
\\& \stackrel{\eqref{eq:ma527l}}{\geq}
- \frac{1}{l} \left( \log P_Z(\vr z) + n \Delta_\pi(k,n) \right)
,
\end{split}\end{equation}
where (a) is because the empirical distribution maximizes the joint distribution of the vector; the expression following (a), where the maximization is over all $k$-letter distributions $\pi$, could be considered an alternative definition of  $H(\tilde{\vr Z}_{l,k})$ (see \selector{\cite[\S 9.1.4]{YL_PhdThesis}}{\S\ref{sec:def_emp_entropy}}). Transition (b) holds because extending the vector reduces its probability (see also the definition of $\pi_k(\vr z^{j})$ \eqref{eq:ma520}).

Finally, in step (c), let us use Fano's inequality (see \eqref{eq:maR_UB} in the proof of Theorem~\ref{theorem:Cifb_UB}):
\begin{equation}\label{eq:ma8812}
R_\tsubs{IFB}^* = R_\tsubs{IFB} (1 - \epsilon_\tsubs{IFB}) - \frac{1}{k} h_b(\epsilon_\tsubs{IFB}) \leq \log |\mathcal{X}| - \frac{1}{k} H(\tilde{\vr Z}_{l,k})
.
\end{equation}
Combining the above yields
\begin{equation}\begin{split}\label{eq:ma566}
R_\tsubs{IFB}^*
& \stackrel{\eqref{eq:ma8812}}{\leq}
\log |\mathcal{X}| - \frac{1}{k} H(\tilde{\vr Z}_{l,k})
\\ & \stackrel{\eqref{eq:ma603}}{\leq}
\log |\mathcal{X}| + \frac{1}{k l} \left( \log P_Z(\vr z) + n \Delta_\pi(k,n) \right)
\\ & \leq
\frac{n}{k l} \left( \log |\mathcal{X}| + \frac{1}{n} \log P_Z(\vr z) \right)
\\ & \qquad + \frac{kl - n}{kl} \log |\mathcal{X}| + \Delta_\pi(k,n)
\\ & \stackrel{\eqref{eq:ma512}}{\leq}
\frac{n}{k l} \Remp + \frac{k}{n} \log |\mathcal{X}| + \Delta_\pi(k,n)
\\ & \stackrel{\eqref{eq:ma520k}}{\leq}
\frac{n}{k l} \left( R_\tsubs{U}^* + \delta_n^* \right) + \frac{k}{n} \log |\mathcal{X}| + \Delta_\pi(k,n)
\\& \leq
R_\tsubs{U}^* + \delta_n^* + \frac{k}{n} \log |\mathcal{X}| + \Delta_\pi(k,n)
.
\end{split}\end{equation}

Since this holds for any noise sequence and any pair $E,D$,
\begin{equation}\label{eq:ma623}
\Delta^*(n,k)
\stackrel{\eqref{eq:madef_IFB_redundancy_finite_nk}}{\leq}
\max_{\vr z, E, D} \left( R_\tsubs{IFB}^* - R_\tsubs{U}^* \right)
\stackrel{\eqref{eq:ma8812}}{\leq}
\Delta_\pi(k,n) + \delta_n^* + \frac{k}{n} \log |\mathcal{X}|
.
\end{equation}
This proves the upper bound of Theorem~\ref{theorem:mod_additive_redundancy}.
\endofproof

\section{Discussion and extensions}\label{sec:ma_discussion}
The model presented in \selector{this paper}{Chapter~\ref{chap:univ_problem_setup}} supplies the first definition of a ``universal communication system'', and the results \onlyphd{of this chapter} indicate that such universal communication with feedback is possible in the non trivial example of the modulo additive channel with an individual state sequence.

\subsection{Alternative definitions of universality}\label{sec:alt_definitions}
The IFB comparison class was chosen as the perhaps simplest and most intuitive comparison class for universal communication. However, it has several drawbacks:
\begin{enumerate}[(1)]
\item The reference system is limited in terms of complexity, feedback, etc.
\item On the other hand, universality is only achieved at ultra-high values of the transmission length $n$. Similar issues exist with Lempel-Ziv universal source coding\onlyphd{ (see \S\ref{sec:um_discussion_asymptotics})}.
\item The definition motivates learning $k$-periodic structures in the channel, which is counter intuitive. This may be solved e.g. by starting the reference system at an arbitrary time rather than at time $1$, or by using structures that are not periodic such as finite state machines \cite{MisraPorosityISIT12}.
\item While the IFB capacity is limited by the ``collapsed channel capacity'', it usually falls short of it. Furthermore, had the channel been a stochastic memoryless one, a rather large block size would be needed for the IFB system in order to yield a small error probability. A possible solution is to define the collapsed channel capacity itself as a target rate, but it is not clear how this should be defined for channels with memory.
\end{enumerate}

\subsubsection{Possible enhancements of the IFB class}
Since the reference system enjoys the advantage of being designed for the specific noise sequence, this advantage is compensated by imposing some restrictions on the reference system, which are not imposed on the universal system. This is similar to what is done in universal source coding and universal prediction, when the comparison class is too rich. The definition of $\Cifb$ limits the reference system in several factors, where the universal system is not restricted. Namely its complexity, the use of feedback, common randomness and rate adaptivity. Relaxing any of these factors, may generate a higher value of the target rate as an alternative to $\Cifb$, which may still be universally attainable.
\onlyphd{While there are reasons to consider such extensions, in \selector{our}{my} opinion, due to the convergence rate issues (Section~\ref{sec:ma_univ_redundancy}) the right direction would be to simplify rather than enrich the comparison class.}

Some potential variations are given below:
\begin{enumerate}[(a)]
\item Randomness: allowing the reference system the use of common randomness.
\item Rate adaptivity: allowing rate adaptivity in various levels. Error detection and \textit{automatic repeat request} (ARQ) can be considered a very basic level of adaptivity.
\item Complexity: definition of the encoder/decoder as finite state machines rather than block encoders/decoders.
\item Feedback: allowing the use of (a possibly limited amount of) feedback for the reference system.
\end{enumerate}

The first two extensions (a),(b) are trivial, and were not pursued here in order to simplify the presentation. Misra and Weissman presented \cite{MisraPorosityISIT12} a class of finite state machine encoders and decoders with feedback, termed the \emph{FS class}, that includes all the enhancements above, and had shown that for the modulo-additive channel, the maximum rate achieved by the reference class is at most $R = (1-\rho(\vr z)) \log |\mathcal{X}|$, so the current result on universality would hold also with respect to this enhanced class. Furthermore, they show that, unlike the IFB class (\S\ref{sec:IFB_capacity_discussion}), the FS class achieves the rate $R$ when the complexity is allowed to grow. Notwithstanding these results, the IFB class is still of interest due to its simplicity, which allows simple analysis and consideration of more complex channel models \selector{\cite{YL_UnivCommMemory}}{(as in Chapter~\ref{chap:univ_vectormem})}.

Below, these extensions are briefly discussed. Although Misra and Weissman already extended the results in the context of the modulo-additive channel, it is interesting to consider these extensions for more general channel models.

\textbf{Common randomness}:
Allowing the reference system the use of common randomness does not change the results, as long as the common randomness is independent of the noise sequence and/or the block number. This is because the IFB capacity would still be upper bounded by the collapsed channel capacity. This holds also for channels with fading memory \selector{\cite{YL_UnivCommMemory}}{(Chapter~\ref{chap:univ_vectormem})}, where the collapsed channel capacity is used as a bound for the IFB rate.

\textbf{Rate adaptivity}:
The IFB system may be allowed to choose the transmission rate adaptively at the decoder. A simple form of rate adaptivity is error detection, i.e. the decoder is allowed to choose between rate $R$ and rate $0$. In the later case, decoding errors are ignored. On the other hand, the IFB rate is defined in an effective way, considering how many blocks were actually decoded. Under suitable definitions\onlyphd{ (see Section~\ref{sec:good_put_bound})}, the effective rate of the IFB system would still be bounded by the collapsed channel capacity, so the results easily extend. Note that allowing error detection effectively models a block coding system using \textit{automatic repeat request} (ARQ). When rate adaptation is considered, for a fair comparison, the decision on the rate must be made at the decoder based on the received sequence alone, rather than be given to the decoder.

\textbf{Complexity}:
In order to achieve competitively universal communication, it is essential that \emph{both} the reference encoder and the decoder be limited in some way, assuming they are designed knowing the channel. Consider, for example, the modulo additive channel. If the encoder is not limited, then it can transmit data at the maximum rate $\log |\mathcal{X}|$ bits/channel use, by uncoded transmission and subtraction of the noise sequence at the encoder. In this case, the decoder does nothing essentially, so restrictions on the decoder will not be helpful. Conversely, if the decoder is not limited, the encoder can transmit the message un-coded and the noise sequence can be canceled at the decoder, so limitations on the encoder would not help. As mentioned, an extension to finite state machines with feedback (FS-class) has already been shown \cite{MisraPorosityISIT12}. An interesting issue for further study is the universality with respect to the FS-class in general channel models.

\textbf{Feedback}:
Several types of feedback may be considered:
\begin{enumerate}
\item Feedback inside the block, i.e. where the state is reset from block to block. Because the collapsed channel is a channel with memory, feedback can increase its capacity. The increase in capacity is obtained by changing the input distribution (prior) in response to feedback, yielding information on the channel state. Hence, in order to complete in this case, the universal system would also need to adapt its input distribution per symbol based on feedback.\onlyphd{\footnote{Consider for example a memoryless channel, where the channel law at each time instance is one of the two channels of Example~\ref{example:prediction_channel1}, and the sequence of channel states has some correlation, i.e. state $0$ appears more frequently after state $0$ and vice versa. Then, a reference system with feedback may guess the next state and optimize the prior for it, while a system which uses a fixed prior over large blocks can only approach a fraction of the rate achievable by the reference system.}}
Hence, the universal systems presented here and in \selector{\cite{YL_UnivCommMemory}}{Chapter~\ref{chap:univ_vectormem}} are not suitable for this setting. However, for the modulo-additive channel, feedback does not increase capacity, because the best input distribution is uniform regardless of any knowledge on channel state (in other words, as easy to see, the bound based on Fano's inequality \eqref{eq:maR_UB} would hold regardless of feedback), and in this particular case, the results do extend to the case of feedback inside the block (see also \cite{MisraPorosityISIT12}).
\item Feedback between blocks, i.e. encoder of block $b$ receives a message from decoder of block $b-1$. This kind of feedback effectively increases the block size of the IFB system, as it allows it to keep track of the block index to some extent by passing it back and forth between the encoder and the decoder, through the channel in one way and the feedback link in the other way. Of course, this cannot be continued when the number of bits required to represent the block index is larger than $k \log |\mathcal{X}|$. In the modulo-additive channel, knowledge of the block index yields the maximum capacity of $\log |\mathcal{X}|$. It is interesting to note that, while such feedback seems to considerably strengthen the IFB system, Misra and Weissman \cite{MisraPorosityISIT12} showed that the rate of the FS-class is limited in spite of feedback. This is because the restriction is on the number of states rather than on the block length.
\item It is possible to allow the reference system the use of asymptotically zero-rate feedback, which does not considerably increase the effective block length and cannot considerably increase the collapsed channel capacity, and is comparable with the amount of feedback used by the universal system.
\end{enumerate}

\subsubsection{An alternative comparison class}\label{sec:ma_alt_class_iid}
As mentioned, a relatively short block size, limits the IFB class from attaining the collapsed channel capacity. This gap is not utilized in the current bounds. The collapsed channel capacity bound would still hold, if the reference encoder and decoder were allowed to encode multiple blocks together, but treat each block in the same way.

One option to define this class is to limit the encoder to a random encoder over the entire transmission length $n$, with an i.i.d. prior of choice (alternatively, i.i.d. in blocks) and limit the decoder to use a memoryless decoding metric (or more generally, alpha decoding, i.e. type-based decoding). Another similar way is to let the encoder and decoder be general but randomly permute the inputs and outputs of the channel. As before, the reference encoder and decoder are limited, but are designed based on full channel knowledge. For the modulo-additive channel, it is easy to see that in both cases, the reference rate would be limited to $\log |\mathcal{X}| - H(\tilde{\vr Z}_{b,k})$. It is more interesting to discuss these classes in the case of general channels -- see \selector{\cite{YL_UnivCommMemory}}{Section~\ref{sec:um_alt_comparison_class}}. Note that although these reference systems would fail for the password channel defined in Example~\ref{example:password_channel}, it is possible to devise an alternative example, showing that universal communication with respect to these classes over general channels is not possible (see Appendix~\ref{sec:password_channel_iid_example}).

\subsection{Other comments}
Theorem~\ref{theorem:adaptive_L}, connecting the transmission rate to the compression rate of the noise sequence is reminiscent of Ahlswede's channel coding scheme with feedback \cite{Ahlswede71}. This scheme sends information by iteratively compressing the receiver's uncertainty with regard to the transmitted message. Indeed, Ooi \cite{Ooi} used this scheme in order to achieve adaptive communication over compound channels, including compound finite state channels. Ooi assumes a compound channel, i.e. probabilistic with unknown parameters, and varies the rate by changing the transmission length, while here an individual noise sequence is considered and the rate is varied by changing the number of bits transmitted. Using a variable block length is a simpler, particular case, that can be obtained by transmitting a single block, in the scheme presented here. Adapting Ooi's scheme to the individual noise sequence channel seems complicated while using random coding yields a simple proof for the current result.

The result of Theorem~\ref{theorem:adaptive_L} is also closely related to Ziv's result \cite{ZivUniversal} regarding universal decoding over compound finite state channels. If Theorem~\ref{theorem:adaptive_L} is particularized to the non-adaptive case, then it can be proven and generalized by the tools used there. The decoder in Ziv's paper uses joint Lempel-Ziv parsing and yields a decoding metric which generalizes in a sense the metric used here, for channels which are not necessarily memoryless\onlyphd{ (see also Section~\ref{sec:examples_compression_condLZ})}. Theorem~2 and particularly Lemma~1 there, relate the size of the error sets $M_0, M_u$ defined there, for the maximum likelihood decoder designed for the finite state channel, and the universal decoder. This relation indicates the rate that can be achieved with a given error probability is asymptotically the same. Furthermore, the only assumption used about the reference maximum likelihood decoder is that it uses a finite state metric (see the proof of Lemma~1 there), and thus the IFB decoder falls into this class.

\onlypaper{
In a previous paper \cite{YL_individual_full} a different framework, termed ``individual channels'' was considered, in which no relation between the input and output of the channel is assumed a-priori, and the communication rate is given as a function of the input and output sequences (see also \cite[Part 1]{YL_PhdThesis}). As an example, the empirical mutual information $\hat I(\vr x, \vr y)$ is shown to be achievable. The current achievability result (Theorem~\ref{theorem:adaptive_L}) can be stated in these terms by saying that the rate function $\Remp(\vr x, \vr y) = \log | \mathcal{X} | - \frac{1}{n} L(\vr y - \vr x)$ is asymptotically adaptively achievable  (i.e. by an adaptive rate system). Note that there is no need to assume that the channel is truly modulo-additive to show this. It is also possible to show \cite[Thm~10.2]{YL_PhdThesis} that all achievable rate functions that depend only on the noise sequence $\Remp(\vr x, \vr y) = R(\vr y - \vr x)$, are asymptotically of this form, i.e. given a system attaining the rate $\Remp(\vr x, \vr y) = R(\vr y - \vr x)$ for each $\vr x^n, \vr y^n$ and where the channel input $\vr x^n$ is uniformly distributed (see the definitions therein), there exists a source encoding scheme with encoding lengths $L(\vr z)$ such that asymptotically $R(\vr z^n) = \log | \mathcal{X} | - \tfrac{1}{n} L(\vr z^n)$.
}

In previous works \cite{Ofer_ModuloAdditive,Eswaran}, rates which reflect the average channel behavior such as $1 - \hat H(\vr z^n)$ were termed ``empirical capacity''  mainly based on the similarity to the capacity expressions for memoryless channels. The term is not completely justified, since clearly this is not the maximum communication rate. The value $\Cifb$ seems to be a better candidate to describe the modulo-additive channel's ``empirical capacity'', although as discussed above, other interesting definitions can be suggested. Note that there is no fixed order between $\Cifb$ and the rate $1 - h_b(\hat \epsilon)$, where $\hat \epsilon$ is the empirical frequency of '1'-s in the sequence (defined in \cite{Ofer_ModuloAdditive}). For example for $\vr z = 0,1,0,1,0,...$, the relation is $0 = 1 - h_b(\hat \epsilon) < \Cifb=1$, while in Example~\ref{example:half_random_sequence} the order is inverse $0 = \Cifb < 1 - h_b(\hat \epsilon) = 1 - h_b\left(\frac{1}{4} \right)$. On the other hand the relation $1 - h_b(\hat \epsilon) \leq 1 - \rho(\vr z)$ always holds,\footnote{This can be shown by block to variable encoding to rate $h_b(\hat \epsilon_i)$ where $\hat \epsilon_i$ is the empirical probability of $1$-s in the block, and the convexity of $h_b(\cdot)$} so the rates achieved by the scheme described here are asymptotically better than the previously achieved rates \cite{Ofer_ModuloAdditive}.

The current results assume the noise sequence is fixed and unknown, and do not extend to the case where the noise sequence is determined by an adversary (i.e. $z_i$ is a function of $x_1^{i-1}$), and the reference class is aware of the adversary strategy. To see this, it is easy to design an adversary that identifies the codebook used by the reference encoder, and locks the channel (by choosing the noise sequence randomly) once a different channel input appears.

\onlypaper{
\section{Conclusion}
This paper considered target rates for universal systems with feedback and focused on the modulo additive channel. The notion of the \textit{iterated finite block capacity}, denoted $\Cifb$, was defined for a vector channel, as the highest rate achievable by encoders and decoders that may be designed for the particular relation that exists between the input and output, yet are constrained to be of finite block length and use the same scheme over each block. The IFB capacity $\Cifb$ was used as a target communication rate to be achieved without any prior knowledge of the channel, using feedback. It was shown that $\Cifb$ cannot be achieved universally for completely general input-output relations, however for the modulo-additive channel with an individual noise sequence, it can be achieved universally without knowing the noise sequence.
Specifically, it was shown that $\Cifb \leq (1-\rho) \log |\mathcal{X}|$, where $\rho$ is the finite state compressibility of the noise sequence, and a universal system with feedback attaining a rate of at least $(1-\rho) \log |\mathcal{X}|$ was presented. This result is relatively simple due to the properties of the modulo additive channel. In a follow-up paper \cite{YL_UnivCommMemory} the result is extended to more general channels.
}

\fi 

\ifx\compileAppendix\flagTrue 

\onlypaper{
\appendix{}
}

\subsection{Proof of Theorem~\ref{theorem:Cifb_UB}}\label{sec:proof_theorem_Cifb_UB}
Suppose a given $E,D$ achieve rate $R$ and average error probability $\epsilon$ over $b$ blocks of size $k$. Let us adopt the definitions of $\tilde{\vr X}_{b,k}$, $\tilde{\vr Z}_{b,k}$ and $\tilde{\vr Y}_{b,k}$ from Section~\ref{sec:IFB_UB}, and likewise define ${\msg}$ and ${\hat \msg}$ to be random variables generated by selecting the block index uniformly over $1, \ldots, b$ and taking the respective encoded/decoded (resp.) messages, i.e. ${\msg} = \msg_U$, ${\hat \msg} =\hat \msg_U $, where $U \sim U\{1, \ldots, b\}$. See Fig.\ref{fig:collapsed_channel_modadditive}. Then
\begin{equation}\begin{split}
\frac{1}{b} \sum_{i=1}^b \Pr(\hat \msg_i \neq \msg_i)
& =
\sum_{i=1}^b \Pr(\hat \msg_i \neq \msg_i) \Pr(U = i)
\\& =
\Pr({\hat \msg} \neq {\hat \msg})
\leq
\epsilon
.
\end{split}\end{equation}
The rate $R$ is now bounded by the entropy of $\tilde{\vr Z}_{b,k}$. By Fano's inequality
\begin{equation}
H(\msg | \hat \msg) \leq h_b(\epsilon) + \epsilon \log M
.
\end{equation}
Therefore by the information processing inequality
\begin{equation}\begin{split}
I(\tilde{\vr X}_{b,k}; \tilde{\vr Y}_{b,k})
& \geq
I(\msg; \hat \msg)
=
H(\msg) - H(\msg | \hat \msg)
\\& \geq
\log M - (h_b(\epsilon) + \epsilon \log M)
.
\end{split}\end{equation}
On the other hand
\begin{equation}\begin{split}
I(\tilde{\vr X}_{b,k}; \tilde{\vr Y}_{b,k})
& =
H(\tilde{\vr Y}_{b,k}) - H(\tilde{\vr Y}_{b,k} | \tilde{\vr X}_{b,k})
\\&=
H(\tilde{\vr Y}_{b,k}) - H(\tilde{\vr Z}_{b,k})
\\& \leq
\log |\mathcal{X}|^k - H(\tilde{\vr Z}_{b,k})
.
\end{split}\end{equation}
Combining the two:
\begin{equation}
(1-\epsilon) \log M - h_b(\epsilon) \leq I(\tilde{\vr X}_{b,k}; \tilde{\vr Y}_{b,k}) \leq k \log |\mathcal{X}| - H(\tilde{\vr Z}_{b,k})
.
\end{equation}
Therefore
\begin{equation}\label{eq:maR_UB}
R \leq \frac{1}{k} \log M \leq (1-\epsilon)^{-1} \left[ \log |\mathcal{X}| - \frac{1}{k} H(\tilde{\vr Z}_{b,k}) + \frac{1}{k} h_b(\epsilon) \right]
.
\end{equation}
If $R$ is an achievable rate then by Definition~\ref{def:IFB_rate}, for any $\epsilon > 0$ there exist $k > 0$ such that (\ref{eq:maR_UB}) holds for this $k$ and $b$ large enough. Therefore taking $\liminf_{b \to \infty}$ on both sides yields:
\begin{equation}\label{eq:maR_UB2}
R \leq (1-\epsilon)^{-1} \left[ \log |\mathcal{X}| - \frac{1}{k} \limsup_{b \to \infty} H(\tilde{\vr Z}_{b,k}) + \frac{1}{k} h_b(\epsilon) \right]
.
\end{equation}

Next, let us relate $H(\tilde{\vr Z}_{b,k})$ to the finite state compressibility (see \eqref{eq:ma444}-\eqref{eq:ma446} in Section~\ref{sec:IFB_UB}). There exists a finite state machine $\tilde{F}$ with $s_k = \mathcal{X}^{k-1} \cdot k$ states that compresses the sequence $\vr z_{1}^{kb}$ to at most $b \cdot (H(\tilde{\vr Z}_{b,k}) +  1)$ bits. This state machine implements a block to variable encoder tuned to the empirical distribution and is structured as follows: its state space includes a counter from $1$ to $k$ which counts the index inside the block, and a memory of $k-1$ input characters. When the counter reaches $k$ the machine outputs an encoded string, and the counter returns to $1$. In the other counter states the machine emits the empty string. The encoded string is generated by a simple block to variable encoder optimized to compress the random variable $Z_{k,b}$ to its minimum average length (e.g. a Huffman encoder, although a simple encoder using lengths $\lceil \log (\Pr(Z_{k,b})^{-1}) \rceil$ is sufficient for this purpose), and therefore its average encoded length for $\tilde{\vr Z}_{b,k}$ is at most $H(\tilde{\vr Z}_{b,k}) +  1$ \cite[Section 5.4]{CoverThomas_InfoTheoryBook}. The encoding length is therefore:
\begin{equation}\begin{split}
& \sum_{i=1}^b |F(\vr z_i^{[k]})|
\\& =
\sum_{\tilde{\vr z} \in \mathcal{X}^k} \sum_{i=1}^b \Ind \left( (\vr z_i^{[k]})_{i=1}^b = \tilde{\vr z} \right) \cdot |F(\tilde{\vr z})|
\\&=
\sum_{\tilde{\vr z} \in \mathcal{X}^k} b \cdot  \Pr(Z_{k,b} = \tilde{\vr z}) \cdot |F(\tilde{\vr z})|
\\& \leq
b (H(\tilde{\vr Z}_{b,k}) +  1)
.
\end{split}\end{equation}
Therefore for $n= b k$
\begin{equation}\begin{split}\label{eq:marhoF_ub_by_H}
\rho_{\mathcal{F}(s_k)} (\vr z_1^n)
& \leq
\rho_{\tilde{F}}(\vr z_1^n)
=
\frac{1}{n \log |\mathcal{X}|} |F(\vr z_1^n)|
\\& \leq
\frac{1}{n \log |\mathcal{X}|} b (H(\tilde{\vr Z}_{b,k}) +  1)
\\& =
\frac{1}{k \log |\mathcal{X}|} (H(\tilde{\vr Z}_{b,k}) +  1)
.
\end{split}\end{equation}
The condition $n=b k$ may be relaxed and the inequality may be applied to any finite $n$, taking $b = \lfloor \frac{n}{k} \rfloor$ (since if the last block is unfinished it will not contribute to the length, and the normalization by $n > b k$ will only decrease the LHS). Now,
\begin{equation}\begin{split}
\limsup_{n \to \infty} \rho_{\mathcal{F}(s_k)} (\vr z_1^n)
& \leq
\limsup_{n \to \infty} \rho_{\tilde{F}} (\vr z_1^n)
\\& \leq
\limsup_{b \to \infty} \frac{1}{k \log |\mathcal{X}|} (H(\tilde{\vr Z}_{b,k}) +  1)
\\&=
\frac{1}{k \log |\mathcal{X}|} (\limsup_{b \to \infty}  H(\tilde{\vr Z}_{b,k}) +  1)
,
\end{split}\end{equation}
and
\begin{equation}\begin{split}
\rho(\vr z)
& =
\lim_{s \to \infty} \limsup_{n \to \infty} \rho_{\mathcal{F}(s)} (\vr z_1^n)
\\& \leq
\limsup_{n \to \infty} \rho_{\mathcal{F}(s_k)} (\vr z_1^n)
\leq
\frac{1}{k \log |\mathcal{X}|} (\limsup_{b \to \infty}  H(\tilde{\vr Z}_{b,k}) +  1)
.
\end{split}\end{equation}
Combining the above with (\ref{eq:maR_UB2}) yields:
\begin{equation}\begin{split}\label{eq:maR_UB3}
& \forall \epsilon: \exists k:
\\&
R
\leq
(1-\epsilon)^{-1} \left[ \log |\mathcal{X}| - \frac{1}{k} \limsup_{b \to \infty} H(\tilde{\vr Z}_{b,k}) + \frac{1}{k} h_b(\epsilon) \right]
\\& \leq
(1-\epsilon)^{-1} \left[ \log |\mathcal{X}| - \log |\mathcal{X}| \rho(\vr z) + \frac{1}{k} + \frac{1}{k} h_b(\epsilon) \right]
.
\end{split}\end{equation}

Since the $k$ obtaining the requirements of Definition~\ref{def:IFB_rate} may be small, the factor $\frac{1}{k}$ on the RHS makes the bound loose. To tighten the bound the following argument is used: choose a number $j > 0$. If there exist $E,D$ with block size $k$ and average error probability $\epsilon$ over $b$ large enough which divides by $j$, then by treating at each consecutive $j$ blocks as a new block (and forming the encoder and decoder with block size $j \cdot k$ by using $j$ times the original encoder and decoder), then by the union bound if $\epsilon_i$ denote the error probabilities over the blocks $i \in \{ 1, \ldots, b \}$, the error probabilities of the aggregate encoder and decoder will satisfy $\epsilon_i' \leq \sum_{d=1}^j \epsilon_{(i-1)j +d}$, and therefore the average error probability will be $\epsilon' = \frac{1}{b/j} \sum_{i=1}^{b/j} \epsilon_i' \leq \frac{j}{b} \sum_{i=1}^{b} \epsilon_i = j \cdot \epsilon$. The conclusion is that if the requirements of Definition~\ref{def:IFB_rate} are met for a certain $\epsilon, k$, they are also met for $j \cdot \epsilon, j \cdot k$. Therefore:
\begin{equation}\begin{split}\label{eq:maR_UB4}
& \forall j,\epsilon: \exists k:
\\ &
R
\leq
(1-j \epsilon)^{-1} \left[  (1 - \rho(\vr z))\log |\mathcal{X}| + \frac{1}{j k} (1 + h_b(j \epsilon)) \right]
.
\end{split}\end{equation}

Note that Definition~\ref{def:IFB_rate} requires the rate to be achievable for any $\epsilon > 0$, and therefore it is possible to take $\epsilon \ntoinfty 0$. By choosing for each $j$, $\epsilon = \frac{1}{j^2}$, denoting $k_j$ as any $k$ that satisfies (\ref{eq:maR_UB4}) for this $j$, and taking the limit $j \to \infty$ yields:
\begin{equation}\begin{split}\label{eq:maR_UB5}
R
& \leq
\lim_{j \to \infty} \Bigg\{ \left(1-\frac{1}{j} \right)^{-1} \Big[  (1 - \rho(\vr z))\log |\mathcal{X}|
\\ & \qquad +
 \frac{1}{j k_j} (1 + h_b(j^{-1})) \Big] \Bigg\}
\\& =
 (1 - \rho(\vr z)) \log |\mathcal{X}|
,
\end{split}\end{equation}
which by Definition~\ref{def:IFB_capacity} proves the theorem. \endofproof

\onlypaper{
\subsection{Proof of Theorem~\ref{theorem:adaptive_L}}\label{sec:modadditive_scheme_and_theorem}
In this section, it is shown that the rate $\log | \mathcal{X} | - \frac{1}{n} L(\vr z)$ can be attained for a wide class of source encoders. Notice that this result is derived for a more general case in \cite[\S 10.5]{YL_PhdThesis}. A rough proof outline appears in Section~\ref{sec:overview}.

\subsubsection{The adaptive communication scheme}\label{sec:modadditive_scheme}
The scheme applies repeated ``rateless'' transmissions: fix a value $K$ of the number of information bits per block. Using the common randomness, generate a random codebook of $\exp(K)$ words chosen independently and distributed uniformly over $\mathcal{X}^n$ which is known at the encoder and decoder. In each rateless block $b=1,2,\ldots$, the encoder sends $K$ bits to the decoder, by sending the respective symbols from codeword indexed by those $K$ bits. Note that at each block different symbols from the codebook are sent. The block terminates when a termination condition is satisfied at the decoder. Then, the decoder stores the decoded bits and indicates this to the encoder, through the feedback link (a 0-1 feedback is sufficient), and a new block, conveying $K$ new bits, begins. The last block is potentially not decoded, if the termination condition is not satisfied at the last symbol.

The decoding and termination rule are specified next. Suppose that the current symbol number is $i$ and the block number is $b$. The last symbol of the previous block (number $b-1$) was sent at symbol $j$ ($j=0$ if $b$ is the first block). Let $\vr {\hat x}_1^j$ denote the transmit sequence that follows from the previous decisions made by the decoder (i.e. is composed of the symbols from the codebook matching the decoded bits at each previously decoded block), and let $\vr x_{j+1}^i(m)$ denote the transmitted symbols matching codeword $m$ ($m=1,\ldots,\exp(K)$). $\vr {\hat z}^i(m)$ defined below is the decoder's hypothesis on the noise sequence $\vr z^i$:
\begin{equation}
\vr {\hat z}^i(m) = \vr y^i - (\vr {\hat x}_1^j, \vr x_{j+1}^i(m))
.
\end{equation}
Take $\vr {\hat z}^j = \vr y^j - \vr {\hat x}_1^j$ to be the $j$ length prefix of $\vr {\hat z}^i(m)$ (which is independent of $m$). The decoder calculates the following condition for all $m = 1,\ldots,\exp(K)$:
\begin{equation}\label{eq:materm_condition}
L_T(\vr {\hat z}^i(m)) - L_S(\vr {\hat z}^j) \leq \lfloor (i-j) \cdot \log |\mathcal{X}| -\log \left( \frac{n}{\epsilon} \right) - K \rfloor
.
\end{equation}
It announces the end of the block and decodes the bits matching codeword index $m$ if the termination condition is satisfied with respect to codeword $m$ (where ties can be broken arbitrarily), and does not terminate the block if the condition fails for all codewords.

Regarding the termination condition (\ref{eq:materm_condition}) note that the RHS starts from a negative value and increases linearly at a rate of $\log |\mathcal{X}|$ bits per symbol, while the LHS starts from a non-negative value, but for a compressible noise sequence, it is expected to increase at a rate slower than $\log |\mathcal{X}|$ bits per symbol, therefore if the noise sequence is compressible and the block length $n$ is large enough, the condition will eventually be met.

\subsubsection{Proof of the theorem}\label{sec:proof_L}
In order to prove Theorem~\ref{theorem:adaptive_L} it is shown that the scheme above achieves an error probability of at most $\epsilon$, and if an error does not occur, the number of bits decoded (determined by the number of blocks sent), approaches $\Remp$ for a suitable choice of $K$.

Let us begin by bounding the error probability. First let us calculate the probability that the decoder decides in favor of an incorrect codeword at any given symbol $i$ (where again $j$ denotes the end of the previous block), by using a property of the sequential encoder. Consider a sequence $\vr z^i$ of length $i$ which is fed into the sequential source encoder in two stages: first, the first $j$ symbols are fed (and the encoder has emitted $L_S(\vr z^j)$ bits), and then the rest $i-j$ symbols are fed and the encoding is terminated. Between the $j$-th and the $i$-th symbol, the encoder has emitted $L_T(\vr z^i) - L_S(\vr z^j)$ additional bits, which can be used to uniquely decode $\vr z_{j+1}^i$ when $\vr z^j$ is given (since the entire encoded stream can be generated from the first $L_S(\vr z^j)$ bits plus these additional bits, and used to decode $\vr z^i$). Therefore the number of sequences $\vr z_{j+1}^i$ for which $L_T(\vr z^i) - L_S(\vr z^j) \leq d$ (where $d \in \mathbb{N}$) is upper bounded by $\exp(d)$ (since they are in effect encoded by $d$ bits).

Since the codewords are independent, given the transmitted symbols, the other codewords in the codebook over the period of the current block are independent sequences uniformly drawn from $\mathcal{X}^{i-j}$. Therefore the hypothesized tail of the sequence $\vr {\hat Z}_{j+1}^i(m) = \vr Y_{j+1}^i - \vr X_{j+1}^i(m)$ for any fixed $m$ is also uniformly distributed (over the common randomness). Since there are at most $\exp(d)$ sequences that satisfy $L_T(\vr z^i) - L_S(\vr z^j) \leq d$, the probability that a particular sequence will satisfy the condition is at most
\begin{equation}
\frac{\exp(d)}{|\mathcal{X}|^{i-j}}
,
\end{equation}
and therefore by the union bound, the probability that any of the competing sequences will satisfy the condition is at most
\begin{equation}
\frac{\exp(d) \exp(K)}{|\mathcal{X}|^{i-j}} = \exp(d + K - (i-j) \log |\mathcal{X}|)
.
\end{equation}
Substituting the value of $d$ given by the termination condition $d = \lfloor (i-j) \cdot \log |\mathcal{X}| -\log \left( \frac{n}{\epsilon} \right) - K \rfloor \leq (i-j) \cdot \log |\mathcal{X}| -\log \left( \frac{n}{\epsilon} \right) - K$, the error probability per symbol is at most $\exp(-\log \left( \frac{n}{\epsilon} \right)) = \frac{\epsilon}{n}$, therefore by the union bound over $n$ symbols, the probability of any error occurring during the decoding process is at most $\frac{\epsilon}{n} \cdot n = \epsilon$.

Next, let us analyze the rate achieved by the scheme. The analysis assumes no decoding errors occur. Denote the number of decoded blocks by $B$ (so potentially there are $B+1$ blocks, if the last block is not decoded). The proof is based on bounding the value of $L(\vr z)$ based on the number of blocks. $\vr z$ denotes the true noise sequence.

Suppose a block was decoded in symbol $i$ and the previous block ended at symbol $j$. By choosing $K$ (or $n$) large enough it can be guaranteed that decoding never happens at the first symbol of any block, therefore $i > j+1$. By the assumption that no decoding errors occurred the sequence $\vr {\hat z}^j$ is identical to $\vr z^j$. In symbol $i-1$ the decoding condition was not met for any codeword, including the correct one, for which $\vr {\hat z}^i(m) = \vr z^i$. Therefore it holds, with respect to the true noise sequence, that:
\begin{equation}
L_T(\vr z^{i-1}) - L_S(\vr z^{j}) > (i-1-j) \log |\mathcal{X}| -\log \left( \frac{n}{\epsilon} \right) - K
.
\end{equation}
This is an inverted version of condition (\ref{eq:materm_condition}). Note that the floor operator $\lfloor \cdot \rfloor$ is not needed here since the LHS is an integer.

Using monotonicity of $L_T$ and the bounded difference $L_T - L_S$ the following telescopic series is lower bounded:
\begin{equation}\begin{split}\label{eq:ma02}
L_T(\vr z^{i}) - L_T(\vr z^{j})
&=
L_T(\vr z^{i}) - L_S(\vr z^{j}) - [L_T(\vr z^{j}) - L_S(\vr z^{j})]
\\& \geq
L_T(\vr z^{i}) - L_S(\vr z^{j}) - \Delta_L^{\max}(n)
\\& \geq
L_T(\vr z^{i-1}) - L_S(\vr z^{j}) - \Delta_L^{\max}(n)
\\& >
(i-1-j) \log |\mathcal{X}| -\log \left( \frac{n}{\epsilon} \right)
\\ & \qquad - K - \Delta_L^{\max}(n)
,
\end{split}\end{equation}
where $\Delta_L^{\max}(n) = \max \{\Delta_L(l) \}_{l=1}^n$. By the same argument, this bound is true also for the undecoded block (with $i-1=n$). Taking $j_b$ ($b=1,\ldots,B$) to be the symbol in which block $b$ ended, and adding $j_0=0$ and $j_{B+1}=n$ the following bound is obtained by summing (\ref{eq:ma02}) over $B+1$ blocks (including the undecoded one, which is taken as a block of length 0 if the last block is decoded):
\begin{equation}\begin{split}\label{eq:ma01}
L_T(\vr z)
& =
L_T(\vr z^{j_{B+1}}) - L_T(\vr z^{j_{0}})
\\& =
\sum_{b=1}^{B+1} \left[ L_T(\vr z^{j_b}) - L_T(\vr z^{j_{b-1}}) \right]
\\& >
\sum_{b=1}^{B+1} \left[ (j_b-1-j_{b-1}) \log |\mathcal{X}| -\log \left( \frac{n}{\epsilon} \right) - K - \Delta_L^{\max}(n)
 \right]
\\& =
n \log |\mathcal{X}| - (B+1) \left( K + \log |\mathcal{X}| + \log \left( \frac{n}{\epsilon} \right) +  \Delta_L^{\max}(n) \right)
.
\end{split}\end{equation}
The actual rate achieved by the scheme is
\begin{equation}
R_{act} = \frac{B K}{n}
.
\end{equation}
Extracting $B$ from (\ref{eq:ma01}) and calculating $R_{act}$ yields:
\begin{equation}\begin{split}\label{eq:maract_LB0}
R_{act}
&=
\frac{B K}{n}
\\& \geq
\frac{K}{n} \cdot \left(
\frac{n \log |\mathcal{X}|  - L_T(\vr z)}{K + \log |\mathcal{X}| + \log \left( \frac{n}{\epsilon} \right) +  \Delta_L^{\max}(n)} - 1 \right)
\\& =
\left(1 + \frac{\log (|\mathcal{X}| n/\epsilon) +  \Delta_L^{\max}(n)}{K} \right)^{-1} \Remp(\vr z) - \frac{K}{n}
\\& \stackrel{(a)}{\geq}
\left(1 - \frac{\log (|\mathcal{X}| n/\epsilon) +  \Delta_L^{\max}(n)}{K} \right) \Remp(\vr z) - \frac{K}{n}
\\& \stackrel{(b)}{\geq}
\Remp(\vr z)
\\& \qquad - \left[ \frac{\log |\mathcal{X}| \cdot (\log (|\mathcal{X}| n/\epsilon) +  \Delta_L^{\max}(n))}{K}  + \frac{K}{n} \right]
,
\end{split}\end{equation}
where (a) is because $\forall x \geq 0: (1+x)^{-1} \geq 1-x$, and (b) is because $\Remp(\vr z) \leq \log|\mathcal{X}|$.
To choose the value of $K$ that approximately minimizes the overhead term in the lower bound, the following lemma is used:
\begin{lemma}\label{lemma:ab_bound}
For $a>0, b>0$ with $b \leq a$
\begin{equation}
r = \min_{k \in \mathbb{N}} \left( \frac{a}{k} + b k \right) \leq 3 \sqrt{ab}
.
\end{equation}
\end{lemma}

\textit{Proof:} It is easy to see by derivation that the minimizer over $x \in \mathbb{R}$ of $\frac{a}{x} + b x$ is $x^* = \sqrt{\frac{a}{b}}$. Choosing $k^* = \lceil x^* \rceil$ yields $k^* \in \mathbb{N}$ and since $\sqrt{\frac{a}{b}} \leq k^* \leq \sqrt{\frac{a}{b}} +1$:
\begin{equation}\begin{split}
\frac{a}{k^*} + b k^*
& \leq
\frac{a}{\sqrt{\frac{a}{b}}} + b \left(\sqrt{\frac{a}{b}} +1 \right)
\\& =
2\sqrt{ab} + b
=
2\sqrt{ab} + \sqrt{b \cdot b}
\stackrel{b \leq a} \leq
3\sqrt{ab}
.
\end{split}\end{equation}
\endofproof

Applying the lemma to the choice of $K$ in (\ref{eq:maract_LB0}) yields:
\begin{equation}\begin{split}\label{eq:maract_LB}
& R_{act}
\geq
\Remp(\vr z)
\\ &
- \underbrace{3 \sqrt { \frac{\log |\mathcal{X}|}{n} \cdot \left[ \log(n) + \log \left(\frac{|\mathcal{X}|}{\epsilon}\right) +  \Delta_L^{\max}(n) \right]}}_{\delta_n}
,
\end{split}\end{equation}
where by assumption (A) of Theorem~\ref{theorem:adaptive_L}, $\delta_n \arrowexpl{n \to \infty} 0$. \endofproof

} 

\subsection{Proof of Theorem \ref{theorem:adaptive_rho}}\label{sec:proof_rho}
\onlyphd{
According to Section~\ref{sec:examples_compression_modadditive}, the compression length $L_{78}(\vr z)$ of the LZ78 algorithm \cite{LZ78} fulfils the requirements of Theorem~\ref{theorem:adaptive_L}, and may be substituted there. Hence the rate function $\Remp(\vr z) = \log |\mathcal{X}| - \frac{1}{n} L_{78}(\vr z)$ is attainable \eqref{eq:A3551}, up to $\delta_n$ defined in the theorem.
}

\onlypaper{
To prove Theorem~\ref{theorem:adaptive_rho}, it is first shown that LZ77 \cite{LZ77} and LZ78 \cite{LZ78} fulfil the requirements of Theorem~\ref{theorem:adaptive_L}. Both algorithms operate by creating a dictionary from previous symbols in the string, compressing a new substring to a tuple containing its location in the dictionary, plus, possibly one additional symbol. In LZ77 the dictionary consists of all substrings that begin in a window of specified length before the first symbol that was not encoded yet. LZ78 parses the string $\vr z$ into phrases. Each phrase is a substring which is not a prefix of any previous phrase, but can be generated from concatenating a previous phrase with one additional symbol. The dictionary contains all phrases.

It is easy to make sure that $L_T$ is monotonous (Assumption~(B) of Theorem~\ref{theorem:adaptive_L}). This depends on the way the last phrase in the string is treated, which does not affect the asymptotical performance. Recall that in LZ compression, in which the new data bits are gathered, and encoded to produce a tuple, once they comprise a phrase that had not appeared before. The last phrase may be an incomplete substring of a string in the dictionary, and therefore does not naturally terminate by this rule. For example, in the following parsed sequence $[1,0,11,110,00,110]$, the last phrase $110$ had appeared before and therefore would not naturally produce a tuple. There are various ways to treat this last phrase. If, for example, the last phrase is sent without coding, then $L_T$ will not be monotonous, since adding more symbols to $\vr z$ that will terminate the phrase and may result in a shorter compression. For example, the addition of either $0$ or $1$ to the sequence above, would generate a phrase that had not appeared before. A simple treatment is to encode the last phrase similarly to other phrases, i.e. refer to one of the phrases in the dictionary which is a prefix of the remaining substring (in the example, refer to the previous appearance of $110$), and always give the length of the last substring, or equivalently the length of the block, at the end. This way the compression length associated with the last substring does not decrease when the substring is extended.

In order to bound $L_T(\vr z) - L_S(\vr z)$ (Assumption~(A)), it is required to bound the tuple which encodes the last phrase. In LZ78 this tuple carries an index to a previous phrase, plus a new symbol. The number of previous phrases is bounded by $n$ (a coarse bound, but sufficient for the current purpose), and therefore \cite[Lemma 13.5.1]{CoverThomas_InfoTheoryBook} its encoding will be of length $\log n + \log \log n + 1$, and the length of the tuple will be $\log n + \log \log n + c$ (where $c$ is a constant accounting also for rounding, encoding of the additional symbol, etc). Therefore, if the encoder ends the block with an indication of its length then $\Delta^{\max}_{LZ78}(n) = \Delta_{LZ78}(n) \leq 2 \log n + 2 \log \log n + c$. In LZ77 this tuple carries a pointer to the window and a length (i.e. two numbers bounded to $\{1,\ldots,n\}$). Therefore after adding an indication of the length at the termination, $\Delta^{\max}_{LZ77}(n) = \Delta_{LZ77}(n) \leq 3 \log n + 3 \log \log n + c$. In both cases $\Delta^{\max}_{LZ}(n) = O(\log n)$ and the requirement is satisfied. Therefore the compression length $L_{78}(\vr z)$ may be substituted in Theorem~\ref{theorem:adaptive_L}.
} 

The rest of the proof deals with analyzing and bounding the overheads related to the achievability of $\Remp$, and the difference between the LZ compression length and the finite state compressibility, in order to show that they tend to $0$ with $n$. Recall the definitions of finite state compressibility \eqref{eq:ma444}-\eqref{eq:ma446} in Section~\ref{sec:IFB_UB}.

A result by Lempel and Ziv \cite[Theorem 2 (item ii)]{LZ78} shows that for every finite $s$
\begin{equation}
\rho_{78}(\vr z_1^n) \defeq \frac{1}{n \log | \mathcal{X} |} L_{78}(\vr z_1^n) \leq \rho_{\mathcal{F}(s)}(\vr z_1^n) + \delta_s(n)
,
\end{equation}
where $\delta_s(n) \arrowexpl{n \to \infty} 0$. By Theorem~\ref{theorem:adaptive_L} for any $\epsilon>0$, the system attains the rate
\begin{equation}\begin{split}\label{eq:ma596}
R
& \geq
\Remp(\vr z) - \delta_n
\\& =
\log | \mathcal{X} | \left(1 - \frac{1}{n \log | \mathcal{X} |} L_{78}(\vr z_1^n) \right) - \delta_n
\\& =
(1 -\rho_{78}(\vr z_1^n))\log | \mathcal{X} |  - \delta_n
\\& \geq
 (1 - \rho_{\mathcal{F}(s)}(\vr z_1^n) - \delta_s(n)) \log | \mathcal{X} | - \delta_n
.
\end{split}\end{equation}

Choose a small $\tilde{\delta}$. Since $\lim_{n \to \infty} \delta_n = 0$ it is possible to find $n_1^*$ large enough so that for any $n > n_1^*$, $\delta_n \leq \tilde{\delta}$. By the definition $\rho(\vr z) = \rho(\vr z^\infty) = \lim_{s \to \infty} \limsup_{n \to \infty} \rho_{\mathcal{F}(s)} (\vr z_1^n)$, it is possible to find $s$ large enough such that $\limsup_{n \to \infty} \rho_{\mathcal{F}(s)} (\vr z_1^n) \leq \rho(\vr z) + \tilde{\delta}$. For this value of $s$, because $\lim_{n \to \infty} \delta_s(n) = 0$, it is possible to find $n_2^*$ large enough so that for any $n > n_2^*$, $\delta_s(n) \leq \tilde{\delta}$. For the same $s$, find $n > n_1^*, n_2^*$ so that $\rho_{\mathcal{F}(s)} (\vr z_1^n) \leq \limsup_{n' \to \infty} \rho_{\mathcal{F}(s)} (\vr z_1^{n'}) + \tilde{\delta} \leq \rho(\vr z) +  2 \tilde{\delta}$. Writing (\ref{eq:ma596}) for these $s,n$ yields:
\begin{equation}\begin{split}
R
& \geq
(1 - \rho_{\mathcal{F}(s)}(\vr z_1^n) - \tilde{\delta})\log | \mathcal{X} |  - \tilde{\delta}
\\& \geq
 (1 - \rho(\vr z) - 3 \tilde{\delta})\log | \mathcal{X} | - \tilde{\delta}
\\& =
 (1 - \rho(\vr z)) \log | \mathcal{X} | - (3 \log | \mathcal{X} |  + 1) \cdot \tilde{\delta}
.
\end{split}\end{equation}
Therefore the requirements of Theorem~\ref{theorem:adaptive_rho} are satisfied by substituting $\tilde{\delta} = (3 \log | \mathcal{X} |  + 1)^{-1} \delta$. \endofproof

\textit{Proof of Corollary \ref{corollary:univ}:}
The corollary follows directly from the definition, by application of Theorem~\ref{theorem:adaptive_rho} and Theorem~\ref{theorem:Cifb_UB}.

\textit{Proof of Corollary \ref{corollary:ergodic}:}
Suppose the sequence $\vr z$ is drawn by a stationary ergodic source.
The mutual information rate is $\overline{I}(\vr X; \vr Y) = \overline{H}(\vr Y) - \overline{H}(\vr Y | \vr X) \leq \log |\mathcal{X}| - \overline{H}(\vr Z)$, and to obtain an equality, the capacity is obtained by a uniform i.i.d. prior, which maximizes $\overline{H}(\vr Y)$. Hence the capacity is $C = \log |\mathcal{X}| - \overline{H}(\vr Z)$.
It was shown \cite[Theorem 4]{LZ78}  that the finite state compressibility equals the entropy rate of the source, with probability one. The proposed communication system would asymptotically attain the communication rate $C$, without prior knowledge of the noise distribution.

\subsection{Proof of Corollary~\ref{corollary:mod_additive_redundancy}}\label{sec:proof_of_corollary_mod_additive_redundancy}
The target is to find the required $n$ such that $\Delta^* \leq \delta \log |\mathcal{X}|$, based on the bounds of Theorem~\ref{theorem:mod_additive_redundancy}. The lower bound of Theorem~\ref{theorem:mod_additive_redundancy} on $\Delta^*$ yields a lower bound on $n^*$ (converse) and the upper bound on $\Delta^*$ yields an upper bound on $n^*$ (achievability).

\subsubsection{Converse}
According to the lower bound \eqref{eq:ma1063b}, either $\tau \leq \frac{ |\mathcal{X}| }{k}$,  or $\left\lfloor  \log \left( k \tau \right)\frac{1}{ \log |\mathcal{X}|} \right\rfloor \frac{\log |\mathcal{X}|}{2 k} \leq \delta \log |\mathcal{X}|$, which combined with $\left\lfloor x \right\rfloor \geq x - 1$ yields, after rearrangement, $\log \left( k \tau \right) \leq (2 k \delta +1) \log |\mathcal{X}|$, i.e. $\tau \leq \frac{1}{k} |\mathcal{X}|^{2 k \delta +1}$. This condition on $\tau$ is always less strict than the former, and because at least one of the conditions should hold, the second always holds. Translating to a condition on $n$ yields:
\begin{equation}\label{eq:ma553}
n = \frac{|\mathcal{X}|^k}{\tau} \geq \frac{1}{|\mathcal{X}|} \cdot k |\mathcal{X}|^{(1-2\delta) k}
.
\end{equation}

\subsubsection{Achievability}
Let us find an $n$ for which the upper bound is at most $\delta \log |\mathcal{X}|$. Define $g(\tau) = \tau \log \left( \frac{1}{\tau} \right)$. Assuming $\tau \leq \frac{1}{2 k}$, then $g(\tau)$ is monotonically increasing, $g(\tau) \geq \tau \log (2 k)$, and $\frac{k}{4} \tau^2 \leq \frac{1}{8} \tau$. Thus:
\begin{equation}\begin{split}\label{eq:ma558}
& \frac{\tau}{2} \log \left( \frac{1}{\tau} \right) + \left( \frac{k}{4} \tau^2 + \tau \right) \log e
\\& \leq
\half g(\tau)  + \left( \frac{1}{8}  + 1 \right) \tau \log e
\\& \leq
\half g(\tau)  + \frac{9}{8} \cdot \frac{g(\tau)}{\log (2 k)}  \log e
\\& =
\left(\half   + \frac{9 \log e}{8 \log (2 k)}  \right) g(\tau)
\\& \leq
3 g(\tau)
.
\end{split}\end{equation}
The same assumption $\tau \leq \frac{1}{k}$ leads to $n \geq k |\mathcal{X}|^k$ and thus
\begin{equation}\label{eq:ma569}
\frac{k}{n} \leq |\mathcal{X}|^{-k}
,
\end{equation}
and
\begin{equation}\begin{split}\label{eq:ma573}
\delta_n^*
& \leq
4 \sqrt{\frac{\log |\mathcal{X}| \cdot \log \left( k^2 |\mathcal{X}|^{2k+1} \right)}{k |\mathcal{X}|^k}}
\\& \stackrel{(a)}{\leq}
4 \sqrt{\frac{ 5 (\log |\mathcal{X}|)^2}{|\mathcal{X}|^k}}
\\& \leq
10 \log |\mathcal{X}| \cdot |\mathcal{X}|^{-k/2}
,
\end{split}\end{equation}
where (a) is because $\log k \leq (k-1) \log e$, so $k^2 \leq \log e^{2(k-1)} \leq |\mathcal{X}|^{3(k-1)}$. Combining \eqref{eq:ma558}, \eqref{eq:ma569}, \eqref{eq:ma573} with \eqref{eq:ma506} yields:
\begin{equation}\begin{split}\label{eq:ma585}
\Delta_{+}
&\leq
3 g(\tau) +
|\mathcal{X}|^{-k} \log (e |\mathcal{X}|) + 10 \log |\mathcal{X}| \cdot |\mathcal{X}|^{-k/2}
\\&=
3 g(\tau) +
\left( |\mathcal{X}|^{-k/2} \left(1 + \frac{\log (e)}{\log |\mathcal{X}|} \right) + 10 \right) \log |\mathcal{X}| \cdot |\mathcal{X}|^{-k/2}
\\& \leq
3 g(\tau) +
\left( 2^{-1/2} \left(1 + \frac{\log (e)}{\log 2} \right) + 10 \right) \log |\mathcal{X}| \cdot |\mathcal{X}|^{-k/2}
\\& \leq
3 g(\tau) +
12 \log |\mathcal{X}| \cdot |\mathcal{X}|^{-k/2}
.
\end{split}\end{equation}
Thus, to guarantee $\Delta^* \leq \delta \cdot \log |\mathcal{X}|$ is it enough if $\tau \leq \frac{1}{k}$ and
\begin{equation}\label{eq:ma600}
\tau \leq g^{-1} \left( \tfrac{1}{3} (\delta - 12 \cdot |\mathcal{X}|^{-k/2}) \cdot \log |\mathcal{X}| \right)
,
\end{equation}
i.e. it is required that
\begin{equation}\label{eq:ma600a}
\tau \leq \min \left[ g^{-1} \left( \tfrac{1}{3} (\delta - 12 \cdot |\mathcal{X}|^{-k/2}) \cdot \log |\mathcal{X}| \right), 1/k \right]
,
\end{equation}
and equivalently
\begin{equation}\label{eq:ma603j}
n = \frac{|\mathcal{X}|^k}{\tau} \geq \frac{|\mathcal{X}|^k}{\min \left[ g^{-1} \left( \tfrac{1}{3} (\delta - 12 \cdot |\mathcal{X}|^{-k/2}) \cdot \log |\mathcal{X}| \right), 1/k \right]}
.
\end{equation}
\endofproof

\subsection{Proof of Lemma~\ref{lemma:bound_on_HZ_test_channel}}\label{sec:bound_on_HZ_test_channel}
In this section, Lemma~\ref{lemma:bound_on_HZ_test_channel} from Section~\ref{sec:proof_theorem_mod_additive_redundancy_reverse}, regarding the entropy of the noise distribution of the test channel defined there, is proven. See page \pageref{page:definition_of_test_channel} for the definition of the test channel.

The entropy of each prefix, conditioned on the past is $\log |\mathcal{X}|^{k-d} = (k-d) \log |\mathcal{X}| = k \cdot \overline H_1$, where $\overline H_1$ is the asymptotical entropy rate per symbol. The entropy of the suffix, given the past, changes over time. When choosing the $i$-th noise sequence (of length $k$), at most $i-1$ different prefixes already appeared. Therefore, the probability that the $i$-th prefix equals one of the previous ones is at most $\frac{i-1}{|\mathcal{X}|^{k-d}}$. Let us define $i^*_d \defeq |\mathcal{X}|^{k-d}$ and consider first the case $i \leq i^*_d$. In this case, the entropy of the suffix, given all previous symbols, is zero with probability at most $\frac{i-1}{|\mathcal{X}|^{k-d}}$, and $\log |\mathcal{X}|^{d}$ with probability at least $1-\frac{i-1}{|\mathcal{X}|^{k-d}}$, and is therefore at least $\left( 1-\frac{i-1}{|\mathcal{X}|^{k-d}} \right) \cdot d \cdot \log |\mathcal{X}|$. Formally, define $P_z^{(i)} \defeq (\vr Z_{i}^{[k]})_{1}^{k-d}, S_z^{(i)} \defeq (\vr Z_{i}^{[k]})_{k-d+1}^{k}$ as the $i$-th prefix and suffix, and $F_i = \Ind \left[ \bigcup_{j=1}^{i-1} \left\{ P_z^{(i)} = P_z^{(j)}\right\} \right]$ as a flag indicating whether $P_z^{(i)}$ appeared before. Then the entropy of the suffix given the past is:
\begin{equation}\begin{split}\label{eq:ma795}
& H(S_z^{(i)} | \vr Z_1^{(i-1)k + k-d})
 =
H(S_z^{(i)} | \vr Z_1^{(i-1)k}, P_z^{(i)})
\\& \stackrel{(a)}{=}
H(S_z^{(i)} | \vr Z_1^{(i-1)k}, P_z^{(i)}, F_i)
\\& =
H(S_z^{(i)} | \vr Z_1^{(i-1)k}, P_z^{(i)}, F_i=0) \cdot \Pr(F_i = 0)
\\ & \qquad +
H(S_z^{(i)} | \vr Z_1^{(i-1)k}, P_z^{(i)}, F_i=1) \cdot \Pr(F_i = 1)
\\& =
d \cdot \log |\mathcal{X}| \cdot \Pr(F_i = 0)
\\& \geq
d \cdot \log |\mathcal{X}| \cdot \left( 1-\frac{i-1}{|\mathcal{X}|^{k-d}} \right)
,
\end{split}\end{equation}
where (a) is because $F_i$ is a function of $\vr Z_1^{(i-1)k}, P_z^{(i)}$. Therefore
\begin{equation}\begin{split}\label{eq:ma970}
H(\vr Z_i^{[k]} | \vr Z^{(i-1)k})
&=
H(P_z^{(i)} | \vr Z^{(i-1)k}) + H(S_z^{(i)} | P_z^{(i)}, \vr Z^{(i-1)k})
\\& \geq
d \cdot \log |\mathcal{X}| \cdot \left( 1-\frac{i-1}{|\mathcal{X}|^{k-d}} \right) + k \cdot \overline H_1
.
\end{split}\end{equation}
For $i \geq i^*_d$ simply bound:
\begin{equation}\label{eq:ma970a}
H(\vr Z_i^{[k]} | \vr Z^{(i-1)k}) \geq H(P_z^{(i)} | \vr Z^{(i-1)k}) = k \cdot \overline H_1
.
\end{equation}
Notice that $H_1$ determines the asymptotical entropy rate of the sequence, and matches the bound on the universal system and the rate of the IFB system.

For $i \leq i^*_d$:
\begin{equation}\begin{split}\label{eq:ma971}
H(\vr Z^{ik})
&=
\sum_{j=1}^i H(\vr Z_j^{[k]} | \vr Z^{(j-1)k})
\\& \stackrel{\eqref{eq:ma970}}{\geq}
\sum_{j=1}^i \left( d \cdot \log |\mathcal{X}| \cdot \left( 1-\frac{j-1}{|\mathcal{X}|^{k-d}} \right) + k \cdot \overline H_1 \right)
\\& =
d \cdot \log |\mathcal{X}| \cdot \left( i -\frac{(i-1)i}{2 |\mathcal{X}|^{k-d}} \right) + i k \cdot \overline H_1
\\& =
d \cdot \log |\mathcal{X}| \cdot i \underbrace{\left( 1 - \frac{i-1}{2 |\mathcal{X}|^{k-d}} \right)}_{\geq \half} + i k \cdot \overline H_1
\\& \geq
\half i \cdot d \cdot \log |\mathcal{X}| + i k \cdot \overline H_1
.
\end{split}\end{equation}
The above implies that the entropy, at times $n=ik \leq i^*_d k$, is bounded above the straight line with slope
\begin{equation}\label{eq:ma840}
\overline H_0 = \frac{d}{2 k} \cdot \log |\mathcal{X}| + \overline H_1
.
\end{equation}
See Fig.\ref{fig:test_channel_noise_entropy}. For $i \geq i^*_d$,
\begin{equation}\begin{split}\label{eq:ma1034}
H(\vr Z^{ik})
&=
H(\vr Z^{i^*_d k}) +  \sum_{t=i^*_d+1}^i H(\vr Z_t^{[k]} | \vr Z^{(t-1)k})
\\& \stackrel{\eqref{eq:ma970a}, \eqref{eq:ma971}}{\geq}
k |\mathcal{X}|^{k-d} \overline H_0 + k (i - |\mathcal{X}|^{k-d}) \overline H_1
.
\end{split}\end{equation}
and in general
\begin{equation}\label{eq:ma1042}
H(\vr Z^{ik}) \geq i k \cdot \overline H_1 + \min(i, |\mathcal{X}|^{k-d}) k (\overline H_0 - \overline H_1)
.
\end{equation}
Consider now $H(\vr Z^n)$ for $n$ that does not, in general, divide by $k$. Inside the block of length $k$, the per-symbol conditional entropy $H(\vr Z_n | \vr Z^{n-1})$ is $\log |\mathcal{X}|$ during the prefix, and then increases at a smaller or equal rate during the suffix. Therefore the entropy $H(\vr Z^n)$ is concave during the block (Fig.\ref{fig:test_channel_noise_entropy}). Because the entropy at block edges is bounded above straight lines \eqref{eq:ma1042}, the entropy inside the block is bounded by these lines as well, i.e. \eqref{eq:ma1042} can be extended to:
\begin{equation}\begin{split}\label{eq:ma1045cpy}
H(\vr Z^{n}) \geq n \cdot \overline H_1 + \min(n, k |\mathcal{X}|^{k-d}) (\overline H_0 - \overline H_1)
,
\end{split}\end{equation}
which proves the lemma.
\endofproof

\subsection{Proof of Lemma~\ref{lemma:univ_distribution_multiple_k}}\label{sec:proof_lemma_univ_distribution_multiple_k}
For the sake of brevity, as long as a single value of $k$ is discussed, let $\mathcal{M} \defeq \mathcal{X}^k$ denote the super-alphabet of length $k$ and $m = |\mathcal{X}|^k$ denote its size. Let $\pi(\cdot)$ define a distribution over $\mathcal{M}$. The Dirichlet$(\half,\ldots,\half)$ density over the set of distributions is defined as:
\begin{equation}
w_k(\pi) = \exp(-C_m) \prod_{a \in \mathcal{M}} \pi(a)^{-1/2}
,
\end{equation}
where
\begin{equation}\label{eq:ma527}
C_m = \log \left( \Gamma(1/2)^m / \Gamma(m/2) \right)
,
\end{equation}
and for a $l$-length vector $\vr a \in \mathcal{M}^l$, let $\pi(\vr a) = \prod_{i=1}^l {\pi(a_i)}$ be the probability given to $\vr a$ by the i.i.d. distribution $\pi(\cdot)$. Let
\begin{equation}\label{eq:ma6664}
P_k(\vr a) = \int_{\Delta_{\mathcal{M}}} \pi(\vr a)  w_k(\pi) d \pi
,
\end{equation}
and define the weighted average of all probabilities given to $\vr a$ by i.i.d. distributions $\pi(\vr a)$, where the integral is over the unit simplex $\Delta_{\mathcal{M}} = \{ \pi: \forall a \in \mathcal{M}: \pi(a) \geq 0, \sum_{a \in \mathcal{M}} \pi(a) = 1 \}$.
By well known results of Shtarkov, which are detailed in Lemma~1 in Xie and Barron's paper \cite{XieBarron00}, it holds that:
\begin{equation}\begin{split}\label{eq:ma6612}
\log \frac{\max_{\pi} \pi(\vr a)}{P_k(\vr a)} & \leq \frac{m-1}{2} \log \frac{l}{2\pi} + C_m + \left( \frac{m^2}{4 l} + \frac{m}{2} \right) \log e
\onlypaper{\\&} \defeq r_{lk}
.
\end{split}\end{equation}
Note that the terms that do not scale with $n$ are usually ignored, because $m$ is considered fixed, however here they matter, because the question would be how fast $m$ (equivalently $k$) may grow with $n$. Thus for any $\pi$:
\begin{equation}\label{eq:ma6667}
\pi(\vr a) \leq P_k(\vr a) \exp(r_{lk})
.
\end{equation}
The same inequality would hold when marginalizing the above to any parts of $\vr a$ (i.e. summing over the remaining elements of $\vr a$). Using this observation, let us set $l = \lceil n/k \rceil$. Then substituting in \eqref{eq:ma6664}, $\vr a = \vr z_1^{lk}$ yields
\begin{equation}\label{eq:ma6665}
P_k(\vr z^{lk}) = \int_{\Delta_{\mathcal{M}}} \pi_k(\vr z^{lk})  w_k(\pi) d \pi
,
\end{equation}
and summing both sizes with respect to $\vr z_{n+1}^{lk}$, yields
\begin{equation}\label{eq:ma6666}
P_k(\vr z^{n}) = \int_{\Delta_{\mathcal{M}}} \pi_k(\vr z^{n})  w_k(\pi) d \pi
.
\end{equation}
Furthermore, by \eqref{eq:ma6667}
\begin{equation}\label{eq:ma6668}
\forall \pi: \pi_k(\vr z^{n}) \leq P_k(\vr z^{n}) \cdot \exp(r_{lk})
.
\end{equation}
Let us now bound $r_{lk}$. Following Xie and Barron's \cite[Remark 7]{XieBarron00}, using $\Gamma(1/2)=\sqrt{\pi}$ and Stirling's approximation $\Gamma(m/2) \geq \sqrt{2\pi} (m/2)^{\tfrac{m-1}{2}} e^{-\tfrac{m}{2}}$ yields from \eqref{eq:ma527}:
\begin{equation}\label{eq:ma562}
C_m \leq \frac{m-1}{2} \log \pi - \half \log 2 - \frac{m-1}{2} \log(m/2)
+ \frac{m}{2} \log e
,
\end{equation} 
and from \eqref{eq:ma6612}:
\begin{equation}
r_{lk} \leq \frac{m-1}{2} \log \left( \frac{l}{m} \right) + \left( \frac{m^2}{4 l} + m \right) \log e - \half \log 2
.
\end{equation}
Note that $r_{lk}$ is always positive, even when $l < m$. When $l < m$, the second factor dominates, and the normalized loss $\frac{r_{lk}}{l}$ does not tend to zero. Therefore it is not useful to consider $m$ in this region. Assuming $l \geq m$ (note that since $m \geq 2$ this also implies $n \geq k$), and substituting $l=\lceil n/k \rceil \leq n$, $m = |\mathcal{X}|^k$, yields
\begin{equation}\begin{split}\label{eq:ma577}
r_{lk}
& \leq
\frac{|\mathcal{X}|^k-1}{2} \log \left( \frac{\lceil n/k \rceil}{|\mathcal{X}|^k} \right) + \left( \frac{k |\mathcal{X}|^{2k}}{4 n} + |\mathcal{X}|^k \right) \log e
\\ & \qquad - \half \log 2
\\ & \leq
\frac{|\mathcal{X}|^k}{2} \log \left( \frac{n}{|\mathcal{X}|^k} \right) + \left( \frac{k |\mathcal{X}|^{2k}}{4 n} + |\mathcal{X}|^k \right) \log e
.
\end{split}\end{equation}
Now let
\begin{equation}\label{eq:ma583}
P_Z(\vr z) = \sum_{k=1}^{\infty} 2^{-k} \cdot P_k(\vr z)
,
\end{equation}
then from \eqref{eq:ma6668}
\begin{equation}\label{eq:ma6668b}
\forall \pi: \pi_k(\vr z^{n}) \leq P_k(\vr z^{n}) \cdot \exp(r_{lk}) \leq \frac{P_Z(\vr z)}{2^{-k}} \exp(r_{lk})
,
\end{equation}
and thus
\begin{equation}\begin{split}\label{eq:ma6668c}
\forall \pi: \frac{1}{n} \log \pi_k(\vr z^{n}) \leq \frac{1}{n} \log P_Z(\vr z) + \frac{1}{n} \left( k \log(2) + r_{lk} \right)
.
\end{split}\end{equation}
The factor $\frac{1}{n} \left( k \log(2) + r_{lk} \right)$ can be coarsely bounded by \eqref{eq:ma577}:
\begin{equation}\begin{split}\label{eq:ma604}
\frac{k \log(2) + r_{lk} }{n}
& \leq
\frac{|\mathcal{X}|^k}{2 n} \log \left( \frac{n}{|\mathcal{X}|^k} \right)
\\& \qquad + \left( \frac{k |\mathcal{X}|^{2k}}{4 n^2} + \frac{|\mathcal{X}|^k}{n} + \frac{k}{n} \right) \log e
\\& =
\frac{\tau}{2} \log \left( \frac{1}{\tau} \right) + \left( \frac{k}{4} \tau^2 + \tau + \frac{k}{n} \right) \log e
\\& \defeq
\Delta_\pi
,
\end{split}\end{equation}
with $\tau \defeq \frac{|\mathcal{X}|^k}{n} \leq 1$.
Combining this bound with \eqref{eq:ma6668c} yields the result of the lemma.
\endofproof

\subsection{A password channel for i.i.d. distributions}\label{sec:password_channel_iid_example}
As noted in Sections~\ref{sec:IFB_capacity_discussion}, \ref{sec:ma_alt_class_iid}, even limiting the reference class it to i.i.d. input distributions would not solve the ``password'' problem, and therefore universality is not possible even with respect to such encoders, for general channels. To see this, consider the following example, where the channel identifies the input distribution of the encoder. This is a variation of the ``password channel'' (Example~\ref{example:password_channel}).

\begin{example}\label{example:password_channel_iid_example}
The channel class is a class of binary input-output channels, parameterized by a single a parameter $p \in [0,1]$. For each value of $p$, the channel is defined as follows:
\begin{itemize}
\item At each symbol $k$ in time, if the normalized number of ones at the input $\vr x_1^{k}$ is not within a range of thresholds $[L_{k,p}, H_{k,p}]$, then from this time on, the channel ``locks'' and the output is $y_i=0, \forall i \geq k$. Otherwise, the channel is noise free and the output equals the input $y_k = x_k$.
\item The threshold sequences $L_{k,p}, H_{k,p}$ are computed such that, if the input is i.i.d. $\Ber(p)$, then with high probability $1-\epsilon_0$, the thresholds will not be crossed during any of the $n$ symbols (i.e. the channel will not lock). Clearly, as $k$ increases, the thresholds will converge to $p$.
\end{itemize}
\end{example}

Thus, the channel ``identifies'' a certain input probability. Notice that all the channels are causal and deterministic, and they allow communication at a rate of approximately $h_b(p)$. The ``memoryless'' reference schemes mentioned in Section~\ref{sec:ma_alt_class_iid} can communicate over this channel using a $\Ber(p)$ input distribution and approach this rate, with a small error probability. But a universal communication over the class is impossible. Until the channel locks, nothing can be inferred about $p$ from the channel output. Therefore the transmit distribution of the universal scheme until the lock time is independent of $p$. On the other hand, any given input sequence, will ``lock'' some of the channels in the class. Therefore any operation of the universal system is bound to cause some of the channels to lock, and achieve an asymptotically zero rate.

\fi 

\ifx\PhdMode\undefined  


\begin{thebibliography}{10}
\providecommand{\url}[1]{#1}
\csname url@samestyle\endcsname
\providecommand{\newblock}{\relax}
\providecommand{\bibinfo}[2]{#2}
\providecommand{\BIBentrySTDinterwordspacing}{\spaceskip=0pt\relax}
\providecommand{\BIBentryALTinterwordstretchfactor}{4}
\providecommand{\BIBentryALTinterwordspacing}{\spaceskip=\fontdimen2\font plus
\BIBentryALTinterwordstretchfactor\fontdimen3\font minus
  \fontdimen4\font\relax}
\providecommand{\BIBforeignlanguage}[2]{{%
\expandafter\ifx\csname l@#1\endcsname\relax
\typeout{** WARNING: IEEEtran.bst: No hyphenation pattern has been}%
\typeout{** loaded for the language `#1'. Using the pattern for}%
\typeout{** the default language instead.}%
\else
\language=\csname l@#1\endcsname
\fi
#2}}
\providecommand{\BIBdecl}{\relax}
\BIBdecl

\bibitem{Ofer_ModuloAdditive}
O.~Shayevitz and M.~Feder, ``Achieving the empirical capacity using feedback:
  Memoryless additive models,'' \emph{IEEE Trans. Information Theory}, vol.~55,
  no.~3, pp. 1269 --1295, Mar. 2009.

\bibitem{HanVerdu}
S.~Verd\'u and T.~Han, ``A general formula for channel capacity,'' \emph{IEEE
  Trans. Information Theory}, vol.~40, no.~4, pp. 1147 --1157, Jul. 1994.

\bibitem{Lapidoth_AVC}
A.~Lapidoth and P.~Narayan, ``Reliable communication under channel
  uncertainty,'' \emph{IEEE Trans. Information Theory}, vol.~44, no.~6, pp.
  2148--2177, Oct. 1998.

\bibitem{Eswaran}
K.~Eswaran, A.~Sarwate, A.~Sahai, and M.~Gastpar, ``Zero-rate feedback can
  achieve the empirical capacity,'' \emph{IEEE Trans. Information Theory},
  vol.~58, no.~1, Jan. 2010.

\bibitem{YL_individual_full}
Y.~Lomnitz and M.~Feder, ``Communication over individual channels,'' \emph{IEEE
  Trans. Information Theory}, vol.~57, no.~11, pp. 7333 --7358, Nov. 2011.

\bibitem{LZ78}
J.~Ziv and A.~Lempel, ``Compression of individual sequences via variable-rate
  coding,'' \emph{IEEE Trans. Information Theory}, vol.~24, no.~5, pp. 530 --
  536, Sep. 1978.

\bibitem{FederMerhav98}
N.~Merhav and M.~Feder, ``Universal prediction,'' \emph{IEEE Trans. Information
  Theory}, vol.~44, no.~6, pp. 2124--2147, Oct. 1998.

\bibitem{FederLapidoth_UnivDecod98}
M.~Feder and A.~Lapidoth, ``Universal decoding for channels with memory,''
  \emph{IEEE Trans. Information Theory}, vol.~44, no.~5, pp. 1726--1745, Sep.
  1998.

\bibitem{YL_UnivCommMemory}
\BIBentryALTinterwordspacing
Y.~Lomnitz and M.~Feder. (2012, Jan.) Universal communication part {II}:
  channels with memory. arXiv:1202.0417 [cs.IT]. Submitted to IEEE-IT.
  [Online]. Available: \url{http://arxiv.org/abs/1202.0417}
\BIBentrySTDinterwordspacing

\bibitem{FederMerhav93}
N.~Merhav and M.~Feder, ``Universal schemes for sequential decision from
  individual data sequences,'' \emph{IEEE Trans. Information Theory}, vol.~39,
  no.~4, pp. 1280 --1292, Jul. 1993.

\bibitem{MisraPorosityISIT12}
V.~Misra and T.~Weissman, ``The porosity of additive noise sequences,'' in
  \emph{IEEE Int. Symp. Information Theory (ISIT)}, 2012.

\bibitem{KrichevskyTrofimov81}
R.~E. Krichevsky and V.~K. Trofimov, ``The performance of universal encoding,''
  \emph{IEEE Trans. Information Theory}, vol.~27, no.~2, pp. 199--207, Mar.
  1981.

\bibitem{YL_PhdThesis}
Y.~Lomnitz, ``Universal communication over unknown channels,'' Ph.D.
  dissertation, Tel Aviv University, Aug. 2012, available online
  \url{http://www.eng.tau.ac.il/~yuvall/publications/YuvalL_Phd_report.pdf}.

\bibitem{YL_ModuloAdditiveEilat}
Y.~Lomnitz and M.~Feder, ``Communicating over modulo-additive channels with
  compressible individual noise sequence,'' in \emph{26-th IEEE Convention of
  Electrical and Electronics Engineers in Israel (IEEEI)}, Nov. 2010.

\bibitem{LZ77}
J.~Ziv and A.~Lempel, ``A universal algorithm for sequential data
  compression,'' \emph{IEEE Trans. Information Theory}, vol.~23, pp. 337--343,
  Sep. 1977.

\bibitem{Ahlswede71}
R.~Ahlswede, ``A constructive proof of the coding theorem for discrete
  memoryless channels with feedback,'' in \emph{Proceedings of the Sixth Prague
  Conference on Information Theory, Statistical Decision Functions, and Random
  Processes}, 1971, pp. 39--50.

\bibitem{Ooi}
J.~Ooi, ``A framework for low-complexity communication over channels with
  feedback,'' Ph.D. dissertation, MIT, Cambridge, MA, 1997.

\bibitem{ZivUniversal}
J.~Ziv, ``Universal decoding for finite-state channels,'' \emph{IEEE Trans.
  Information Theory}, vol.~31, no.~4, pp. 453--460, Jul. 1985.

\bibitem{CoverThomas_InfoTheoryBook}
T.~M. Cover and J.~A. Thomas, \emph{Elements of Information Theory}.\hskip 1em
  plus 0.5em minus 0.4em\relax John Wiley \& sons, 1991.

\bibitem{XieBarron00}
Q.~Xie and A.~Barron, ``Asymptotic minimax regret for data compression,
  gambling, and prediction,'' \emph{IEEE Trans. Information Theory}, vol.~46,
  no.~2, pp. 431 --445, Mar. 2000.

\end{thebibliography}

\end{document}
\fi 